\documentclass[longauth]{aa} 

\usepackage{graphicx}
\usepackage{txfonts}
\usepackage{mathabx}
\usepackage[normalem]{ulem}
\usepackage{color}
\usepackage[dvipsnames]{xcolor}
\usepackage{soul}

\def \MJ{M$_{\mathrm{Jup}}$}
\def \MN{M$_{\mathrm{Nep}}$}
\def \ME{M$_{\Earth}$}
\def \RJ{R$_{\mathrm{Jup}}$}
\def \RN{R$_{\mathrm{Nep}}$}
\def \RE{R$_{\Earth}$}
\def \RS{R$_{\odot}$}
\def \msol{M$\mathrm{_\odot}$}
\def \lsol{L$\mathrm{_\odot}$}

\def \kms{km\,s$^{-1}$}
\def \ms{m\,s$^{-1}$}
\def \1s{$1\,\sigma$}

\def \t0{T$_0$}

\begin{document} 

   \title{TOI-1759~b: a transiting sub-Neptune around a low mass star characterized with SPIRou\thanks{ Based on observations obtained at the Canada-France-Hawaii Telescope (CFHT) which is operated from the summit of Maunakea by the National Research Council of Canada, the Institut National des Sciences de l'Univers of the Centre National de la Recherche Scientifique of France, and the University of Hawaii. Based on observations obtained with SPIRou, an international project led by Institut de Recherche en Astrophysique et Plan\'etologie, Toulouse, France.} and TESS}
   \titlerunning{Detection and characterization of TOI-1759~b}
   \author{ E. Martioli 
            \inst{1,2} \and
    G. H\'ebrard \inst{1,3} \and
    P. Fouqu\'e \inst{4} \and   
    Artigau, \'E.\inst{5} \and
    J.-F. Donati \inst{4} \and
    C. Cadieux \inst{5} \and
    S. Bellotti \inst{4} \and
    A. Lecavelier des Etangs \inst{1} \and
    R. Doyon \inst{5} \and
    J.-D. do Nascimento Jr. \inst{10} \and
    L. Arnold \inst{7} \and
    A. Carmona \inst{6} \and
    N. J. Cook \inst{5} \and
    P. Cortes-Zuleta \inst{8} \and 
    L. de Almeida \inst{2,10} \and
    X. Delfosse \inst{6} \and
    C. P. Folsom \inst{36,37} \and
    P.-C. K\"{o}nig \inst{9,1} \and
    C. Moutou \inst{4} \and
    M. Ould-Elhkim \inst{4} \and
    P. Petit \inst{4} \and
    K. G. Stassun \inst{13} \and
    A. A. Vidotto \inst{30} \and
    T. Vandal \inst{5} \and
    B. Benneke \inst{35} \and
    I. Boisse \inst{8} \and
    X. Bonfils \inst{6} \and
    P. Boyd \inst{16} \and
    C. Brasseur \inst{17} \and
    D. Charbonneau \inst{19}\and
    R. Cloutier \inst{19} \and
    K. Collins \inst{29}\and
    P. Cristofari \inst{4}\and
    I. Crossfield \inst{32} \and
    R. F. D\'{i}az \inst{34} \and
    M. Fausnaugh \inst{18} \and
    P. Figueira \inst{25,26} \and
    T. Forveille \inst{6} \and
    E. Furlan \inst{31} \and
    E. Girardin \inst{15} \and
    C. L. Gnilka \inst{20,31} \and
    J. Gomes~da~Silva \inst{26} \and    
    P.-G. Gu \inst{28} \and  
    P. Guerra \inst{14} \and
    S. B. Howell \inst{20} \and
    G. A. J. Hussain \inst{12} \and
    J. M. Jenkins \inst{20} \and
    F. Kiefer \inst{22} \and
    D. W. Latham \inst{19} \and
    R. A. Matson \inst{27} \and
    E. C. Matthews \inst{33} \and
    J. Morin \inst{11} \and
    R. Naves \inst{14} \and
    G. Ricker \inst{18} \and
    S. Seager \inst{18,23,24} \and
    M. Takami \inst{28} \and
    J. D. Twicken \inst{29,20} \and
    A. Vanderburg \inst{18} \and
    R. Vanderspek \inst{18} \and
    J. Winn \inst{21}
    }
   \institute{
   \inst{1} Institut d'Astrophysique de Paris, CNRS, UMR 7095, Sorbonne Universit\'{e}, 98 bis bd Arago, 75014 Paris, France, \email{martioli@iap.fr} \\
   \inst{2} Laborat\'{o}rio Nacional de Astrof\'{i}sica, Rua Estados Unidos 154, 37504-364, Itajub\'{a} - MG, Brazil \\
   \inst{3} Observatoire de Haute Provence, St Michel l'Observatoire, France \\
   \inst{4} Institut de Recherche en Astrophysique et Plan\'{e}tologie, Universit\'{e} de Toulouse, CNRS, IRAP/UMR 5277, 14 avenue Edouard Belin, F-31400, Toulouse, France \\
   \inst{5} Universit\'{e} de Montr\'{e}al, D\'{e}partement de Physique, IREX, Montr\'{e}al, QC H3C 3J7, Canada \\
   \inst{6} Universit\'{e} Grenoble Alpes, IPAG, 38000, Grenoble, France; CNRS, IPAG, 38000, Grenoble \\
   \inst{7} Canada-France-Hawaii Telescope, CNRS, 96743 Kamuela, Hawaii, USA \\
   \inst{8} Aix Marseille Univ, CNRS, CNES, LAM, Marseille, France \\
   \inst{9} European Southern Observatory, Karl-Schwarzschild-Straße 2, 85748 Garching, Germany  \\
   \inst{10} Departamento de F\'{i}sica Te\'{o}rica e Experimental, Universidade Federal do Rio Grande do Norte, Natal, RN 59072-970, Brazil \\
   \inst{11} Universit\'e de Montpellier, CNRS, LUPM,34095 Montpellier, France \\
   \inst{12} Science Division, Directorate of Science, European Space Research and Technology Centre (ESA/ESTEC), Keplerlaan 1, 2201 AZ, Noordwijk, The Netherlands \\
   \inst{13} Department of Physics and Astronomy, Vanderbilt University, Nashville, TN 37235, USA \\
   \inst{14} Observatori Astron\`{o}mic Albany\`{a}, Cam\'{i} de Bassegoda S/N, Albany\`{a} 17733, Girona, Spain \\
   \inst{15} Grand Pra Observatory, 1984 Les Haud\`{e}res, Switzerland \\
   \inst{16} Astrophysics Science Division, NASA Goddard Space Flight Center, Greenbelt, MD 20771, USA \\
   \inst{17} Space Telescope Science Institute, 3700 San Martin Drive, Baltimore, MD, 21218, USA \\
   \inst{18} Department of Physics and Kavli Institute for Astrophysics and Space Research, Massachusetts Institute of Technology, Cambridge, MA 02139, USA \\
   \inst{19} Center for Astrophysics | Harvard \& Smithsonian, 60 Garden Street, Cambridge, MA, 02138, USA \\
   \inst{20} NASA Ames Research Center, Moffett Field, CA 94035, USA \\
   \inst{21} Department of Astrophysical Sciences, Peyton Hall, 4 Ivy Lane, Princeton, NJ 08544, USA \\
   \inst{22} LESIA, Observatoire de Paris, Universit\'{e} PSL, CNRS, Sorbonne Universit\'{e}, Universit\'{e} de Paris, 5 place Jules Janssen, 92195 Meudon, France \\
   \inst{23} Department of Earth, Atmospheric and Planetary Sciences, Massachusetts Institute of Technology, Cambridge, MA 02139, USA \\
   \inst{24} Department of Aeronautics and Astronautics, MIT, 77 Massachusetts Avenue, Cambridge, MA 02139, USA \\
   \inst{25} European Southern Observatory, Alonso de Cordova 3107, Vitacura, Santiago, Chile \\
   \inst{26} Instituto de Astrof\'{i}sica e Ci\^{e}ncias do Espa\c{c}o, Universidade do Porto, CAUP, Rua das Estrelas, 4150-762 Porto, Portugal \\
   \inst{27} U.S. Naval Observatory, Washington, D.C. 20392, USA \\
   \inst{28} Institute of Astronomy and Astrophysics, Academia Sinica 11F of Astronomy-Mathematics Building No.1, Sec. 4, Roosevelt Rd, Taipei 10617, Taiwan, R.O.C. \\
   \inst{29} SETI Institute, 189 Bernardo Ave., Suite 200, Mountain View, CA 94043, USA \\
   \inst{30} Leiden Observatory, Leiden University, PO Box 9513, 2300 RA Leiden, The Netherlands \\
   \inst{31} NASA Exoplanet Science Institute, Caltech IPAC, 1200 E. California Blvd., Pasadena, CA 91125, USA \\
   \inst{32} Physics \& Astronomy Department, University of Kansas, Lawrence, KS, USA \\
   \inst{33} Observatoire de l’Universit\'e de Gen\`eve, Chemin Pegasi 51, 1290 Versoix, Switzerland \\
   \inst{34} International Center for Advanced Studies (ICAS) and ICIFI (CONICET), ECyT-UNSAM, Campus Miguelete, 25 de Mayo y Francia, (1650), Buenos Aires, Argentina \\
   \inst{35} Institut de Recherche sur les Exoplan\`etes, D\'epartement de Physique, Universit\'e de Montr\'eal, 1375 Avenue Th\'er\`ese-Lavoie-Roux, Montreal, QC H2V 0B3, Canada \\
   \inst{36} Department of Physics \& Space Science, Royal Military College of Canada, PO Box 17000 Station Forces, Kingston, ON, Canada K7K 0C6 \\
   \inst{37} Tartu Observatory, University of Tartu, Observatooriumi 1, T\~{o}ravere, 61602 Tartumaa, Estonia \\ 
   }
    
   \date{Received October 28, 2021; accepted February 02, 2022}

  \abstract { 
  We report the detection and characterization of the transiting sub-Neptune TOI-1759~b, using photometric time-series from the Transiting Exoplanet Survey Satellite (TESS) and near infrared spectropolarimetric data from the Spectro-Polarimètre Infra Rouge (SPIRou) on the Canada-France-Hawaii Telescope (CFHT). 
  TOI-1759~b orbits a moderately active M0V star with an orbital period of $18.849975\pm0.000006$~d, and we measure a planetary radius and mass of $3.06\pm0.22$~\RE\ and $6.8\pm2.0$~\ME. Radial velocities were extracted from the SPIRou spectra using both the cross-correlation function (CCF) and the line-by-line (LBL) methods, optimizing the velocity measurements in the near infrared domain. We analyzed the broadband spectral energy distribution of the star and the high-resolution SPIRou spectra to constrain the stellar parameters and thus improve the accuracy of the derived planet parameters. A least squares deconvolution (LSD) analysis of the SPIRou Stokes~$V$ polarised spectra detects Zeeman signatures in TOI-1759. We model the rotational modulation of the magnetic stellar activity using a Gaussian Process regression with a quasi-periodic covariance function, and find a rotation period of $35.65^{+0.17}_{-0.15}$~d. We reconstruct the large-scale surface magnetic field of the star using Zeeman-Doppler imaging, which gives a predominantly poloidal field with a mean strength of $18\pm4$~G. Finally, we perform a joint Bayesian MCMC analysis of the TESS photometry and SPIRou radial velocities to optimally constrain the system parameters. At $0.1176\pm0.0013$~au from the star, the planet receives $6.4$ times the bolometric flux incident on Earth, and its equilibrium temperature is estimated at $433\pm14$~K.  
  TOI-1759~b is a likely gas-dominated sub-Neptune with an expected high rate of photoevaporation. Therefore, it is an interesting target to search for neutral hydrogen escape, which may provide important constraints on the planetary formation mechanisms responsible for the observed sub-Neptune radius desert.
  }

   \keywords{stars: planetary systems --  stars: individual: TOI-1759 --  stars: magnetic field -- techniques: photometric, radial velocity}

   \maketitle
%
\section{Introduction}

The characterization of transiting planets in the $2-4$~\RE\ regime provides important constraints on the formation and evolution processes responsible for the observed scarcity of planets with radii between $1.7$ and $2.0$~\RE\, also known as the radius gap \citep{FultonAndPetigura2018}. This gap separates two classes of planets, the rocky super-Earths and the lower density sub-Neptunes whose bulk compositions may be primarily composed of rocky cores enveloped in H/He gas \citep{Owen2017,Rogers2021}, or rocky cores plus a comparable mass of water ice \citep{Zeng2019,Venturini2020}. This bimodality of planet compositions around Sun-like stars is likely explained by thermally driven mass loss either via photoevaporation \citep{Lecavelier_2007,Owen2013,Lopez2013} or by the luminosity of the cooling core \citep{Ginzburg2018,Gupta2021}. However, around the lower mass M dwarfs, the dependence of the radius gap on insolation suggests that the gap may be a direct outcome of the planet formation process without the need to invoke a subsequent mass loss process \citep{CloutierMenou2020}. The dominant physics that sculpts the radius gap around M dwarfs remains unknown and requires the detailed characterization of more planets that span the radius gap across a range of host stellar masses. \cite{Loyd2020} estimated that confirming or ruling out photoevaporation as the primary cause of the exoplanet radius gap requires roughly doubling the current sample of well-characterized $<4$~\RE\ planets. Finding additional transiting planets in this size regime and characterizing those planets and their host stars is therefore crucial to understanding the planetary formation process. 

Here we present the detection and characterization of TOI-1759~b, a new sub-Neptune orbiting the high proper motion M0V star TOI-1759 (TIC 408636441, TYC 4266-736-1). TOI-1759 was first identified as a TESS object of interest (TOI) by the Transiting Exoplanet Survey Satellite \citep[TESS,][]{tess_paper}, which detected recurrent transit-like events in its light curve \citep{Jenkins2002,Jenkins2010}. The transit signature was fitted with an initial limb-darkened transit model \citep{Li:DVmodelFit2019} and subjected to a suite of diagnostic tests \citep{Twicken:DVdiagnostics2018}, all of which it passed. The TESS Science Office reviewed the data validation reports and issued an alert of a possible planet candidate \citep{Guerrero2021}. We subsequently followed-up TOI-1759 within the SPIRou Legacy Survey - Follow-up of Transiting Exoplanets \citep[SLS-WP2,][]{donati2020} with the Spectro-Polarimètre Infra Rouge (SPIRou) instrument coupled to the 3.6m Canada-France-Hawaii Telescope (CFHT). The TESS photometry constrains the planet size, orbital inclination and period, while the high-resolution near infrared polarimetric spectra of SPIRou establish its planetary nature and constrain its mass, orbital eccentricity and mean density. They also constrain the properties of the host star, showing that it has a low mass and moderate levels of magnetic activity.

This paper is organized as follows. Sect. \ref{sec:observations} describes the observations. Sect. \ref{sec:reductionandanalysis} describes the data analysis methods employed to obtain the high-resolution template spectrum of TOI-1759, spectropolarimetry, and precise velocimetry with SPIRou. Sect. \ref{sec:star} presents the derivation of the stellar parameters and the characterization of the stellar magnetic field. Sect. \ref{sec:analysis} constrains the planet parameters through a  simultaneous Bayesian MCMC analysis of the photometry and radial velocity data. Sect. \ref{sec:discussion} discusses the insolation and the atmospheric properties of this new planet, and Sect. \ref{sec:conclusions} concludes. 

\section{Observations}
\label{sec:observations}

\subsection{TESS photometry}
\label{sec:tessphotometry}

\begin{figure}
   \centering
       \includegraphics[width=1.0\hsize]{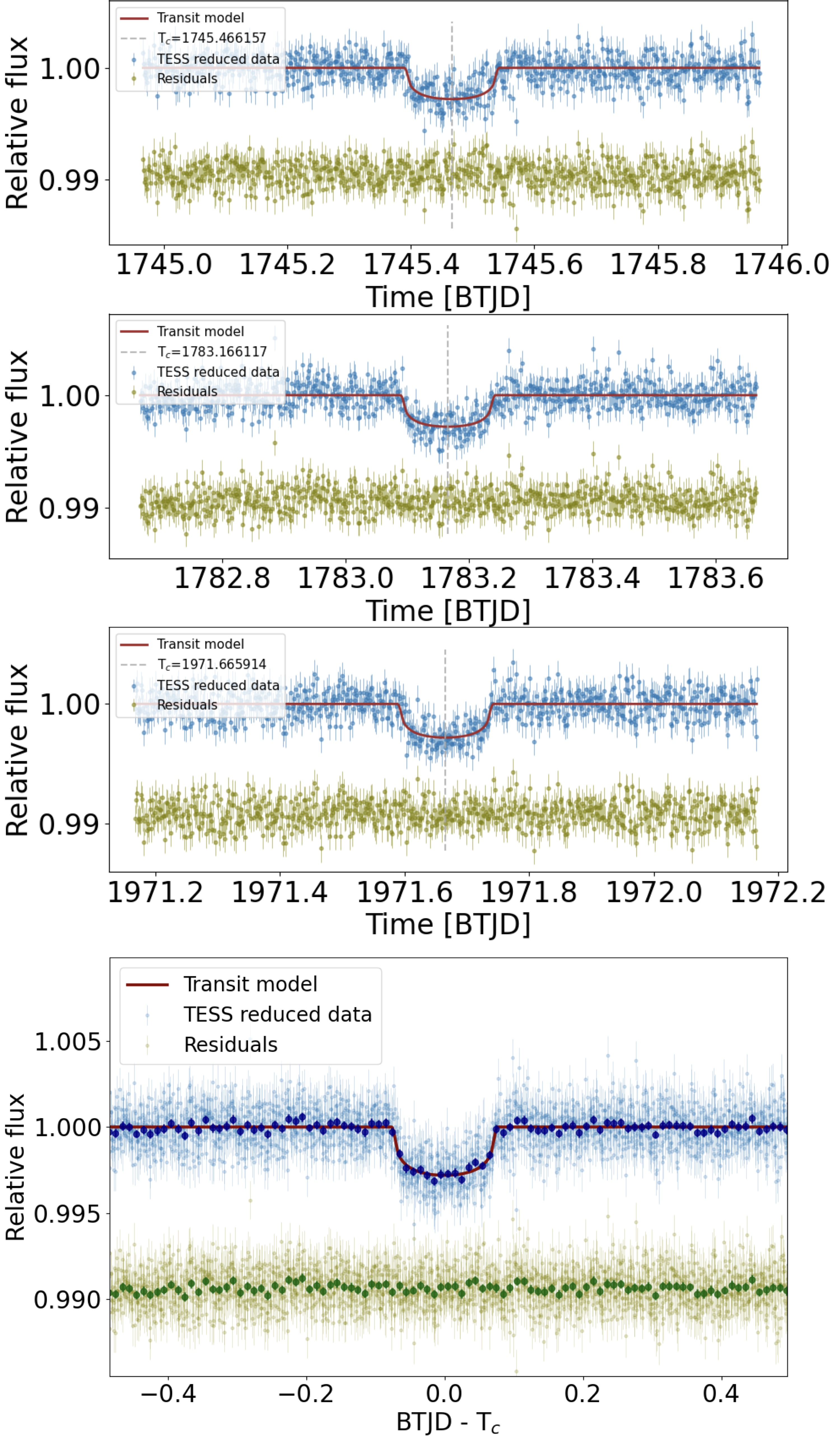}
      \caption{The blue points show the TESS photometry data around the three transits of TOI-1759~b, where the bottom panel shows all data with the times being relative to the central time of each transit. The red lines show the best-fit transit model and the green points show the residuals.
      }
        \label{fig:toi1759_transits_fit}
  \end{figure}

TESS observed TOI-1759 with a cadence of 2 minutes in Sectors 16 and 17 (September to November 2019) and in Sector 24 (April to May 2020), as detailed in Table \ref{tab:tessobservations}. Our analysis uses TESS data products obtained from the Mikulski Archive for Space Telescopes (MAST)\footnote{\url{mast.stsci.edu}}. We used the Presearch Data Conditioning (PDC) flux time series \citep{Smith2012,Stumpe2012,Stumpe2014} processed by the TESS Science Processing Operations Center (SPOC) pipeline \citep{jenkinsSPOC2016,Caldwell2020} versions listed in Table \ref{tab:tessobservations}.  The SPOC pipeline provides a Data Validation (DV) report \footnote{https://outerspace.stsci.edu/display/TESS/2.0+-+Data+Product+Overview} for assessment of the detected transit events. The DV reports for TOI-1759 in sectors 16-24 show the detection of three transit events with depth of $0.27\pm0.01$\% and a period of $37.696\pm0.002$~d. Half period (18.850~d) was also possible from the TESS data alone, and finally turned to be the correct period (see Sect. \ref{sec:groundbasedphotometry}). Figure \ref{fig:toi1759_transits_fit} shows the three TESS transit light curves. The DV reports also include a difference imaging centroid test that locates the origin of transits to within $2\pm5$~arcsec; all stars in the TESS Input Catalog (TIC) within the confusion radius for this test are fainter than $T_{\rm mag}>17$, which is too faint for an eclipsing binary to explain the transit signature. We employed the statistical validation method of \cite{Giacalone2021} to calculate the False Positive Probability (FPP) that the transits of TOI-1759 observed by TESS are of planetary nature. We used the tool \texttt{TRICERATOPS}\footnote{\url{https://github.com/stevengiacalone/triceratops}}, where we obtained a FPP of 0.31\%. \texttt{TRICERATOPS} also considers the flux contribution to the photometry of all sources within a radius of $\sim200$~arcsec surrounding the target to estimate the blended scenario as the origin of the transit events. We obtained an almost null value ($6\times10^{-88}$) for the Nearby False Positive Probability (NFPP). According to the validation criteria of \cite{Giacalone2021}, these values of FPP and NFPP place the candidate planet TOI-1759~b in the ``VALIDATED PLANET'' regime where FPP$<1.5$\% and NFPP$<0.1$\%.

\begin{table*}
\centering
\tiny
\caption{Log of TESS observations.}
\label{tab:tessobservations}
\begin{tabular}{ccccccc}
\hline
TSTART (UTC) & TSTOP (UTC) & Duration (d) & Sector & Camera & CCD & SPOC version \\
\hline
2019-09-12T03:40:24.448 & 2019-10-06T19:42:46.276 & 25.0 & 16 & 2 & 4 & 4.0.28-20200407 \\
2019-10-08T04:26:46.051 & 2019-11-02T04:42:30.214 & 24.7 & 17 & 3 & 4 & 4.0.28-20200407 \\
2020-04-16T06:59:59.027 & 2020-05-12T18:40:29.338 & 26.5 & 24 & 4 & 2 & 4.0.36-20200520 \\
\hline
\end{tabular}
\end{table*} 

\subsection{Ground-based photometry}
\label{sec:groundbasedphotometry}

\begin{figure}
   \centering
   \includegraphics[width=0.9\hsize]{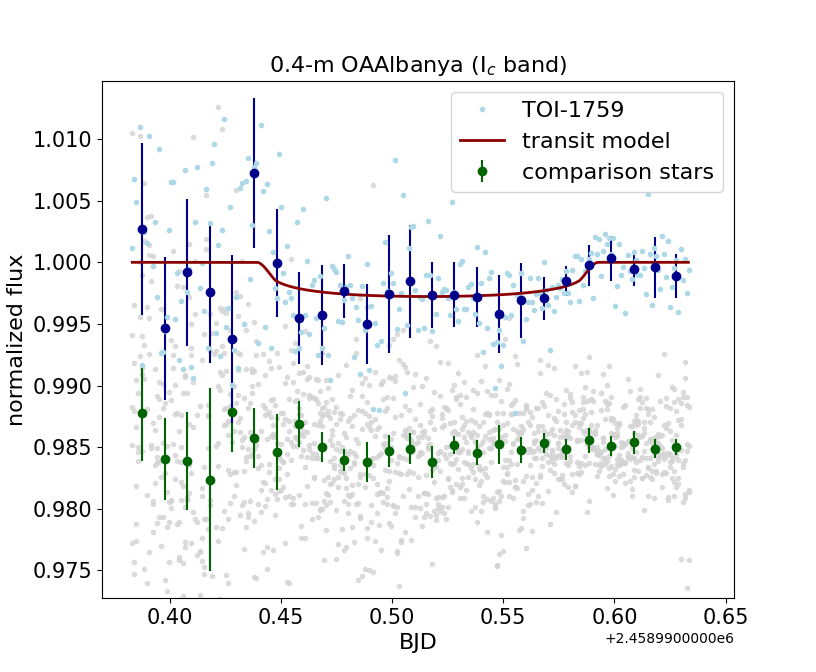}
      \caption{Ground-based I$_{\rm c}$-band differential photometry time series of a transit of TOI-1759~b obtained by the 0.4-m OAAlbanya observatory on May 20, 2020. The light blue points show relative fluxes of TOI-1759 and the dark blue points show weighted average bins with bin size of 0.01~d. The grey and green points show the relative fluxes for the comparison stars that we used in the differential photometry. The red line shows the best-fit transit model obtained from our analysis of the TESS data alone, as presented in Section \ref{sec:analysis}. }
        \label{fig:groundbasedphotometry}
  \end{figure}

We obtained ground-based photometric observations of a single transit of TOI-1759~b on May 20, 2020, using three different telescopes: the 0.3 m OMontcabrer (r band), the 0.4~m RCO (i' band) and the 0.4~m OAAlbanya-0m4 (band I$_{\rm c}$). The latter observed a full transit showing a good agreement with the transit model evidenced mainly during the egress, as illustrated in Fig. \ref{fig:groundbasedphotometry}. These observations are also reported on The Exoplanet Follow-up Observing Program for TESS (ExoFOP-TESS) website\footnote{\url{https://exofop.ipac.caltech.edu/tess/target.php?id=408636441}} by Guerra and Girardin from the TFOP Working Group. Although we detected the transit of TOI-1759~b, we did not included these data in our analysis. However, they are valuable in establishing that the orbital period of TOI-1759~b is 19 days rather than 38 days, which was also compatible with the TESS data alone. 

\subsection{High constrast imaging}

\begin{figure}
   \centering
   \includegraphics[width=0.9\hsize]{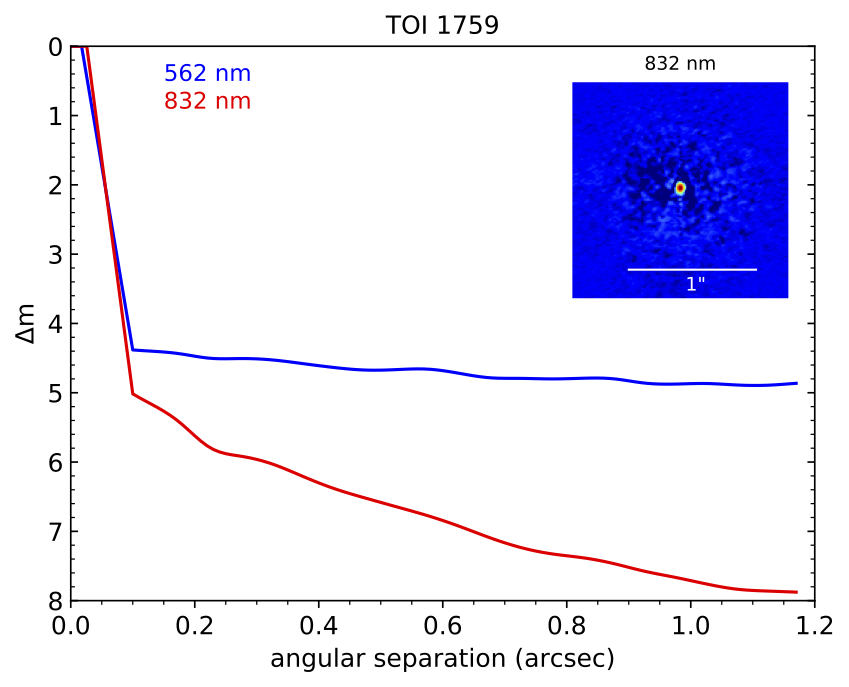}
      \caption{Contrast ratio of TOI-1759 as a function of angular separation at 562~nm (blue line) and at 832~nm (red line) obtained from the `Alopeke/Gemini speckle imaging observations.
       }
        \label{fig:gemini_AO_constrast}
  \end{figure}

High-angular resolution observations can probe close companions within  $\sim1.2$~arcsec that would create a false positive transit signal (if that companion is an eclipsing binary) and which dilute the transit signal and thus yield underestimated planet radii \citep{Ciardi2015}.  TOI-1759 was observed on June 13, 2020 by the `Alopeke dual-channel speckle imaging instrument on Gemini-N (PI: Crossfield) with a pixel scale of 0.01~arcsec/pixel and a full width at half maximum (FWHM) resolution of 0.02~arcsec. `Alopeke provided simultaneous speckle imaging at 562 and 832~nm.  Five sets of $1000\times0.06$~s exposures were taken and processed with the speckle pipeline \citep{Howell2011}, which yielded the 5-sigma sensitivity curves and reconstructed images shown in Fig. \ref{fig:gemini_AO_constrast}.  These observations provide a contrast at an angular separation of 0.5~arcsec of 4.67~mag at 562~nm and 6.58~mag at 832~nm (Fig. \ref{fig:gemini_AO_constrast}). The ExoFOP-TESS website also reports that TOI-1759 was observed with the NIRC2 near-infrared camera and the adaptive optics system of the 10-m Keck II telescope on September 9, 2020 (PI: Gonzales) with a pixel scale of 0.01~arcsec/pixel and a PSF FWHM of 0.05 arcsec, providing a contrast at 0.5~arcsec separation of 6.77~mag in Br$\gamma$. We included the 832~nm contrast curve in the \texttt{TRICERATOPS} calculation to further constrain the FPP, which now gives a value of FPP$<0.03$~\%. Therefore, these high-contrast imaging observations set strong upper limits against any close companion or close-by field star that could significantly contribute to the observed flux of TOI-1759.

\subsection{SPIRou spectropolarimetry}

TOI-1759 was observed by SPIRou\footnote{ \url{http://spirou.irap.omp.eu} and \url{https://www.cfht.hawaii.edu/Instruments/SPIRou/}} under the large program SLS-WP2\footnote{\url{http://spirou.irap.omp.eu/Observations/The-SPIRou-Legacy-Survey}} (id P42, PI: Jean-Fran\c{c}ois Donati). SPIRou is a stabilized high-resolution near-infrared (NIR) spectropolarimeter \citep{donati2020} mounted on the 3.6~m CFHT atop Maunakea, Hawaii. It is designed for high-precision velocimetry to detect and characterize exoplanets and it provides a full coverage of the NIR spectrum from 950~nm to 2500~nm at a spectral resolving power of $\lambda / \Delta \lambda \sim 70000 $. 

A total of 218 spectra of TOI-1759 were obtained on 54 different epochs/nights, spanning 447 days from 2020-06-05T13:16:41 to 2021-08-26T08:51:39. Table \ref{tab:spirouobservations} in Appendix \ref{app:spiroudata} presents the log of our SPIRou observations. These observations were carried out in the circular polarization mode (Stokes~V), where each set of four exposures provides a polarimetric spectrum. Each exposure in the sequence corresponds to a different position of the two rotating Fresnel rhombs, where the sequence number of each exposure is also presented in Table \ref{tab:spirouobservations}. On a few occasions, one exposure needed to be repeated due to passing clouds, implying more than four exposures per sequence. Under these circumstances, we select the set of four exposures with the highest signal-to-noise ratio (SNR). These observations were obtained at an average air mass of 1.47 with a dispersion of 0.12, and with a SNR per spectral element measured at 1670~nm ranging from 43 to 210, with a median of 184. The fourth exposure of one polarimetric sequence on 2021-06-19 has a low SNR, and could not be repeated due to degrading weather conditions. We therefore do not consider this sequence for polarimetry, but do use its three good exposures for spectroscopy.

A set of baseline calibration (flats, darks, comparison and aligns) is obtained in the afternoon and in the morning of each night of observation with SPIRou. In addition, hot stars (A type) are observed nightly as telluric absorption standards. A set of bright inactive cool stars are also regularly observed as constant radial velocity (RV) standards. We make use of these data to calibrate the measurements we extract from the SPIRou spectra, as discussed in more detail in Sect \ref{sec:reductionandanalysis}.

\section{SPIRou data reduction and analysis}
\label{sec:reductionandanalysis}

\subsection{APERO reduction}
\label{sec:aperoreduction}

Our SPIRou data was reduced with the software {\it A PipelinE to Reduce Observations} (APERO\footnote{\url{https://github.com/njcuk9999/apero-drs}} v0.6.132, Cook et al., in prep.). APERO first performs some initial processing of the 4096$\times$4096~pixel images of the HAWAII 4RG$^{\rm \tiny TM}$ (H4RG), applying a series of procedures to correct detector effects, remove background thermal noise, and identify bad pixels and cosmic ray impacts. 

It then uses exposures of a quartz halogen lamp (flat) to calculate the position of the 49 echelle spectral orders. It optimally extracts \citep{horne86} spectra of the two science channels (fibers A and B, fed by two orthogonal polarized beams) and the simultaneous calibration channel (fiber C). This APERO extraction takes into account the asymmetric shape of the instrument profile generated by the pupil slicer. Both a 2-D order by order and 1-D order-merged spectrum are produced for each channel of each scientific exposure. A blaze function is obtained from the flat-field exposures and a master flat is used to do the flux calibration. 

The pixel-to-wavelength calibration is obtained from exposures of both a UNe hollow cathode lamp and a Fabry-Pérot etalon, as described in \citet{Hobson2021}. This provides wavelengths in the rest frame of the observatory, but APERO also calculates the Barycentric Earth Radial Velocity (BERV) and the Barycentric Julian Date (BJD) of each exposure using the code \texttt{barycorrpy}\footnote{\url{https://github.com/shbhuk/barycorrpy}}\citep{Kanodia2018,Wright2014}. These can then be used to reference the wavelength and time to the barycentric frame of the solar system. 

APERO calculates the spectrum of the telluric transmission using a novel technique based on a model obtained from the collection of standard star observations carried out since the beginning of SPIRou operations in 2018 and a fit  made for each individual observation using the principal component analysis (PCA) technique of \cite{artigau14}. APERO also calculates the Stokes V (and where appropriate, Q and U) spectra using the method of \cite {donati1997}, as described in detail in \cite{Martioli2020}.

\subsection{Spectropolarimetry analysis}
\label{sec:spectropolarimetry}

\begin{figure}
   \centering
   \includegraphics[width=0.9\hsize]{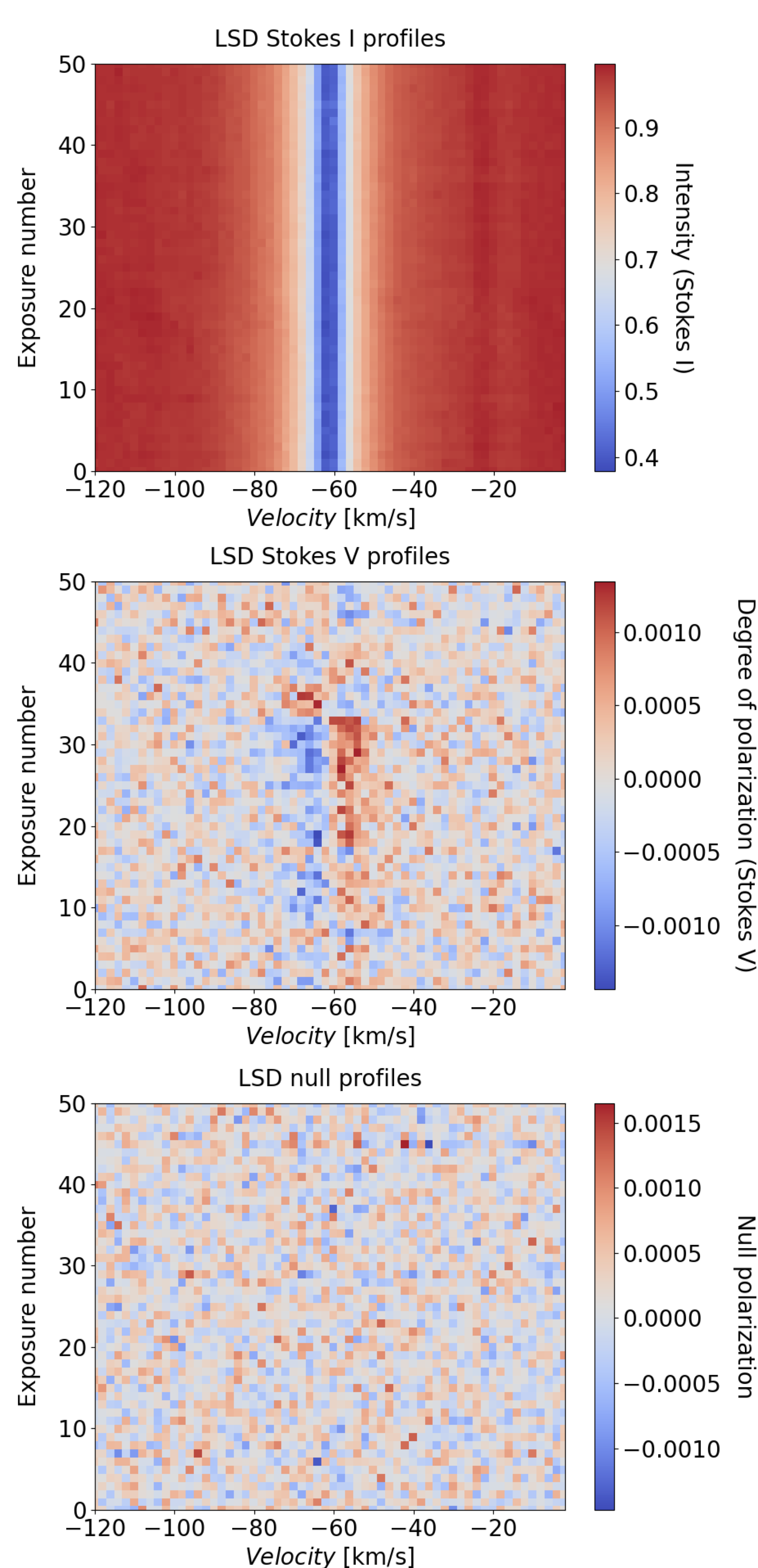}
      \caption{Stokes I (top panel), Stokes V (middle panel), and null polarization (bottom panel) LSD profiles in the TOI-1759 SPIRou time series. 
       }
        \label{fig:spirou-lsd-profiles-timeseries}
  \end{figure}

\begin{figure}
   \centering
   \includegraphics[width=0.9\hsize]{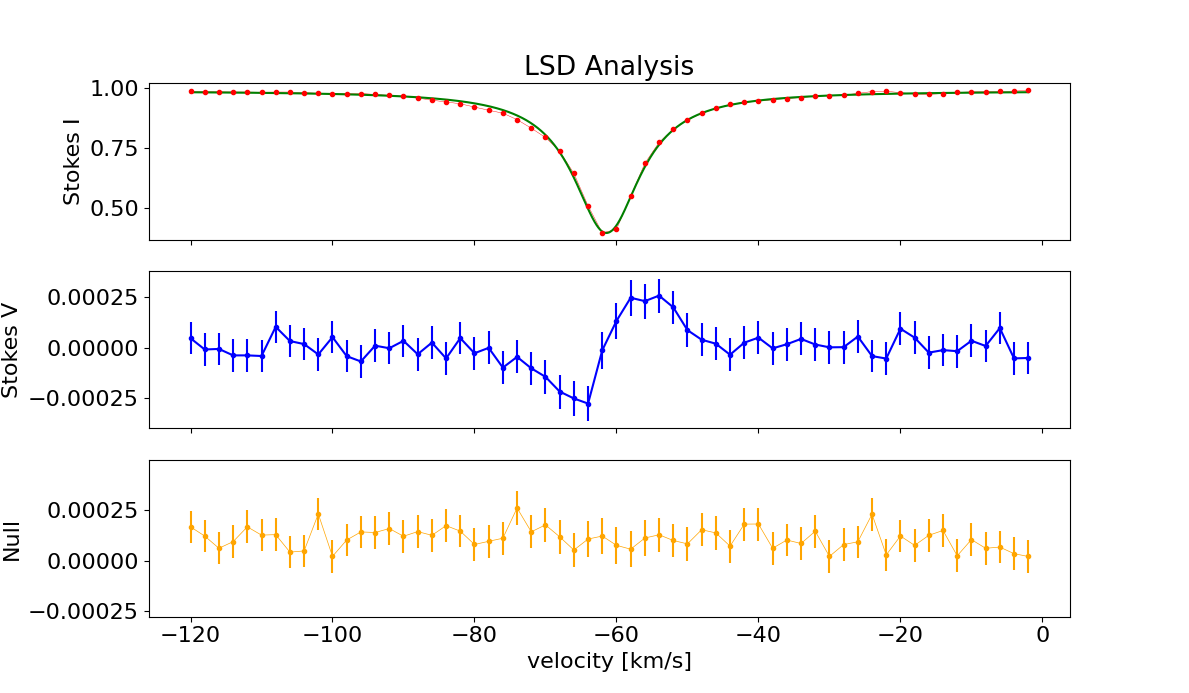}
      \caption{Median of all LSD profiles in the TOI-1759 SPIRou time series. The top panel shows Stokes I LSD (red points) with a Voigt profile model fit (green line); the middle panel shows Stokes V (blue points) and the bottom panel shows the null polarization profile (orange points). 
       }
        \label{fig:spirou-lsd-profiles}
  \end{figure}

We further analyze the SPIRou polarized spectra using the \texttt{spirou-polarimetry}\footnote{\url{https://github.com/edermartioli/spirou-polarimetry}} code. The Stokes I, Stokes V, and null polarization spectra are compressed to one line profile using the least squares deconvolution (LSD) method of \cite{donati1997}. The line mask used in our LSD analysis of TOI-1759 was computed using the VALD catalog \citep{piskunov1995} and a MARCS model atmosphere \citep{Gustafsson2008} with an effective temperature of 4000~K and surface gravity of $\log g=5.0$~dex. We select all lines deeper than 3\% and with a Land\'{e} factor of $g_{\rm eff}>0$, for a total of 2460 atomic lines. Fig. \ref{fig:spirou-lsd-profiles-timeseries} displays the resulting LSD profiles at each observing epoch. Its Stokes V panel shows a significant and time-variable Zeeman signature, indicating the presence of magnetic field in this star, as will be explored in more detail in Sect. \ref{sec:stellaractivity}. Fig. \ref{fig:spirou-lsd-profiles} shows the medians of all profiles in the time series, where the Zeeman signature is clearly evidenced in the 'S' shape of the Stokes V profile.

To check the consistency of our measurements, we also obtained an independent polarimetric reduction and LSD analysis of our SPIRou data using the Libre-Esprit (LE) pipeline \citep{donati1997,donati2020}. The Stokes V profiles obtained from the APERO reduction show an RMS of 0.0039\% (estimated in the outer regions of the profile, $|v - v_{\rm sys}| > 20$~\kms) and a semi-amplitude of the median profile of 0.031\%, thus a signal-to-noise ratio of SNR=8. The LE data show an RMS of 0.0016\% and a semi-amplitude of 0.0061\%, thus a SNR=4. The difference in the amplitude scale of these datasets is due to the different normalization factors adopted in the LSD analysis.  However, the factor of two in detection significance stems from the noise characteristics that result from different reduction methods. A thorough comparison between the two pipelines is beyond the scope of this paper. Nevertheless, we analyzed and compared the results obtained from the two datasets, as will be shown in Sects. \ref{sec:stellaractivity} and \ref{sec:zdianalysis}.

\subsection{Radial velocities}
\label{sec:radialvelocities}

Obtaining precise radial velocities (RV) from the Doppler shift of the stellar spectrum in the NIR is more challenging than in the optical domain. For instance, the NIR spectral domain of SPIRou is largely polluted by telluric absorption and sky emission lines, which affect measurements across the entire observed spectrum. The telluric correction inevitably changes the noise pattern and introduces more noise due to imperfect corrections.  On the other hand, SPIRou's NIR H4RG detector has artifacts such as evolving bad pixels, nonlinearity and persistence, which are not present in the same proportions in CCD detectors. The challenges faced by high-precision velocimetry in the NIR were noted in other instruments with characteristics similar to SPIRou \citep{Figueira2010, Carleo2016, Cale2019, Lafarga2020}. 

Here we use two different methods to measure the radial velocities in the SPIRou data. First, we employ the well-established cross-correlation function (CCF) method of \cite{pepe2002} with several specific data reduction procedures that are required to minimize the aforementioned problems found in the NIR observations, as presented in more detail in Sect. \ref{sec:ccfanalysis}. Then, we use a line-by-line (LBL) method proposed by \cite{Dumusque2018} and adapted by Artigau et al. (submitted), which is summarized in Sect. \ref{sec:lblanalysis}. The latter seems to be more robust to the noise introduced by the telluric correction and also to the other artifacts of NIR observations. 

\subsection{CCF analysis}
\label{sec:ccfanalysis}

\begin{figure*}
   \centering
   \includegraphics[width=0.9\hsize]{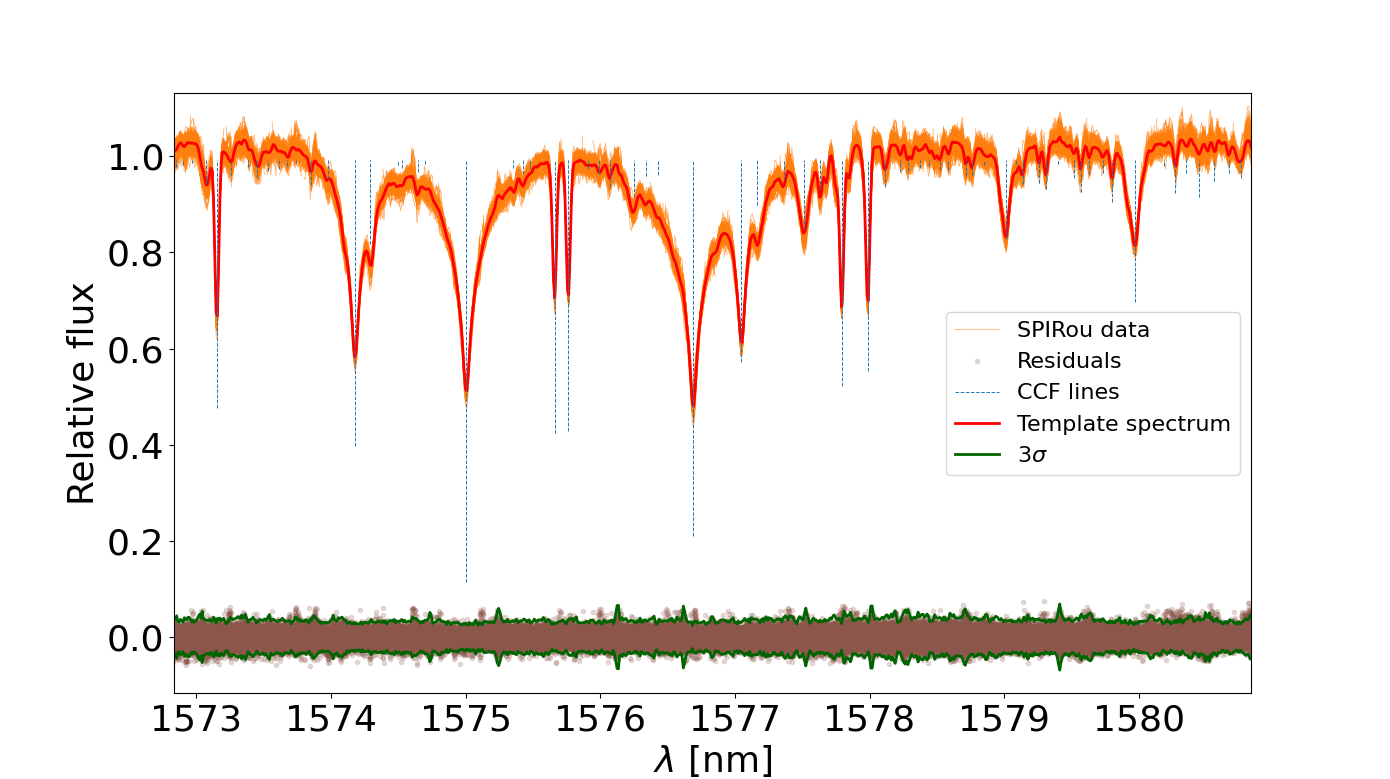}
      \caption{Example of the reduced SPIRou spectra of TOI-1759 in the CCF analysis. Orange lines show the normalized SPIRou spectra for a small range in the H band . The red line shows the template spectrum and the brown points show the residuals. The green lines show the measured $\pm3\sigma$ dispersion of the residuals, which are used by the sigma-clipping algorithm to reject outliers. The blue dashed lines show the central wavelengths of the CCF lines in the star's frame of reference, where the depth of these lines is proportional to the CCF weight.
       }
        \label{fig:spirou-spectrum-template}
  \end{figure*}

\begin{figure}
   \centering
   \includegraphics[width=0.9\hsize]{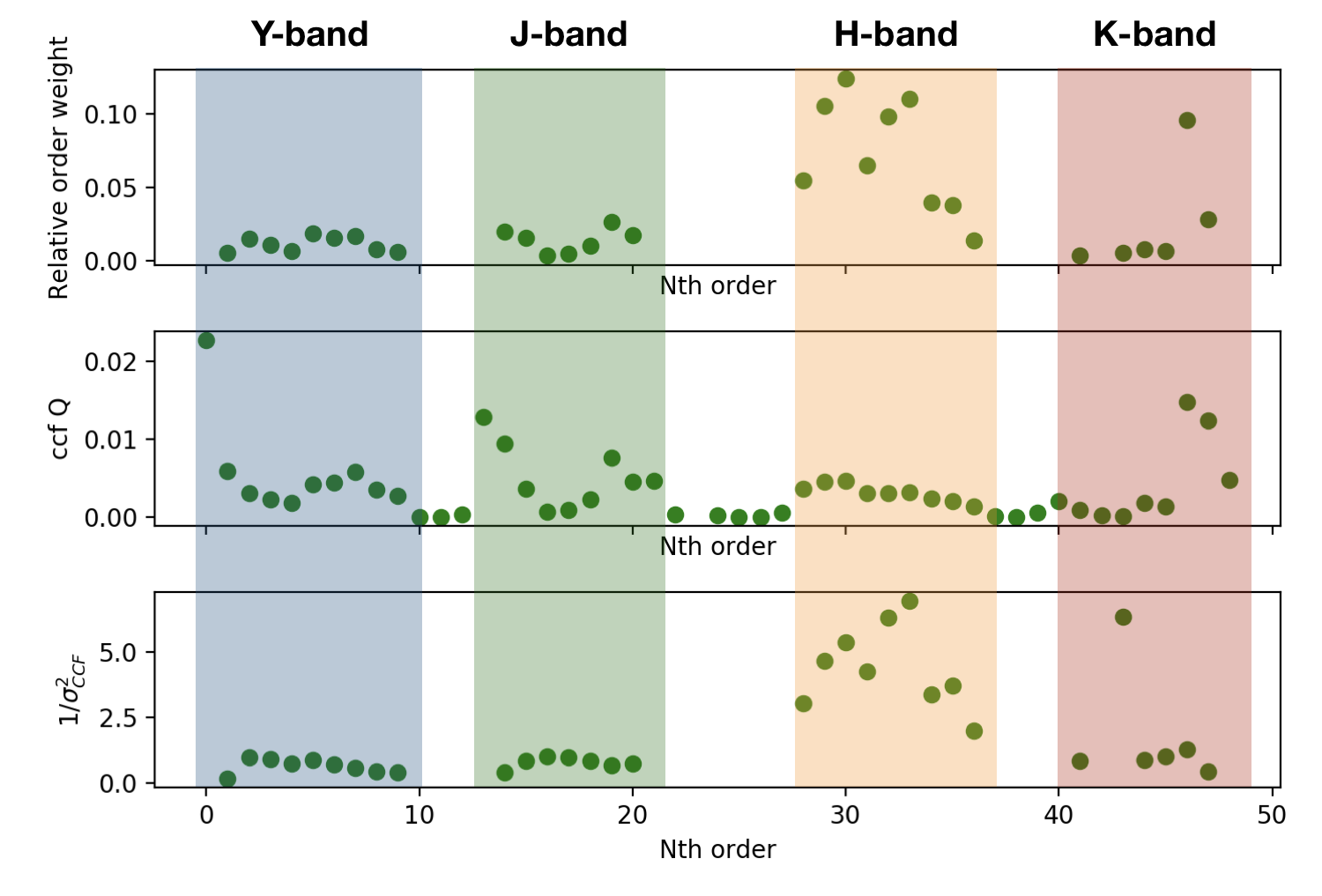}
      \caption{ 
      CCF order weights. The top panel shows the relative weights applied to the CCFs of the SPIRou spectra of TOI-1759 as a function of the order number.  Order numbering increases with wavelength starting at zero. Photometric NIR bands are also marked with different colors and indicated at the top. The middle panel shows the $Q$ factor that quantifies the radial velocity content and the lower panel shows the statistical weight as explained in the text.
       }
        \label{fig:spirou-ccfs-weights}
  \end{figure}

\begin{figure}
   \centering
   \includegraphics[width=0.9\hsize]{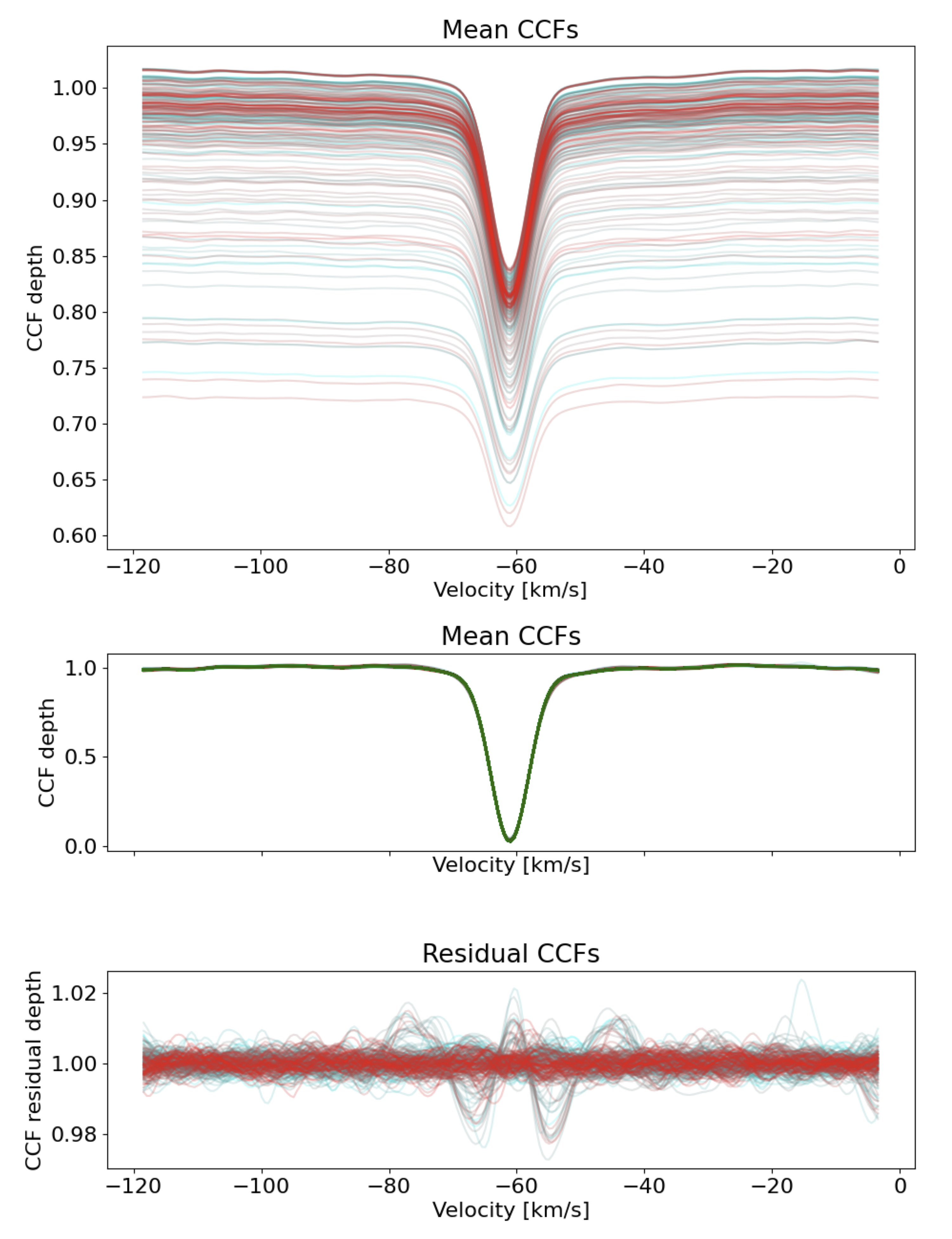}
      \caption{ 
      CCF template matching. The top panel shows the weighted mean CCFs obtained for each spectrum of TOI-1759, where the color code shows the first epoch in dark blue and the last epoch in dark red. The middle panel shows the same CCFs normalized to the polynomial fit to match the template CCF (green line). The lower panel shows the CCFs subtracted by the template CCF. 
       }
        \label{fig:spirou-ccfs}
  \end{figure}

For the CCF analysis we use the package \texttt{spirou-ccf}\footnote{\url{https://github.com/edermartioli/spirou-ccf}}, which implements the CCF method to measure the radial velocities of the SPIRou spectra. The input data that we use in our analysis are the telluric corrected spectra calculated by the APERO pipeline (see Sect. \ref{sec:aperoreduction}).  In the \texttt{spirou-ccf} package, several processing steps are done to minimize the strong systematic effects found in NIR data. Below is a summary of the procedures performed by our CCF analysis:

\begin{enumerate}
    \item\label{item:ccfmask} CCF MASK: select an empirical CCF mask from a repository of masks obtained from observations of bright stars, which in this case is Gl~846, an M0.5V star that almost corresponds to the spectral type of TOI-1759. Mask selection is based on the criterion of proximity to the spectral type of the star. Each mask consists of a set of atomic and molecular lines, where the central wavelengths are obtained from the VALD catalog \citep{piskunov1995} and the line depths are obtained empirically from the template spectra of bright stars observed by SPIRou;
    
    \item AVOID GAPS: mask out sparsely sampled spectral ranges, which are defined as the ranges with ``holes'' (data flagged as \texttt{NaN} due to bad pixels or failed telluric correction) greater than 5~km/s;
    
    \item\label{item:berv} BERV CORRECTION: apply a relativistic BERV correction to the wavelengths;
    
    \item RESAMPLING: re-sample each spectrum to a constant 1.8~\kms\ grid by cubic spline interpolation;
    
    \item \label{item:template} BUILD TEMPLATE SPECTRUM: combine all spectra into a high SNR template spectrum as illustrated in Fig. \ref{fig:spirou-spectrum-template};

    \item \label{item:templatematch} TEMPLATE MATCH: to account for flux variations, we fit a low-order multiplicative polynomial for the flux of each spectrum, $F(\lambda)$, on an order-by-order basis, as follows:
    
    \begin{equation}
        F(\lambda) = c_{0} + c_{1} F_{\rm T}(\lambda) + c_{2} \lambda^{2}
    \end{equation}
    
    where $F_{\rm T}$ is the flux of the template spectrum obtained in step \ref{item:template}, $c_{0}$ is a constant offset, $c_{1}$ is a scaling factor, and $c_{2}$ accounts for smooth wavelength ($\lambda$) dependent variability.

    \item \label{item:sigmaclip} SIGMA-CLIP: apply a $3\sigma$ clip filter with $\sigma$ being the median absolute deviation (MAD) of each spectral element in the time domain. Fig. \ref{fig:spirou-spectrum-template} illustrates the dispersion of residuals that gives $\sigma$; 

    \item  RE-BUILD TEMPLATE SPECTRUM: repeat step \ref{item:template} to obtain an improved template, and step \ref{item:templatematch} to minimize deviations between data obtained at different times and the template; 
    
    \item \label{item:templatecontinuum} CONTINUUM FIT: fit a polynomial to the continuum of each spectral order of the template spectrum by using an iterative sigma-clip algorithm as in the \texttt{IRAF} task \texttt{noao.onedspec.continuum}\footnote{\url{https://astro.uni-bonn.de/~sysstw/lfa_html/iraf/noao.onedspec.continuum.html}};
    
    \item \label{item:normalization} CONTINUUM NORMALIZATION: normalize each spectrum in the time series to the same continuum measured on the template spectrum, that is, the one obtained in step \ref{item:templatecontinuum};
    
    \item\label{item:templateccf} CCF OF TEMPLATE: calculate the order-by-order CCFs between the mask obtained in step \ref{item:ccfmask} and the normalized template spectrum;
    
    \item\label{item:measurevsys} SYSTEMIC RV \& FWHM: measure the systemic velocity ($v_{\rm sys}$) and the FWHM from a Gaussian fit to the mean CCF obtained in step \ref{item:templateccf};
    
    \item\label{item:velrange} SET VELOCITY RANGE: calculate a CCF velocity range as $v_{\rm sys} \pm n \times{\rm FWHM}$, where we set $n=7$;
    
    \item UPDATE CCF MASK: update the CCF mask weights as $w = d / \bar{\sigma}^{2}$, as illustrated in Fig. \ref{fig:spirou-spectrum-template}, where $\bar{\sigma}$ is the mean dispersion in flux within the CCF velocity range (as defined in step \ref{item:velrange}) around the line center and $d$ is the line depth;
    
    \item CCF: calculate the order-by-order CCF of each normalized spectrum in the time series using the same mask obtained in step \ref{item:ccfmask} and the same velocity range obtained in step \ref{item:velrange} for all spectra;
    
    \item \label{item:buildmeanccfs} BUILD MEAN CCFS: calculate a weighted mean of the spectral order CCFs to build a final mean CCF per exposure. Fig. \ref{fig:spirou-ccfs-weights} shows the weights given by $Q / \sigma_{\rm ccf}^{2}$, where $Q=\int{(df(v)/dv)^{2}dv}$ for $f(v)$ being the CCF value at a given velocity $v$, and $\sigma_{\rm ccf}$ is the root mean square (RMS) dispersion of the CCF time series ;    
    
    \item \label{item:buildtemplateccf} BUILD TEMPLATE CCF: calculate a CCF template by the median of the mean CCFs of all exposures as presented in Fig. \ref{fig:spirou-ccfs};
    
    \item CALIBRATE CCFS: apply a polynomial fit to the continuum of each mean CCF to match the template CCF (see Fig. \ref{fig:spirou-ccfs}). The continuum is defined as the points where $|v-v_{\rm sys}| < 1.5\times{\rm FWHM}$; 
    
    \item CCF SIGMA-CLIP: apply a 4$\sigma$ clip filter to remove outliers from the CCF data;
    
    \item REMOVE BAD CCFS 1: exclude the CCF data from spectral orders that have less than 50\% of useful velocity bins, that is, those with more than 50\% of \texttt{NaN} values; 
    
    \item REMOVE BAD CCFS 2: exclude the CCF data from spectral orders that present a velocity shift greater than a given threshold of 3.0~\kms;
    
    \item REBUILD MEAN CCFS AND TEMPLATE CCF: recalculate mean CCFs as in step \ref{item:buildmeanccfs} and a new template CCF as in step \ref{item:buildtemplateccf};
    
    \item CCF TEMPLATE MATCH RVs: calculate the final RV by least-square fitting for the velocity shift that best matches the mean CCF of an individual exposure to the template CCF.
\end{enumerate}

The above procedure is applied to the spectra obtained from the sum of the flux of the two scientific fibers `A + B' of SPIRou. For the simultaneous FP spectrum obtained with the calibration fiber `C', we applied the same procedure described above with the following changes: (a) we replaced the mask in step \ref{item:ccfmask} by a mask containing the Fabry-Perot lines; (b) we did not correct for the BERV (step \ref{item:berv}); (c) we assumed a null systemic velocity; (d) we did not remove the continuum (steps \ref{item:templatecontinuum} and \ref{item:normalization}). The RVs obtained from the simultaneous calibration are compared to the RVs obtained for the same fiber in the FP exposures taken during the night calibration sequence. In this way, we calculated the spectral shift (or instrumental drift) that occurred from the moment the wavelength calibration data were taken until the scientific exposure. This drift is finally used to correct the RVs obtained from the scientific fiber. The drift-corrected RVs of our CCF analysis are listed in the Table \ref{tab:spirouRVSCCF+LBL} of Appendix \ref{app:spiroudata}.  The RMS dispersion of uncorrected CCF RVs is 10.6~\ms, and the RMS of drift-corrected data is 8.4~\ms, showing that the drift correction accounts for an additional noise of about 6.4~\ms\ in our CCF data, assuming uncorrelated Gaussian noise.  
Note that the approach presented here was critical to achieving \ms\ precision. Otherwise, when applying the CCF method in a more standardized way, the SPIRou RVs are completely dominated by systematic errors.

\subsection{LBL analysis}
\label{sec:lblanalysis}

The line-by-line (LBL) method that we used in our analysis will be exhaustively presented in a forthcoming paper by Artigau et al. (submitted).  The method uses an approach similar to that used by \cite{Dumusque2018} which draws on the \citealt{Bouchy2001} formalism, in particular Eq. 3 and 4 therein, and applies it to individual lines rather than the entire spectrum. 
As for the \cite{Bouchy2001} velocity measurement, one must have a noise-less template to compute a per-line velocity. This template is used to compare the residuals between the observed spectrum and the template to the derivative of the template. In practice, we use a high SNR combined spectrum of the star, assuming that any remaining noise contribution will be much smaller than the noise in the observation being considered. For TOI-1759, the template spectrum obtained from our observations does not have a SNR as good as in the templates of standard stars. Therefore, in this case, we use a template of Gl846, which is a standard star of almost the same spectral type as TOI-1759.

The LBL algorithm provides one velocity per line (typically 16\,000 for an M dwarf observed with SPIRou), that must be combined into a single RV measurement. The per-line uncertainties vary from $\sim50$\,m/s for the best lines to tens of km/s for shallow features. In the absence of outlying points, a weighted sum would suffice to retrieve a per-spectrum velocity. As there are high-sigma outliers among lines, due to a number of plausible causes (cosmic rays hitting the array, error in telluric absorption correction), we opt for a finite-mixture model approach to derive a mean spectrum velocity. Lines either belong to a Gaussian distribution around the mean velocity with a sigma derived from the \cite{Bouchy2001} framework, or they belong to a statistically flat distribution of outliers. In representative SPIRou data, 0.2\% of lines are consistent with being outliers. 

As in the CCF method, we also calculated the LBL RVs for the simultaneous calibration fiber to measure and correct for the instrumental drift. The drift-corrected RVs of our LBL analysis are also listed in the Table \ref{tab:spirouRVSCCF+LBL} of Appendix \ref{app:spiroudata}. The RMS of uncorrected and drift-corrected LBL RVs are 9.5~\ms\ and 5.6~\ms, respectively, showing that the instrumental drift contributed about 7.7~\ms\ to the noise in our LBL data. A comparative analysis of the RVs and the drifts obtained by the two methods is presented in Appendix \ref{app:ccfvslbl}.

\section{Stellar characterization} 
\label{sec:star}
We carried out a study to derive the stellar properties and to characterize magnetic activity in TOI-1759, as will be detailed in the next sections. Table \ref{tab:stellarparams} presents a summary of the TOI-1759 stellar parameters.

\begin{table*}
\centering
\caption{Summary of stellar parameters of TOI-1759.}
\label{tab:stellarparams}
\begin{tabular}{lcc}
\hline
Parameter & Value & Ref. \\
\hline
ID (LSPM) &  J2147+6245 & \\
ID (TYC) & 4266-736-1 & \\
ID (TIC) &  408636441 & \\
RA (hh:mm:ss.ss) & 21:47:24.386 & 1 \\
Dec (dd:mm:ss.ss) & +62:45:13.733 & 1 \\
Epoch (TCB) & J2016.0 & 1 \\
proper motion in RA, $\mu_{\alpha}$ (mas\,yr$^{-1}$) & $-173.425\pm0.012$ & 1 \\
proper motion in Dec, $\mu_{\delta}$ (mas\,yr$^{-1}$) & $-10.654\pm0.011$ & 1 \\
parallax (mas) & $24.922\pm0.010$ & 1 \\
distance (parsec)  & $40.12\pm0.02$ & 1 \\
{\it TESS} T (mag)  & $9.928\pm0.007$ &  \\
{\it GAIA} G$_{BP}$ (mag) & $11.7164\pm0.0007$ & 1 \\
{\it GAIA} G (mag)  & $10.8386\pm0.0002$  & 1 \\
{\it GAIA} G$_{RP}$ (mag) & $9.9174\pm0.0005$ & 1 \\
{\it 2MASS} J (mag) & $8.771\pm0.043$  & 2 \\
{\it 2MASS} H (mag) & $8.114\pm0.059$ & 2 \\
{\it 2MASS} K (mag) & $7.930\pm0.020$ & 2 \\
{\it WISE} 1 (mag) & $7.83\pm0.03$ & 3 \\
{\it WISE} 2 (mag) & $7.89\pm0.03$ & 3 \\
{\it WISE} 3 (mag) & $7.8\pm0.3$ & 3 \\
{\it WISE} 4 (mag) & $7.64\pm0.11$ & 3 \\
effective temperature, $T_{\rm eff}$ (K) & $4075\pm75$ & this work (4) \\
effective temperature, $T_{\rm eff}$ (K) & $4036\pm100$ & this work (5) \\
effective temperature, $T_{\rm eff}$ (K) & $4046\pm40$ & this work (6) \\
effective temperature, $T_{\rm eff}$ (K) & $3972\pm40$ & this work (7) \\
Fe metallicity, $[{\rm Fe}/{\rm H}]$ (dex) & $0.0\pm0.3$ & this work (4) \\
metallicity, $[{\rm M}/{\rm H}]$ (dex) & $+0.2\pm0.3$ & this work (5) \\
metallicity, $[{\rm M}/{\rm H}]$ (dex) & $+0.4\pm0.2$ & this work (6) \\
metallicity, $[{\rm M}/{\rm H}]$ (dex) & $+0.1\pm0.2$ & this work (7) \\
surface gravity,$\log g$ (dex) & $5.1\pm0.6$ & this work (5) \\
surface gravity,$\log g$ (dex) & $4.9\pm0.2$ & this work (6) \\
surface gravity, $\log g$ (dex) & $4.5\pm0.2$ & this work (7) \\
bolometric flux, $F_{\rm bol}$ (erg\,s$^{-1}$\,cm$^-2$) & $1.765\pm0.020\times10^{-9}$ & this work (4) \\
star radius, $R_{\star}$ (\RS) & $0.60\pm0.03$ & this work (4) \\
star radius, $R_{\star}$ (\RS) & 0.628$\pm$0.018 & this work (8) \\
star mass, $M_{\star}$ (\msol) & $0.61\pm0.02$ & 9 \\
luminosity, $L_{\star}$ (\lsol) &  0.089$\pm$0.011 & this work \\
rotation period, $P_{\rm rot}$ (d) & $35.65^{+0.17}_{-0.15}$ & this work (10) \\
rotation velocity, $v_{\rm rot}$ (\kms) & $0.85\pm0.04$ & this work (11) \\
age (Gyr) & 3--7 & this work  \\
\hline
\end{tabular}
\tablebib{
(1) \cite{Gaia2021};
(2) \cite{Cutri2003};
(3) \cite{Wright2010};
(4) SED analysis;
(5) spectroscopic analysis from spectral synthesis;
(6) spectroscopic analysis from PHOENIX grid;
(7) spectroscopic analysis from Turbospectrum+MARCS grid;
(8) using the radius-luminosity (M$_{\rm K}$) relationship \citep{Mann2015};
(9) ExoFOP\footnote{\url{https://exofop.ipac.caltech.edu/}};
(10) $P_{\rm rot}$ obtained from $B_\ell$ (see Sect. \ref{sec:stellaractivity});
(11) $v_{\rm rot} =2\pi R_{\star} / P_{\rm rot}$
}
\end{table*}

\subsection{Spectral energy distribution analysis}
\label{sec:sedanalysis}

As a first determination of the basic stellar parameters, we performed an analysis of the broadband spectral energy distribution (SED) of the star together with the {\it Gaia} EDR3 parallax \citep[with no systematic offset applied; see, e.g.,][]{StassunAndTorres2021}, in order to determine an empirical measurement of the stellar radius, following the procedures described in \cite{Stassun2016,Stassun2017,Stassun2018}. We retrieved the $JHK_S$ magnitudes from {\it 2MASS}, the W1--W4 magnitudes from {\it WISE}, and the $G_{\rm BP} \ G \ G_{\rm RP}$ magnitudes from {\it Gaia}. Together, the available photometry spans the stellar SED over the wavelength range 0.4--22~$\mu$m (see Table~\ref{tab:stellarparams}).

\begin{figure}
   \centering
   \includegraphics[width=0.9\hsize]{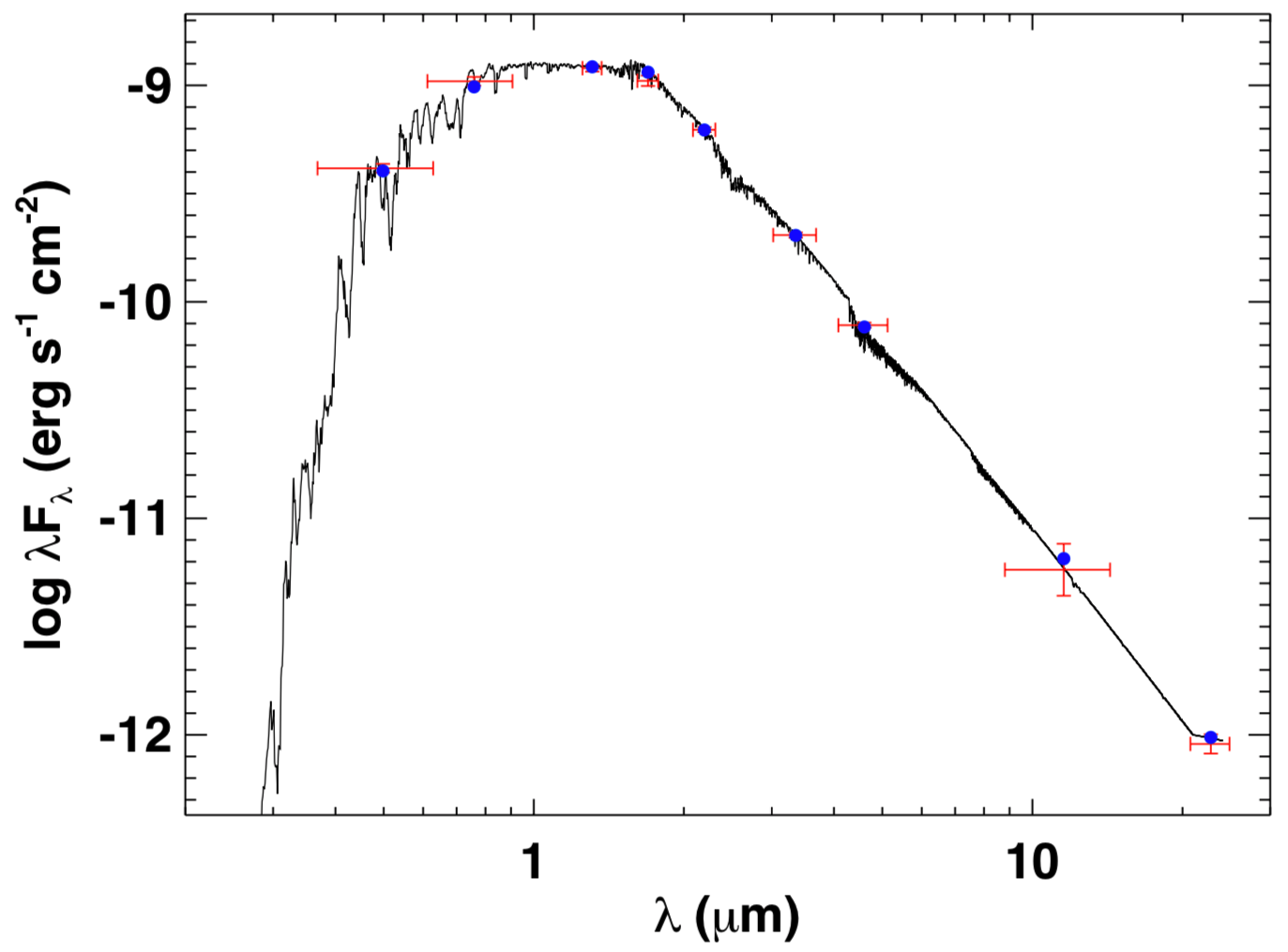}
      \caption{Spectral energy distribution of TOI-1759. Red symbols represent the observed photometric measurements, where the horizontal bars represent the effective width of the passband. Blue symbols are the model fluxes from the best-fit NextGen atmosphere model (black).}
        \label{fig:sed}
  \end{figure}

We performed a fit using NextGen stellar atmosphere models, with the effective temperature ($T_{\rm eff}$) and metallicity ([Fe/H]) as free parameters (the surface gravity, $\log g$, has little influence on the broadband SED). We fixed the extinction, $A_V$, to zero due to the proximity of the system. The resulting fit (Figure~\ref{fig:sed}) has a reduced $\chi^2$ of 1.2, with best-fit $T_{\rm eff} = 4075 \pm 75$~K and [Fe/H] = $0.0 \pm 0.3$. Integrating the model SED gives the bolometric flux at Earth, $F_{\rm bol} =  1.765 \pm 0.020 \times 10^{-9}$~erg~s$^{-1}$~cm$^{-2}$. Taking the $F_{\rm bol}$ and $T_{\rm eff}$ together with the {\it Gaia\/} parallax, gives the stellar radius, $R_\star = 0.60 \pm 0.03$~R$_\odot$.
  
\subsection{Spectral synthesis analysis of the SPIRou template}

\begin{figure}
   \centering
   \includegraphics[width=0.9\hsize]{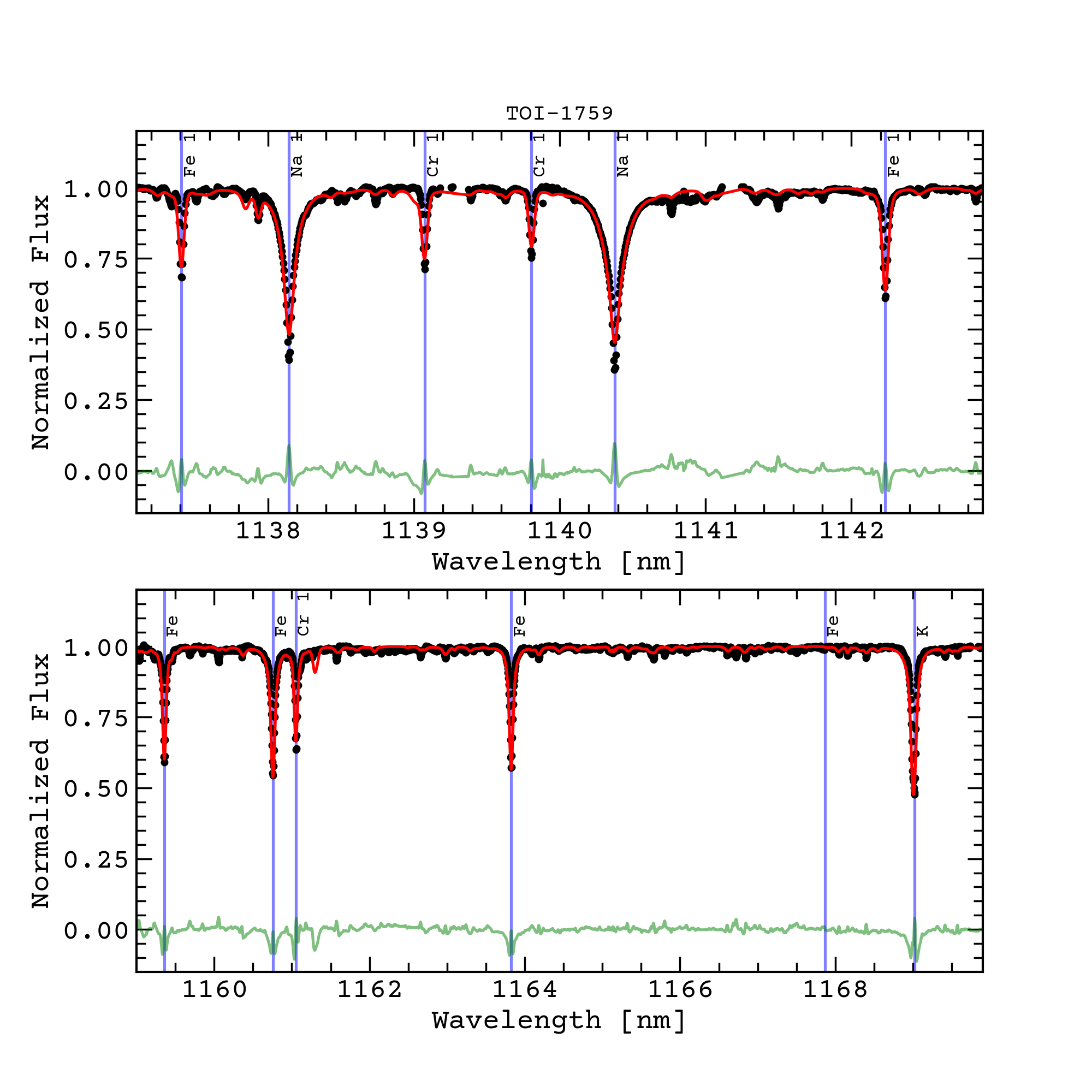}
      \caption{ Black points show the normalized template SPIRou spectrum of TOI-1759 and the red line shows the best-fit synthetic spectrum with $T_{\rm eff} = 4036\pm100$~K, $\log g=5.1\pm0.6$~dex, and $[{\rm M}/{\rm H}]=+0.2\pm0.3$~dex. The blue vertical lines show the positions of the lines of the main chemical species (as indicated on the labels) considered in our analysis. Our analysis included a total of 10242 lines within the spectral range 1137--1169~nm. Solid green lines show the residuals (observed minus synthetic).}
        \label{fig:spectralsynthesis}
  \end{figure}

As a second determination, we analyzed the normalized template SPIRou spectrum of TOI-1759 using the code iSpec\footnote{\url{https://www.blancocuaresma.com/s/iSpec}} \citep{Blanco-Cuaresma2014,Blanco-Cuaresma2019}, and the radiative transfer code SPECTRUM \citep{GrayAndCorbally2014}. In this approach, a grid of synthetic spectra is computed using stellar atmospheric models from MARCS \citep{Gustafsson2008} and a custom line list from the VALD catalog (1100 to  2400 nm) \citep{piskunov1995}.  The solar abundances are from \cite{Asplund2009}. Here, we considered the range between 1100~nm and 1250~nm, where we used a total of 44615 lines to produce the synthetic spectra. The best-fit synthetic spectra were obtained by minimizing the $\chi^2$, which gives us a reliable measurement of the fundamental parameters of TOI-1759. Fig. \ref{fig:spectralsynthesis} shows the SPIRou template spectrum and the best-fit synthetic spectrum of TOI-1759 obtained by our analysis. The best-fit stellar parameters are shown in Table \ref{tab:stellarparams}.  The spectroscopic parameters obtained in this analysis are consistent with those obtained in the SED analysis (see Sect. \ref{sec:sedanalysis}), although the uncertainty on $T_{\rm eff}$ obtained here is larger, mainly due to the presence of activity \citep[e.g.,][]{Nascimento2016}, whose complete characterization would require a detailed analysis beyond the scope of this work.

In addition, we obtained the spectroscopic parameters of TOI-1759 from a spectral characterization tool specifically developed to analyze SPIRou spectra of M dwarfs by Cristofari et al. (submitted), which performs a least-squares search for best-fit parameters in a pre-computed grid of spectra using both a grid of PHOENIX spectra from \cite{Husser2013} and a grid of spectra computed with Turbospectrum \citep{Plez2012} from MARCS \citep{Gustafsson2008} model atmospheres. The PHOENIX grid gives $T_{\rm eff}= 4046\pm40$~K, $\log g=4.9\pm0.2$~dex, and $[{\rm M}/{\rm H}]=+0.4\pm0.2$~dex, and the Turbospectrum+MARCS grid gives $T_{\rm eff}= 3972\pm40$~K, $\log g=4.5\pm0.2$~dex, and $[{\rm M}/{\rm H}]=+0.1\pm0.2$~dex. These results agree with our SED fit and spectral synthesis analyses, showing an improvement in the uncertainty of $T_{\rm eff}$. 

\subsection{Magnetic activity, rotation and age}
\label{sec:stellaractivity}

\begin{figure}
   \centering
       \includegraphics[width=1.0\hsize]{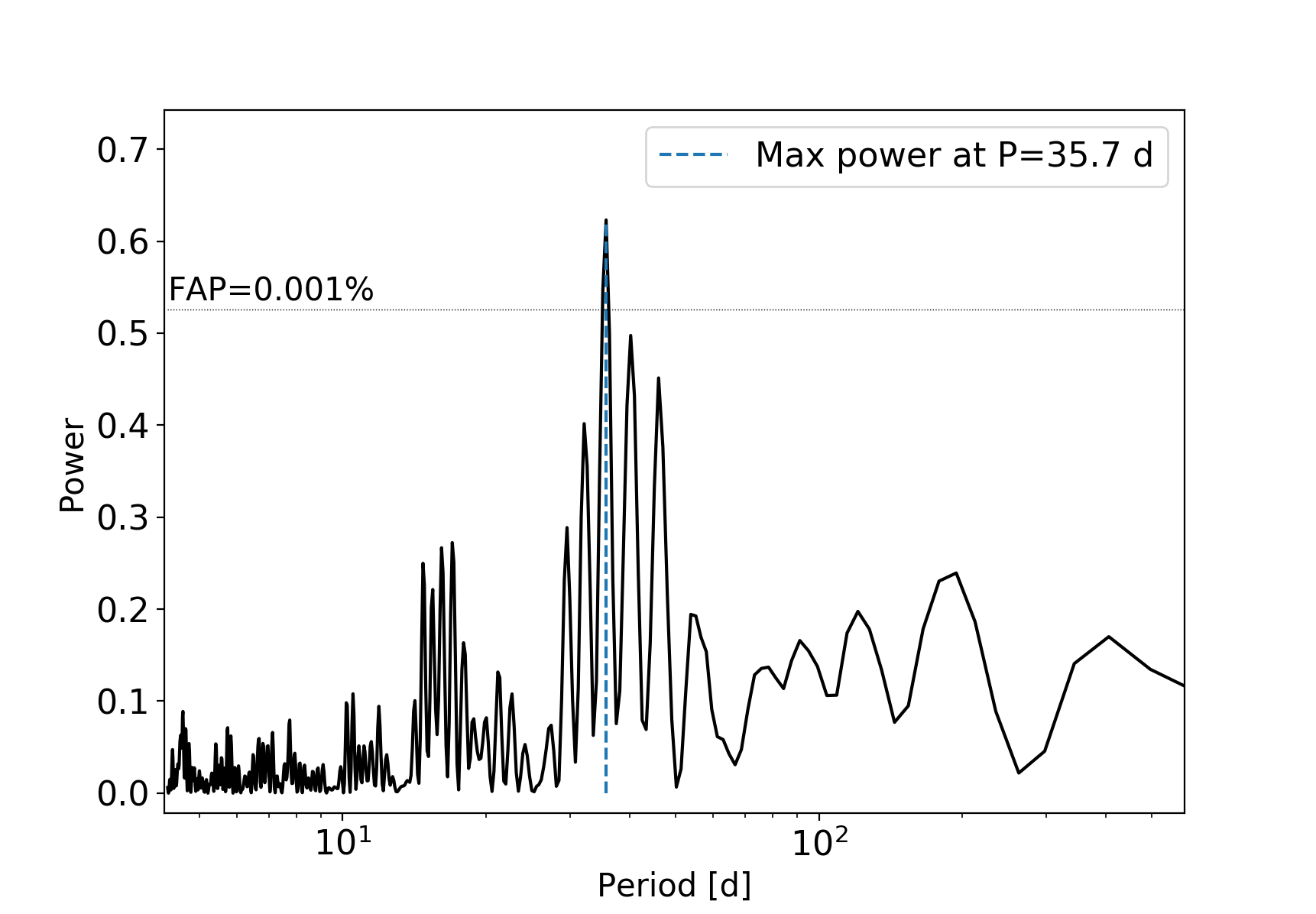}
      \caption{GLS periodogram analysis of the longitudinal magnetic field ($B_\ell$) time series of TOI-1759. The blue dashed line shows the highest power at a period of 35.7~d.
      }
        \label{fig:toi1759blonganalysis}
  \end{figure}

To investigate the magnetic activity in TOI-1759 we calculate the disk-integrated longitudinal magnetic field $B_\ell$ in the LSD profiles of SPIRou following the same prescription as in \cite{donati1997, moutou2020, Martioli2020}. Table \ref{tab:toi1759spiroublongdata} in Appendix \ref{app:spiroudata} presents the values of $B_\ell$ for TOI-1759 from both APERO and LE reductions. The $B_\ell$ corresponds to a measurement of the net magnetic field projected on the line-of-sight direction originated from magnetic regions in the visible hemisphere of the stellar photosphere. Since the spatial distribution of these magnetized features is likely to be heterogeneous, as later confirmed by our ZDI analysis (see Sect. \ref{sec:zdianalysis}), $B_\ell$ is expected to be modulated by the star's rotation, allowing to derive the rotation period if any periodicity is detected in its time series \citep[e.g.,][]{BorraAndLandstreet1980, morin2008, moutou2017, Petit2021}.  Fig. \ref{fig:toi1759blonganalysis} shows the generalized Lomb-Scargle (GLS) periodogram \citep{Zechmeister2009} for the $B_\ell$ data calculated using the \texttt{astropy.timeseries}\footnote{\url{https://docs.astropy.org/en/stable/timeseries/lombscargle.html}} tool. We find the maximum power at a period of 35.7~d with a false alarm probability (FAP) below 0.001\%.

The magnetic features of this star appear to evolve rapidly, as seen in our ZDI analysis (see Sect. \ref{sec:zdianalysis}) showing a change in the magnetic properties between 2020 and 2021. Therefore, this study requires a flexible model to account for the variability of the stellar magnetic field. We employ a Gaussian Process (GP) regression analysis \citep[e.g.,][]{haywood2014,aigrain2015} using the code \texttt{george}\footnote{https://george.readthedocs.io/} \citep{georgecode}, where we assume that the rotational modulated stellar activity signal in $B_\ell$ is quasi-periodic (QP). Thus, we adopt a parameterized covariance function (or kernel) as in \cite{Angus2018}, which is given by:

\begin{equation}
    k(\tau_{ij}) = \alpha^{2} \exp{\left[ -\frac{\tau_{ij}^2}{2l^2} - \frac{1}{\beta^2} \sin^{2}{\left( \frac{\pi \tau_{ij}}{P_{\rm rot}} \right)} \right]} + \sigma^{2} \delta_{ij},
\end{equation}

\noindent where $\tau_{ij} = t_{i} - t_{j}$ is the time difference between data points $i$ and $j$, $\alpha^{2}$ is the amplitude of the covariance, $l$ is the decay time, $\beta$ is the smoothing factor, $P_{\rm rot}$ is the star rotation period, and $\sigma$ is the uncorrelated white noise, which adds a ``jitter'' term to the diagonal of the covariance matrix.  This kernel combines a squared exponential component describing the overall covariance decay and a component that describes the periodic covariance structure, the amplitude of which is controlled by the smoothing factor. Values of $\beta$ around 1, as we find for this object, correspond to a periodic variation without a strong harmonic content. As pointed out by \cite{Angus2018} the flexibility of this model can easily lead to overfitting of the data. Therefore, we adopt a prior distribution for the parameters (see Table \ref{tab:activitygpfitparams}) that restricts the search range to realistic values and avoids overfitting.

\begin{figure*}
   \centering
       \includegraphics[width=1.0\hsize]{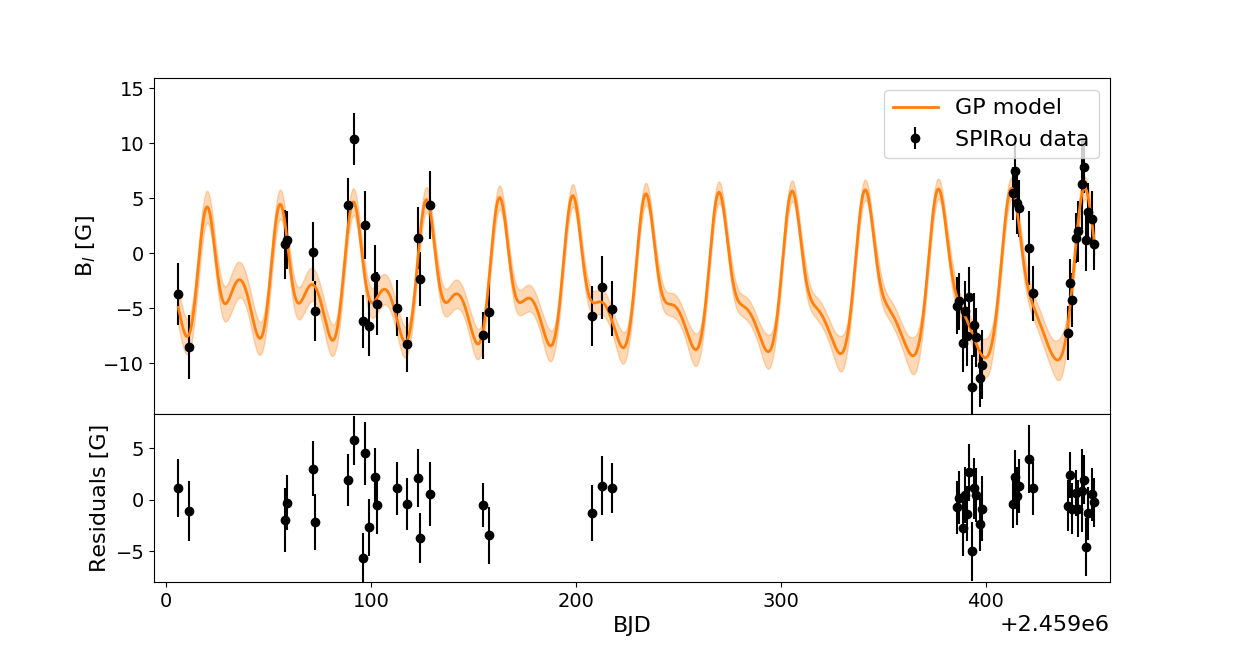}
      \caption{GP analysis of the SPIRou $B_\ell$ data. In the top panel the black points show the observed $B_\ell$ data and the orange line shows the best-fit quasi-periodic GP model. Bottom panel shows the residuals with an RMS dispersion of 2.3~G.}
        \label{Blong_QPgp_fitmodel}
  \end{figure*}

\begin{figure}
   \centering
       \includegraphics[width=0.8\hsize]{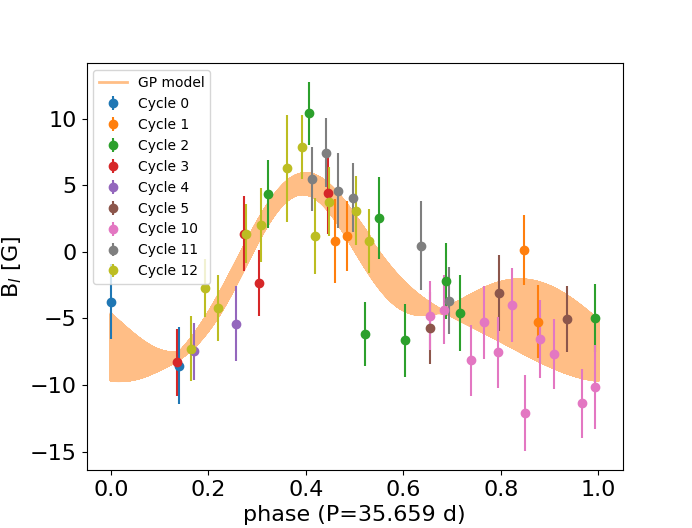}
      \caption{SPIRou $B_\ell$ data phase folded with the best-fit period of 35.7~d. The data are represented by a different color for each rotation cycle, considering the time of the zeroth cycle to be the time of the first SPIRou observation, that is, $2459006.063741$~BJD. The orange shaded region shows the best-fit GP model.}
        \label{fig:blong_phase}
  \end{figure}

We use this GP framework to model the temporal variability of the $B_\ell$ data, where we first fit the GP model parameters by the maximization of the likelihood function \citep{Rasmussen2006, celerite} using the package \texttt{scipy.optimize}, and then we sampled the posterior distribution of the free parameters using a Bayesian Markov chain Monte Carlo (MCMC) framework with the package \texttt{emcee} \citep{foreman2013}. We set the MCMC with 50 walkers, 1000 burn-in samples, and 5000 samples. The results of our analysis are illustrated in Fig. \ref{Blong_QPgp_fitmodel} where we present the $B_\ell$ observed data and the best-fit GP model. Fig. \ref{fig:Blong_QPgp_pairsplot} in Appendix \ref{app:blonggpposteriors} shows the MCMC samples and posterior distributions of the GP parameters. Table \ref{tab:activitygpfitparams} shows the priors and best-fit parameters calculated by the medians and their 0.16 and 0.84 quantiles uncertainties. We performed the same analysis on both APERO and LE $B_\ell$ data, which gives consistent GP parameters within $1\sigma$. The APERO data is noisier, with an RMS of residuals of 2.3~G, while LE data gives an RMS of 1.4~G. However, we adopt the star rotation period obtained with the APERO data, which has a posterior distribution that is more tightly constrained than in the LE data.  In Fig. \ref{fig:blong_phase} we present the $B_\ell$ data phase-folded to the best-fit star rotation period of $P_{\rm rot}=35.7$~d, where we highlight the different rotation cycles by different colors. 

   \begin{table*}
    \caption[]{Best fit parameters of a quasi-periodic GP model obtained in our analysis of the stellar activity in the SPIRou $B_\ell$ data. }
    \label{tab:activitygpfitparams}
    \begin{center}
    \begin{tabular}{cccc}
        \hline
        \noalign{\smallskip}
 Quantity & Priors & \multicolumn{2}{c}{ Fit values }\\
  &  & APERO & Libre-Esprit \\
        \noalign{\smallskip}
        \hline
mean, $\mu$ [G] & $\mathcal{U}(-\infty,+\infty)$ & $-3\pm6$ & $-3\pm3$ \\
white noise, $\sigma$ [G] & $\mathcal{U}(0,+\infty)$ & $0.8^{+0.7}_{-0.5}$  & $0.7\pm0.4$ \\
amplitude, $\alpha$ [G] & $\mathcal{U}(0,+\infty)$ &  $8^{+5}_{-3}$ &  $5^{+3}_{-2}$ \\
decay time, $l$ [d] & $\mathcal{U}(50,1000)$ & $671^{+218}_{-243}$ & $395^{+271}_{-172}$ \\
smoothing factor, $\beta$ & $\mathcal{U}(0.2,1.5)$  & $0.9\pm0.2$  & $0.7^{+0.3}_{-0.2}$ \\
rotation period, $P_{\rm rot}$ [d] & $\mathcal{U}(2,300)$ &  $35.65^{+0.17}_{-0.15}$ &  $35.8\pm0.3$\\
RMS of residuals [G] & & 2.3 & 1.4 \\
$\chi^2$ & & 0.85 & 1.02 \\
        \hline
    \end{tabular}
    \end{center}
  \end{table*}

\begin{figure}
   \centering
       \includegraphics[width=1.0\hsize]{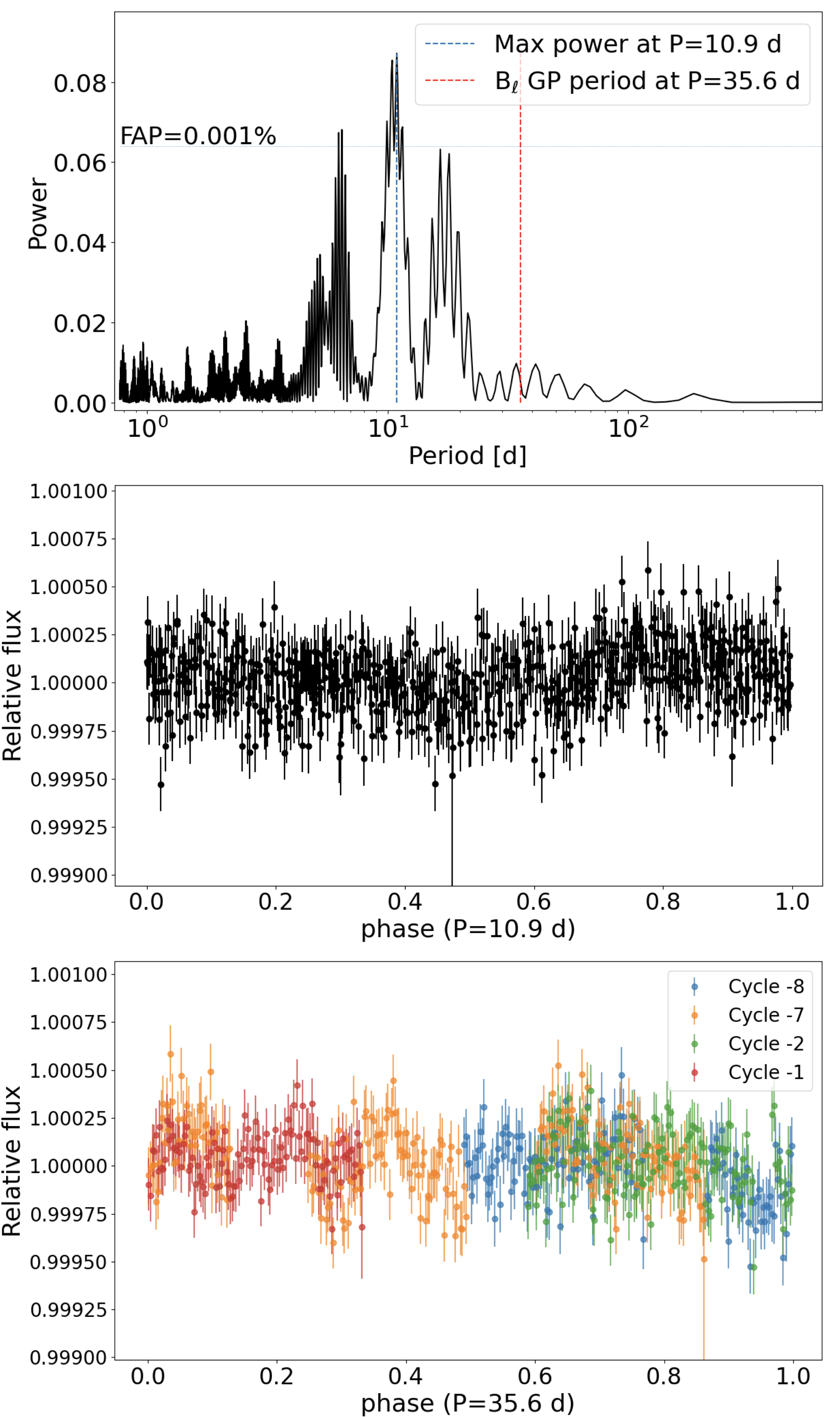}
      \caption{
      GLS periodogram analysis of the TESS photometry data subtracted by the best-fit transit model and binned by weighted average with a bin size of 0.1~d. The top panel shows the GLS periodogram where the maximum power at 10.9~d is marked by the dashed blue line. The dashed red line shows the best-fit star rotation period of 35.7~d obtained in our analysis of the $B_\ell$ data. The middle panel shows the TESS data phase-folded with a period of 10.9~d and the bottom panel also shows TESS data phase-folded with a period of 35.7~d. The latter shows the data represented by a different color for each rotation cycle (as in Fig. \ref{fig:blong_phase}), considering the time of the zeroth cycle to be the time of the first SPIRou observation, that is, $2459006.063741$~BJD. 
      }
        \label{fig:toi1759tessGLSperiodogram}
  \end{figure}

We investigate if the periodicity found in the $B_\ell$ data is also present in the TESS photometry data.  Here we consider the TESS flux data subtracted by the best-fit transit model using the parameters in Table \ref{tab:toi-1759-final-params} and binned by weighted average with a bin size of 0.1~d. Fig. \ref{fig:toi1759tessGLSperiodogram} shows the GLS periodogram with a highest power at a period of 10.9~d and two other smaller peaks at $6.2$~d and $17.2$~d. The rotation period of $P_{\rm rot}=35.7^{+0.17}_{-0.15}$~d obtained from $B_\ell$ does not show a significant peak in the TESS data, although there are significant peaks near its harmonics. However, as illustrated in the bottom panel of Fig. \ref{fig:toi1759tessGLSperiodogram}, the TESS data phase-folded with $P_{\rm rot}=35.7$~d shows that different rotation cycles present some agreement in their overlapping features. This suggests that the surface of TOI-1759 has several small spots rather than large spots at some specific longitudes that would generate a simpler oscillatory modulation in the TESS light curve.

\begin{figure}
   \centering
       \includegraphics[width=1.0\hsize]{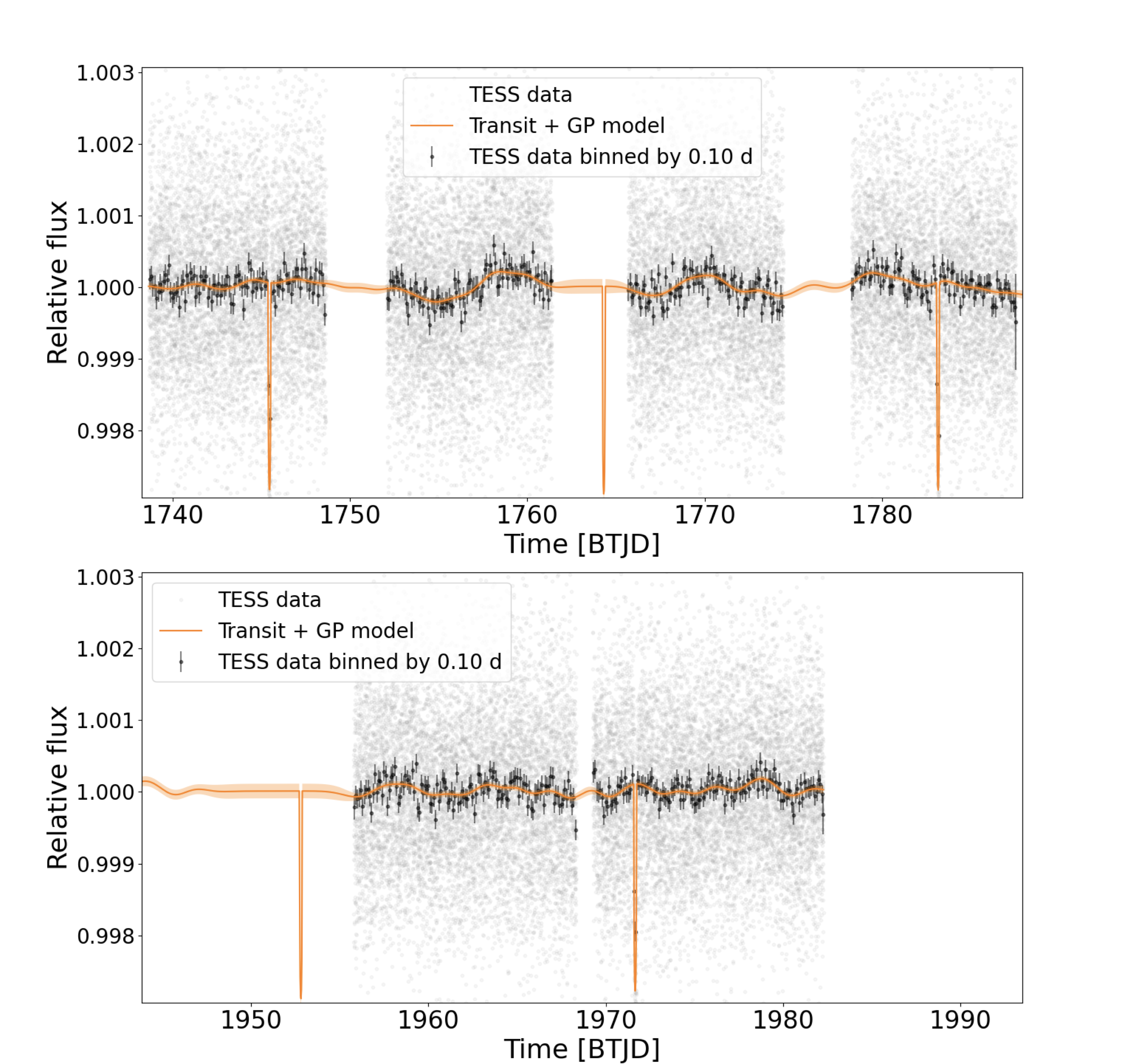}
      \caption{GP analysis of the TESS photometry time series.  The top panel shows the TESS data obtained in Sectors 16 and 17 and the bottom panel shows the data obtained in Sector 24. Black hollow circles show the TESS photometry in its original sampling and the black points show the binned data, where each bin is calculated by the weighted mean within windows of 0.1~d. The orange line and shaded region show the best fit GP model (multiplied by the transit model) and its uncertainty. 
      }
        \label{fig:toi1759tessphotometry}
  \end{figure}

In Fig. \ref{fig:toi1759tessphotometry}, we present the results of our analysis of the TESS photometry data using the same quasi-periodic GP framework, where we assumed the same priors listed in Table \ref{tab:activitygpfitparams}, except for the $P_{\rm rot}$, which we assumed a prior with the value obtained from $B_\ell$, that is, $P_{\rm rot}=\mathcal{N}(35.65,0.17)$. We tried to fit the TESS data alone but it does not place a strong constraint on the rotation period, as we found that the best-fit GP model converges to significantly different values of $P_{\rm rot}$ with a small change in the quality of the fit, depending strongly on the choice of the initial values of $P_{\rm rot}$. 

Finally, we can use the star's rotation period that we obtained from $B_\ell$, $P_{\rm rot}=35.7$~d, to estimate the system's age via empirical gyrochronology relations. For example, we obtain an age of $\approx 2.9$~Gyr via the relations of \citet{Mamajek2008}, although the star is slightly redder than the limits of applicability of those relations. Alternatively, we obtain an age of $\approx 3.9$~Gyr with the M-dwarf relations of \citet{Engle2018}, although the star is slightly hotter than the limits of applicability of those relations.  A recent study of stellar clusters by \cite{Curtis2020} discovered a stalling in the spin-down of cool stars. Therefore, our age derivation above of 2.9--3.9~Gyr may be significantly underestimated. According to the results presented in Fig.~7 of \cite{Curtis2020}, TOI-1759 seems to correspond better to a $6\pm1$~Gyr field star than to the younger members of clusters. Therefore, we conservatively estimate the empirical gyrochronology age of TOI-1759 to be in the range 3--7~Gyr.

\subsection{Magnetic Imaging}
\label{sec:zdianalysis}

We reconstruct the large-scale magnetic field at the surface of TOI-1759 using Zeeman-Doppler Imaging (ZDI). The algorithm models the magnetic topology as a combination of a poloidal and a toroidal component, which are both formulated via spherical-harmonics decomposition \citep{Donati2006}. An iterative comparison between synthetic and observed Stokes~V profiles is performed until the maximum-entropy map at a given reduced $\chi^2$ level is found \citep[for more details see][]{SkillingAndBryan1984,DonatiAndBrown1997, Folsom2018}.

\begin{figure}[ht]
\includegraphics[width=1.0\hsize]{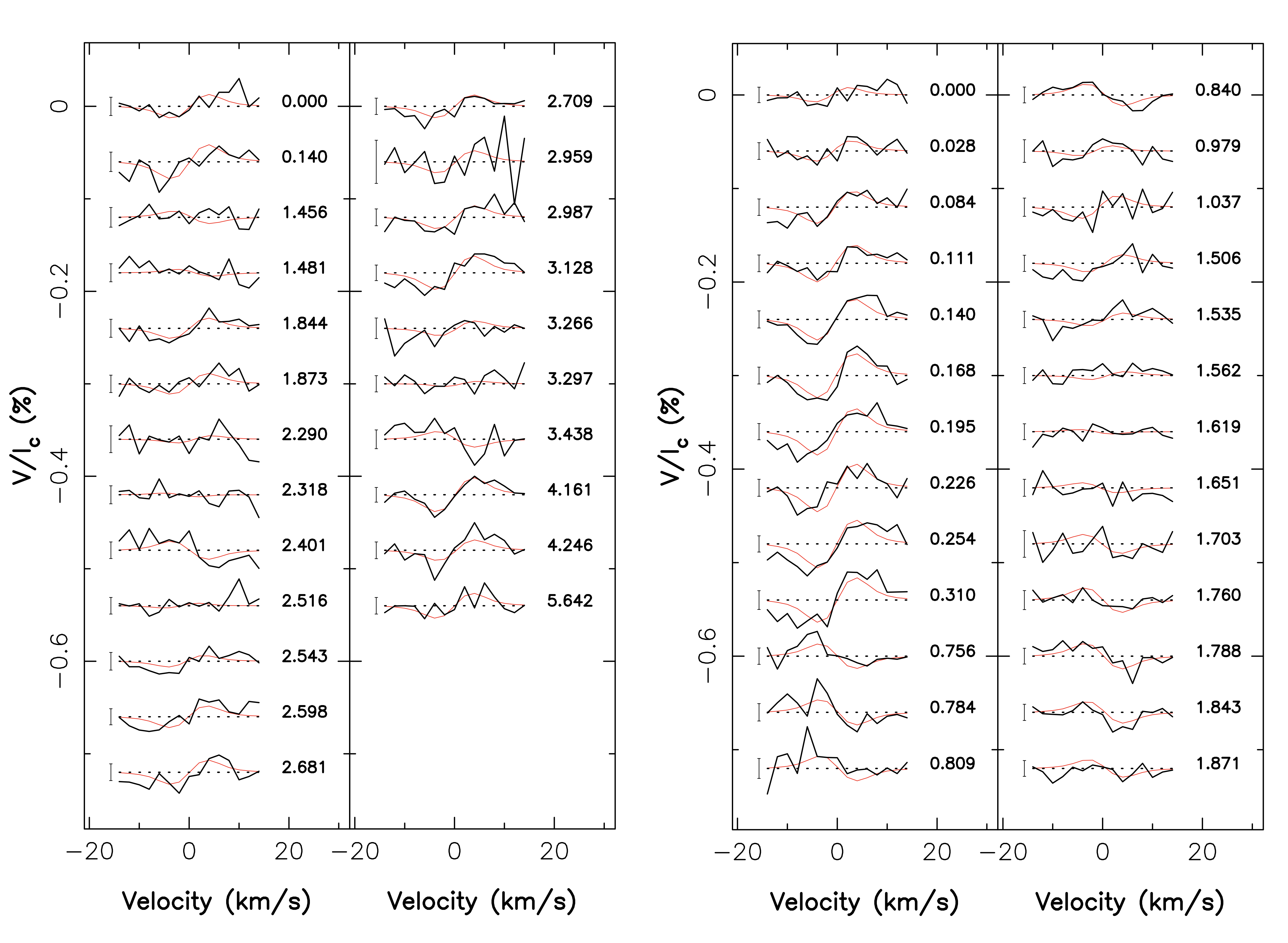}
\caption{Stokes V profiles (black lines) and their model (red lines) obtained with P$_\mathrm{rot}=35.7$~d, $v_\mathrm{eq}\sin(i)$=0.84~\kms\ and $i=80$~deg. The numbers at the right of each profile indicate the rotational cycle relative to the first Julian Date of the time series of each season. The profiles are shifted vertically for better visualization. Left: 2020 time series, Right: 2021 time series.}
\label{fig:toi1759zdistokesVfit}
\end{figure}

We produce two ZDI maps from observations collected almost one year apart: between the 5th June 2020 and 24th December 2020 (23 observations), and between the 20th June 2021 and 26th August 2021 (25 observations). The local Stokes $I$ profiles are truncated at $\pm15$~\kms\ from line center and modeled with a Voigt function defined by a Gaussian and Lorentzian width of 1.2 and 5.1~\kms, respectively. As central wavelength and Land\'e factor, we use the normalization values of 1750~nm and 1.2. We adopt the linear limb darkening coefficient reported in Table \ref{tab:toi-1759-final-params} and let the spherical harmonics expansion reach the 5th degree in $l$, as higher degrees are unnecessary given the low value of the projected velocity \citep{morin2008}. The observed and modeled Stokes~$V$ profiles are shown in Fig.~\ref{fig:toi1759zdistokesVfit}.

\begin{figure}[ht]
\includegraphics[width=1.0\hsize]{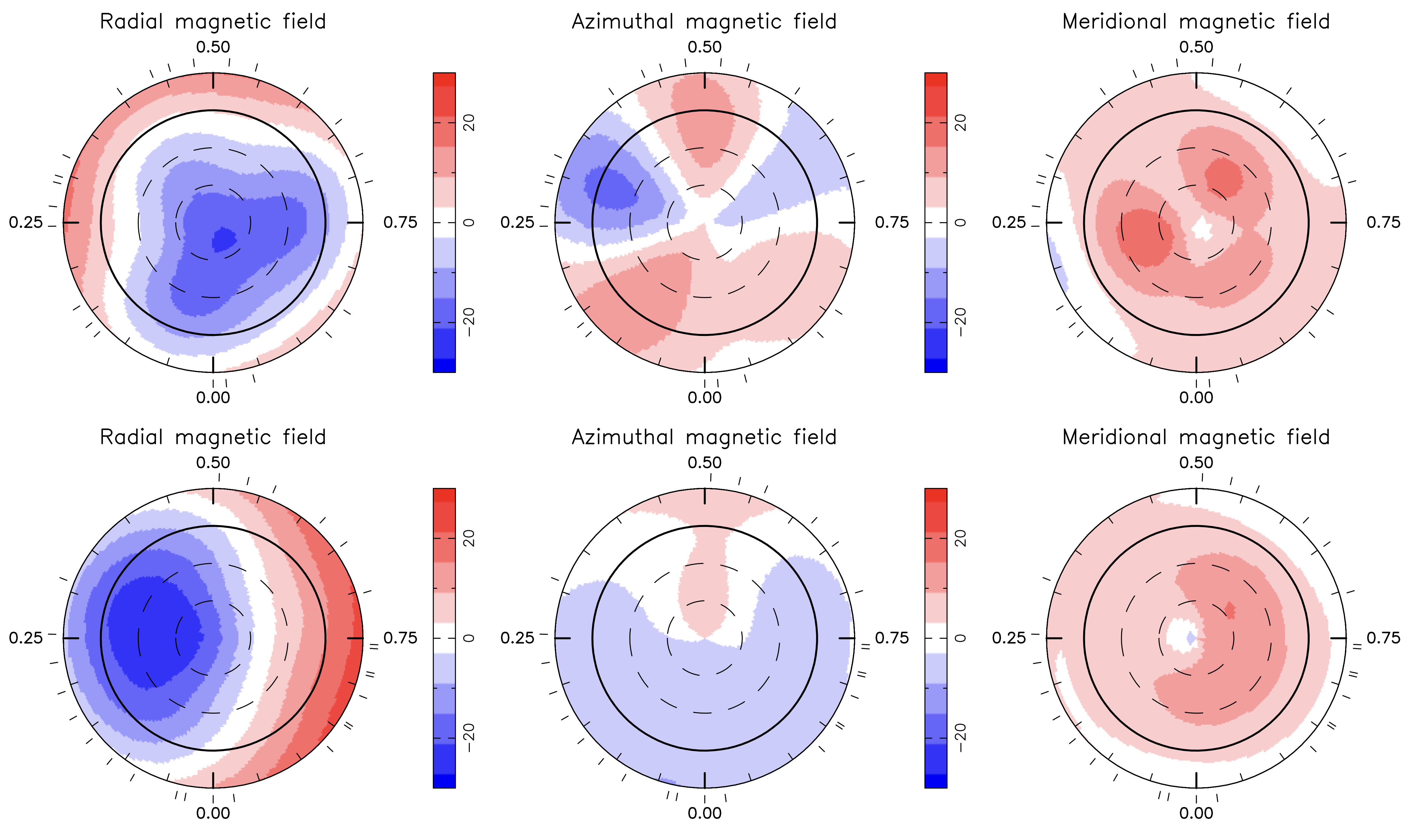}
\caption{ZDI maps of the recovered surface magnetic flux of TOI-1759 obtained from the best-fit of the SPIRou data in 2020 (top panels) and 2021 (bottom panels). We show the radial (left panels), azimuthal (middle panels), and meridional (right panels) components of the magnetic field. The star is shown in a flattened polar projection, with the equator depicted as a bold circle and the 30$^{\circ}$ and 60$^{\circ}$
latitude parallels as dashed lines. Ticks around the star mark the rotational phases of our observations. The magnetic topology is predominantly poloidal, with the axisymmetric mode at an intermediate level.}
\label{fig:toi1759zdimaps}
\end{figure}

The two maps of surface magnetic flux are shown in Fig.~\ref{fig:toi1759zdimaps} and the map characteristics obtained from both APERO and LE datasets are reported in Table~\ref{tab:zdiresults}.  Assuming solid body rotation, P$_\mathrm{rot}=35.7$~d, $v_\mathrm{eq}\sin(i)=0.84$~\kms, and stellar inclination = 80 deg (instead of 90 deg to prevent mirroring effects), we are able to fit the Stokes~$V$ profiles down to a reduced chi-square level of $\chi^2\sim$0.9 for LE and $\chi^2_r\sim$0.4 for APERO. The latter indicates that APERO overestimates the error bars of its LSD profiles. This problem will be addressed in future works, therefore, we adopt here the ZDI results obtained from the LE profiles. 

\begin{table*}[ht]
\caption{Properties of the magnetic field of TOI-1759 obtained from our ZDI analysis of the 2020 and 2021 SPIRou datasets, using both the APERO and Libre-Esprit reduced data. We list the square root of the magnetic energy at the surface, the intensity and tilt of the dipole field, the fractional energy of the poloidal component and the fractional energy of the axisymmetric component of the poloidal field. } 
\label{tab:zdiresults}     
\centering                       
\begin{tabular}{l c c c c}      
\hline     
Quantity &  \multicolumn{2}{c}{APERO} &  \multicolumn{2}{c}{Libre-Esprit}\\ 
         & 2020 & 2021 & 2020 & 2021\\ 
\hline
mean magnetic field, $B_\mathrm{m}$ [G] & $10\pm2$  &  $9\pm2$  &  $17\pm3$    & $18\pm4$  \\
dipole magnetic field, B$_\mathrm{dip}$ [G]   & $-14\pm3$   & $-14\pm3$    & $-23\pm5$    & $-27\pm5$  \\
tilt of the dipole field, i$_\mathrm{dip}$ [deg]     & $23\pm5$    & $50\pm5$ & $22\pm5$    & $46\pm5$    \\
poloidal component   & $99\pm5$\%    & $99\pm5$\%    & $99\pm5$\%    & $97\pm5$\%   \\
axisymmetric component   & $78\pm5$\%    & $46\pm5$\%    & $75\pm5$\%    & $54\pm5$\%  \\
$\chi^2$ & \multicolumn{2}{c}{0.4} &  \multicolumn{2}{c}{0.9}  \\
\hline                                 
\end{tabular}
\end{table*}

The magnetic topology is predominantly poloidal (99\% of the magnetic energy) with the axisymmetric component decreasing from $75\pm5$\% in 2020 to $54\pm5$\% in 2021. We also observed a change in the tilt of the dipole field from $22\pm5$~deg in 2020 to $46\pm5$~deg in 2021.  Our data also suggests a 20\% increase in the intensity of the dipole component between 2020 and 2021, otherwise the intensity of the mean magnetic field remained constant: the mean (B$_\mathrm{m}$) and the dipole (B$_\mathrm{dip}$) field strengths are $17\pm3$~G (2020) and $18\pm4$~G (2021), and $-23\pm5$~G (2020) and $-27\pm5$~G (2021), respectively.  For comparison, the APERO dataset gives no substantial change in the field strengths between the two seasons, however the field strengths are about 1.5 to 2.0 times lower than the LE results, with an agreement within $2\sigma$. This slight disagreement indicates that the different noise characteristics and normalization of the LSD profiles can probably bias the measurements of the field strengths.

To summarise, the magnetic topology of TOI-1759 is characterized by a weak ($<20$~G) and predominantly poloidal field, whose main axis of symmetry shows a variable inclination with time. This agrees with previous results on similar stars, e.g., on Gl~205 whose spectral type, rotation periods and magnetic properties are very similar to those of TOI-1759 \citep{Hebrard2016}.  

\section{Planet characterization}
\label{sec:analysis}

We model the TESS photometry data with a baseline GP component to account for stellar activity as described in Sect. \ref{sec:stellaractivity} multiplied by a transit model calculated using the \texttt{BATMAN} toolkit by \cite{Kreidberg2015}. The SPIRou RV data is modeled by the orbital reflex motion of the star caused by the planet, given by a Keplerian velocity as described in the Sect. 4 of \cite{martioli2010}.  To perform a Bayesian MCMC joint analysis of the photometry and RV datasets, we build a global likelihood function that can be evaluated at each iteration of the MCMC sampler. Our likelihood function is given by the sum of the logarithm of the likelihood (log-likelihood) of the prior probability on the data and on the planet parameters plus the posterior probability of the trial models conditioned to the data. The general form of our likelihood function is given by: 

\begin{equation}
    \ln \mathcal{L} = -\frac{1}{2}\sum_{i=1}^{N}{\left[  \frac{(y_{i} - \mu)^2}{\sigma_{i}^{2}} + \ln{2\pi \sigma_{i}^{2}} \right]},
\end{equation}

\noindent where $N$ is the number of data points (or parameters), $y_{i}$ is a given data point (or parameter) and $\sigma_{i}$ is its Gaussian uncertainty, and $\mu$ is the mean value. We sampled the posterior distribution of the model parameters using the \texttt{emcee} package.  The chain is set with 50 walkers and 5000 MCMC steps of which we discard the first 1000.  To compare different datasets and model assumptions, we calculated the Bayesian Information Criterion (BIC), given by:

\begin{equation}
    {\rm BIC} = k \ln{n} - 2 \ln{\mathcal{\hat{L}}},
\end{equation}

where $k$ is the number of free parameters, $n$ is the number of data points, and $\mathcal{\hat{L}}$ is the likelihood for the best-fit model.

We first fit the transit model to the photometry data in the regions around the three transits of TOI-1759~b observed by TESS, as shown in Fig. \ref{fig:toi1759_transits_fit}. We defined the priors considering ranges that include only realistic values and the initial values were measured directly from the data as in \cite{Martioli2021}. Then, we performed a joint analysis of photometry and RV data, setting the initial model with the planet parameters obtained from the transit analysis and uniform priors for the velocity semi-amplitude of $K_{p}=\mathcal{U}(0,\infty)$~\ms\ and for the systemic velocity of $v_{\rm sys}=\mathcal{U}(-\infty,\infty)$~\ms. Note that in this first analysis we do not include the activity model to the RV data. We first adopt a uniform prior for the orbital eccentricity and longitude of periastron, $e=\mathcal{U}(0,1)$ and $\omega=\mathcal{U}(0,360)$~deg, and then we repeat the analysis assuming a circular orbit ($e=0$). The best-fit eccentricities for CCF and LBL data are $0.5\pm0.2$ and $0.2\pm0.2$, respectively.  However, the BIC values obtained for a circular orbit, ${\rm BIC}_{\rm CCF}=19896$ and ${\rm BIC}_{\rm LBL}=19617$, are lower than those obtained for a non-circular orbit, ${\rm BIC}_{\rm CCF}=19917$ and ${\rm BIC}_{\rm LBL}=19645$. This means that the information provided by the data does not justify increasing the number of model parameters, so we adopt a circular orbit from now on. 

    \begin{figure}
   \centering
       \includegraphics[width=1.0\hsize]{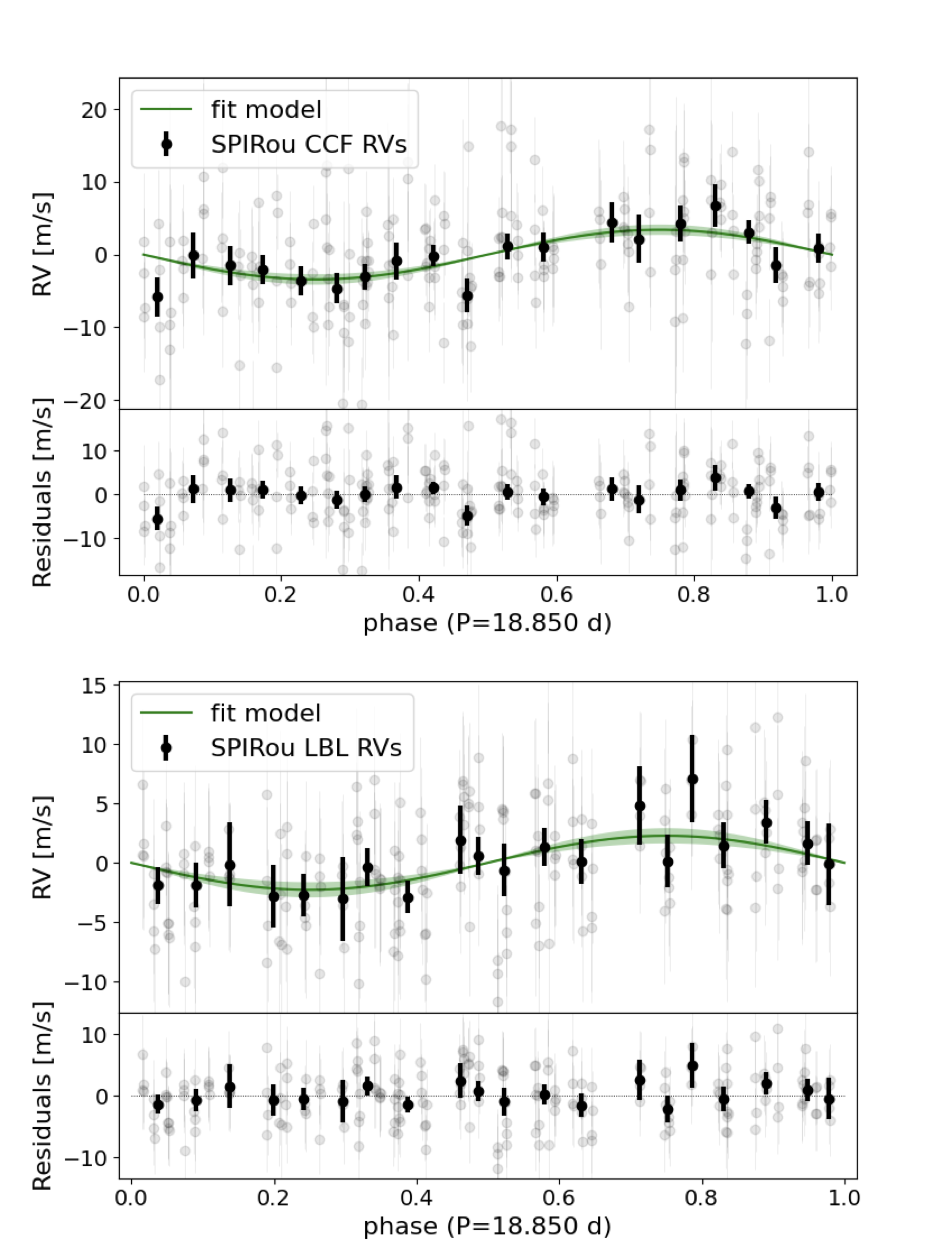}
      \caption{The points in light gray show the SPIRou CCF (upper panel) and LBL (lower panel) RVs phase-folded at an orbital period of 18.850~d. Black points show weighted averages within a bin size of 0.05 ($\sim0.95$~d). The green lines and shaded regions show the best fit models and their uncertainties for the orbit of TOI-1759~b. The residuals are displayed at the bottom of each panel, giving an RMS of 7.7~\ms\ and 4.6~\ms\ for CCF and LBL, respectively.
      } 
        \label{fig:toi1759_rv_fit_phase}
  \end{figure}

To investigate if an activity related signal modulates our RV data, we fit a quasi-periodic GP activity component to the RV data subtracted by the best-fit orbit model obtained above, where we assume a prior for the rotation period of $P_{\rm rot}=\mathcal{N}(35.65,0.17)$~d, as obtained in Sect. \ref{sec:stellaractivity}, and the same priors shown in Table \ref{tab:activitygpfitparams} for the other GP parameters. Then we subtract this GP model from the original RVs and perform again a joint fit of the photometry and RV data.  In appendix \ref{app:orbit+activitygpmodels} we present the results of this analysis, where we show the MCMC samples and posterior distributions for the fit parameters and the RV data and each component of the best-fit model.  Fig. \ref{fig:toi1759_rv_fit_phase} shows the RV data and the best-fit orbit models phase-folded to the orbital period and with t$_0$ being the time of transit (T$_{c}$). Notice that both orbits are in phase with the TESS transits.  The BIC also improves with respect to the solution without a planet for all datasets, showing a consistent preference in favor of the orbit model.  The Appendix \ref{app:periodogram-analysis-rvs} presents a periodogram analysis for the CCF and LBL data, both showing coherent peaks in the orbital period of $P=18.85$~d. These peaks become more relevant when subtracting the GP model.  As an additional test, we performed a joint analysis of the photometry, RV, and $B\ell$ data including the transit model, the RV orbit models and the GP activity model, simultaneously. However, since the GP model is more flexible than the orbit model, the GP tends to overfit the data, resulting in a less significant detection of the RV semi-amplitude. Our bisector analysis (see Appendix \ref{app:bisectoranalysis}) indicates an absence of activity-related signal in the RVs. Therefore, the GP model may actually be related to other spurious signals.

The Table \ref{tab:separateanalysisfitparameters} shows a comparison of the results obtained for the CCF and LBL data, and for a model with and without the GP component.  Note that regardless of the method used (CCF or LBL) or whether or not to include the GP component, the velocity semi-amplitude is constrained and all planet parameters have posterior distributions that are consistent within $1\sigma$. The GP activity component improves the final dispersion of the residuals and the BIC for both datasets. Our MCMC analysis assumes flat priors for the velocity semi-amplitude, $K_{p}=\mathcal{U}(0,100)$~\ms, therefore a null detection should result in a flat probability distribution towards zero. However, as illustrated in Figs. \ref{fig:toi-1759_pairsplot_ccf} and \ref{fig:toi-1759_pairsplot_ccf}, $K_{p}$ has a well-constrained posterior distribution at the upper and lower bounds for both data sets, defining an important constraint for the mass of TOI-1759~b. 

With LBL+GP we obtained the lowest BIC value, with a velocity semi-amplitude of $ K_{p}=2.3\pm0.7$~\ms\ and a RMS of residuals of 4.6~\ms. Therefore, the final fit parameters we adopt are those obtained by the LBL RVs with GP activity model. The best-fit system parameters and derived quantities are summarized in Table \ref{tab:toi-1759-final-params}. With a radius of $3.06\pm0.22$~\RE\ and a mass of $6.8\pm2.0$\ME, TOI-1759~b is confirmed as a planet with a mean density of $1.3\pm0.5$~g\,cm$^{-3}$, therefore it is likely to be a gas-dominated sub-Neptune.

\begin{table*}
\begin{footnotesize}

    \caption[]{Best-fit parameters obtained from our analyses using different combinations of RV data sets (CCF and LBL) and models (with and without a GP activity component). }
    \label{tab:separateanalysisfitparameters}
    \begin{center}
    \begin{tabular}{lcccc}
        \hline
        \noalign{\smallskip}
      parameter   & CCF  &  LBL  &  CCF+GP  &  LBL+GP \\
        \noalign{\smallskip}
        \hline
RV semi-amplitude, $K_{p}$ [m\,s$^{-1}$] & $3.4\pm0.7$ & $2.4\pm0.7$ & $3.2\pm0.7$ & $2.3\pm0.7$ \\
planet mass, $M_{p}$ [\ME] & $10.2\pm2.1$ & $7.2\pm2.0$ & $9.7\pm2.1$ & $6.8\pm2.0$ \\
systemic velocity, $v_{\rm sys}$ [m\,s$^{-1}$] & $-61072.2\pm0.5$ & $-60993.2\pm0.4$ &  $-61072.2\pm0.5$ & $-60993.2\pm0.4$ \\
orbital period, $P$ [d] & $18.849986\pm0.000005$ & $18.849979\pm0.000006$ & $18.849983\pm0.000005$ & $18.849975\pm0.000006$ \\
GP mean, $\mu_{v}$ [m\,s$^{-1}$] & - & - &  $0\pm2$ &  $0^{+2}_{-3}$ \\ 
GP white noise, $\sigma_{v}$  [m\,s$^{-1}$] & - & - & $0.5^{+1.1}_{-0.8}$ & $0.1^{+0.5}_{-0.3}$ \\ 
GP amplitude, $\alpha_{v}$  [m\,s$^{-1}$] & - & - & $5^{+3}_{-2}$ & $3^{+4}_{-2}$\\ 
GP decay time, $l_{s}$ [d] & - & - & $141^{+389}_{-192}$ & $195^{+352}_{-301}$\\ 
GP smoothing factor, $\beta_{s}$  & - & - & $0.22^{+0.13}_{-0.06}$ & $0.3^{+0.5}_{-0.4}$\\ 
GP period, $P_{\rm rot}$  [d] & - & - & $36^{+42}_{-8}$ & $35^{+31}_{-21}$ \\
$\chi^2$   &  1.23 & 0.75 & 0.74 & 0.55 \\
RMS (no planet$^\dag$) [m\,s$^{-1}$] & 9.4 & 5.6 & 7.7 & 4.8 \\
RMS of residuals [m\,s$^{-1}$] & 9.3 & 5.4 & 7.4 & 4.6 \\
$\Delta$BIC$^\S$ (no planet$^\dag$) & 335 & 43 & 246 &  6 \\
$\Delta$BIC$^\S$  & 315 & 35 & 226 &  0 \\
        \hline
    \end{tabular}

\tablebib{
$^\dag$ RMS and BIC calculated for the RV data before subtracting the orbit model of TOI-1759~b; $^\S$ $\Delta$BIC = BIC - BIC$_{\rm min}$, for BIC$_{\rm min}$=19582.
}        
    \end{center}
    \end{footnotesize}
 \end{table*}

\begin{table*}
    \caption[]{Summary of the final fit parameters of TOI-1759 from the joint MCMC analysis of the TESS photometry and LBL SPIRou RVs. }
    \label{tab:toi-1759-final-params}
    \begin{center}
    \begin{tabular}{lccccc}
        \hline
        \noalign{\smallskip}
Parameter & unit & fit value \\
        \noalign{\smallskip}
        \hline

time of conjunction, $T_{c}$ & BJD & $2458745.4661\pm0.0010$ \\

orbital period, $P$ & d & $18.849975\pm0.000006$  \\

normalized semi-major axis, $a/R_{\star}$ & - & $36\pm5$ \\
semi-major axis, $a_{p}$ $^\dag$ & au & $0.1176\pm0.0013$ \\

transit duration, $t_{\rm dur}$ & h & $3.7\pm0.9$ \\

orbital inclination, $i_{p}$ & deg &  $89.2\pm0.5$  \\
impact parameter, $b$ & - & $0.5\pm0.3$ \\

eccentricity, $e$ $^\S$ & - & 0  \\

planet-to-star radius ratio, $R_{p}/R_{\star}$  & - & $0.049\pm0.003$ \\
planet radius, $R_{p}$ & \RJ & $0.279\pm0.020$ \\
planet radius, $R_{p}$ & \RN  & $0.79\pm0.06$ \\
planet radius, $R_{p}$ & \RE & $3.06\pm0.22$ \\

velocity semi-amplitude, $K_{p}$ & m\,s$^{-1}$ & $2.3\pm0.7$ \\

planet mass, $M_{p}$ & \MJ & $0.021\pm0.006$ \\
planet mass, $M_{p}$ & \MN & $0.40\pm0.12$ \\
planet mass, $M_{p}$ & \ME & $6.8\pm2.0$ \\

planet density, $\rho_{p}$ & g\,cm$^{-3}$ & $1.3\pm0.5$ \\

equilibrium temperature, $T_{\rm eq}$ $^{\dag\dag}$ & K & $433\pm14$ \\

linear limb dark. coef., $u_{0}$ & - & $0.4^{+0.5}_{-0.7}$ \\
quadratic limb dark. coef., $u_{1}$ & - & $0.5\pm1.0$ \\

RMS of RV residuals & m\,s$^{-1}$ & 4.6 \\
RMS of flux residuals & ppm & 1093 \\
        \hline
    \end{tabular}
    
\tablebib{
$^\dag$ semi-major axis derived from the fit period and the Kepler's law; $^\S$ assuming a circular orbit, that is, the eccentricity is fixed to zero; $^{\dag\dag}$ assuming a uniform heat redistribution and an arbitrary geometric albedo of 0.1. 
}    
    \end{center}
  \end{table*}

\section{TOI-1759~b as a potential target for atmospheric characterization}
\label{sec:discussion}

We consider the best-fit parameters from our analysis to calculate the habitable zone (HZ) for TOI-1759 using the equations and data from \cite{Kopparapu2014}, which gives an optimistic lower limit (recent Venus) at 0.24~au, and an upper limit (early Mars) at 0.61~au, with the runaway greenhouse limits ($M_{\rm p}=1$~\ME) ranging between 0.31~au and 0.58~au. TOI-1759~b resides at an orbital distance of $0.1176\pm0.0013$~au and receives a flux of 6.4 times the flux incident on Earth, and therefore it is not inside the HZ. We estimate the equilibrium temperature for TOI-1759~b as in \cite{heng2013} assuming a uniform heat redistribution and an arbitrary geometric albedo of 0.1, which gives $T_{\rm eq}=433\pm14$~K, showing that this sub-Neptune is in a temperate region.

With a J magnitude of 8.7 for TOI-1759, the transits of TOI-1759~b can potentially be searched for atmospheric signatures.  To characterize the potential for the detection of atmospheric absorption lines, we can evaluate the ratio between the atmospheric absorption depth and noise in the transit light curve. However, since the noise level will depend on the telescope, instrument, spectral range, etc., and the atmospheric absorption depth will depend on the species, abundances, number and oscillator strength of the searched lines, etc., at this point one can only calculate a relative signal-to-noise ratio of detection (S/N) to be compared with other planets and observed with the same instrument.  

Here, we made the calculation in the J band, which is currently available with space and ground based facilities, and which includes the H$_2$O bands observed in many studies of exoplanet atmospheres \citep[e.g.,][]{Fraine_2014, Benneke_2019, Tsiarias_2019, Mikal-Evans_2021}. The atmospheric absorption depth, which is the fraction of the stellar flux that is absorbed by the atmosphere during transit, is proportional to the area of the absorbing layer, which is given by the scale height of the atmosphere ($H$) times $2\pi R_p$, for $R_p$ being the planet radius, and inversely proportional to the area of the stellar disk ($\pi R_{star}^2$). The noise is simply assumed to be proportional to the square root of the stellar flux given by $F_J \propto 10^{-0.4 m_{\rm J}}$,  where $m_{\rm J}$ is the J magnitude of the star. The atmospheric scale height is given by $kT/\mu g$, where $k$ is the Boltzmann constant, $T$ is the atmosphere temperature, $\mu$ is the mean molar molecular mass, and $g$ the planet gravity in the atmosphere. Finally, with $g\propto M_p/R_p^2$, where $M_p$ is the planet mass, we have a signal to noise ratio $S/N \propto H R_p / R_{star}^2 \sqrt{F_{\rm J}}$, in agreement with the TSM (transmission spectroscopic metric) defined by \cite{Kempton_2018}, \citep[see also,][]{Cointepas_2021}.

\begin{figure}
   \centering
   \includegraphics[width=1\hsize]{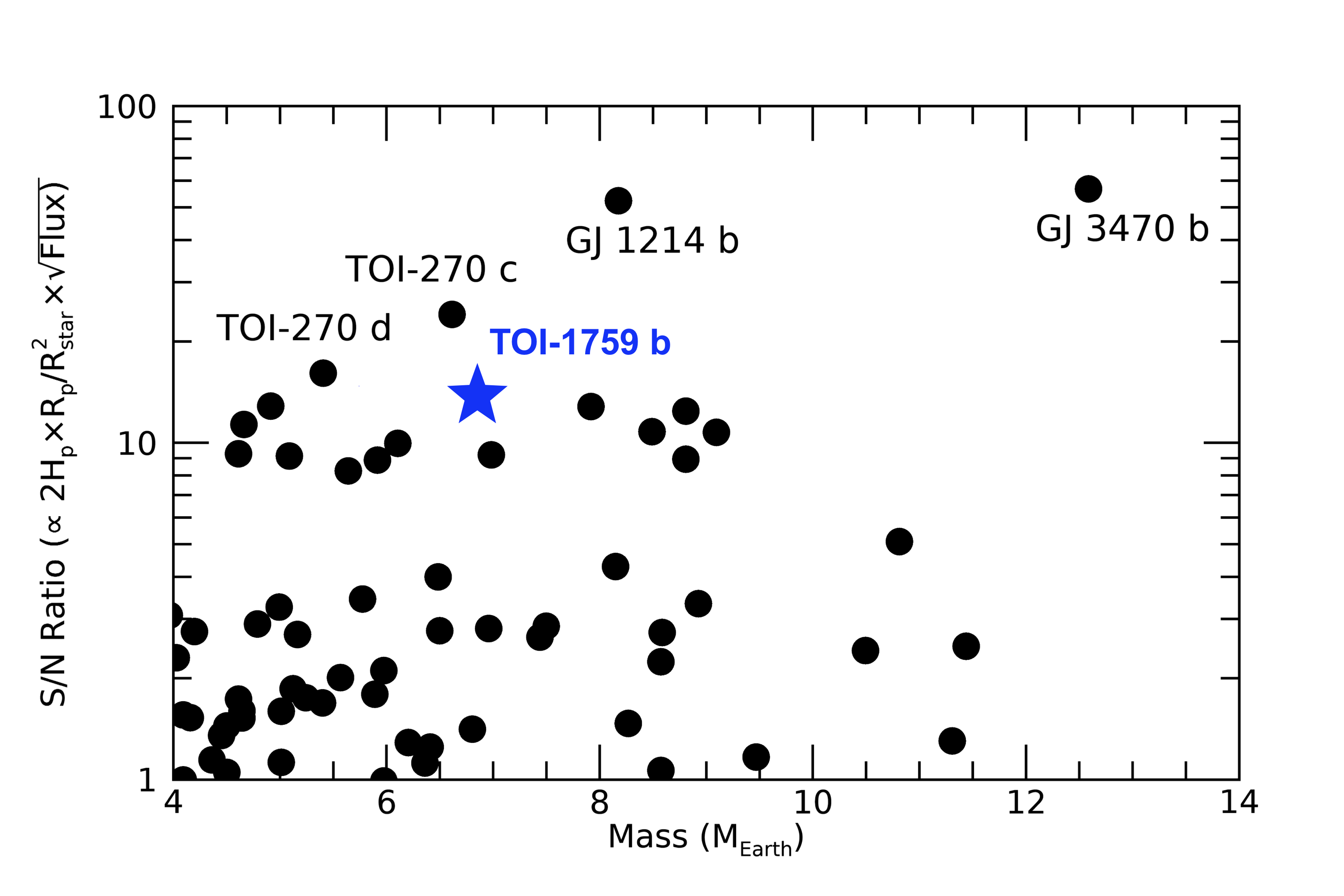}
      \caption{Signal to noise ratio of the atmospheric signatures in the J band as a function of the planetary mass for exoplanets orbiting M~type stars with masses between 8 and 14~Earth mass. The S/N ratios have been normalized to a reference S/N of 100 for the planet AU~Mic~b. In the considered mass range, the best S/N ratios are obtained for the exoplanets GJ1214~b and GJ3470~b with S/N about half the one of AU~Mic~b.}
        \label{fig:SN_vs_Mass}
  \end{figure}

We calculate the S/N ratio expected for all known exoplanets transiting an M-type star, and normalized them to a value of 100 for the best case of AU~Mic~b. We used the catalog of exoplanets given by the Exoplanets Encyclopedia on the 1$^{\mathrm st}$ August, 2021 \citep{Schneider_2011}. For the J magnitudes, we used the tabulated values when available, or calculated theoretical values from the V magnitudes and the stars effective temperatures assuming a black-body spectrum. A star is considered to be an M-type if it is catalogued as such or if its effective temperature is between 2200\,K and 4100\,K.  The result is shown in Fig.~\ref{fig:SN_vs_Mass} where we plotted the S/N ratio of the atmospheric signatures in the J band as a function of the planetary mass for known exoplanets orbiting M~type stars with masses between 4 and 14~\ME. In this mass range, TOI-1759~b is the forth best S/N after GJ3470~b, TOI-270~c and TOI-270~d with a S/N of about 1/4 of GJ3470~b and half the one of TOI-270~c.

In addition to the characterization of the deep atmosphere, TOI-1759~b provides interesting prospect in the search for evaporation signature. This planet shares many similar properties with GJ3470~b (planetary radius, equilibrium temperature, stellar type and effective temperature). GJ3470~b has shown a deep signature of an escaping atmosphere in Lyman-$\alpha$ \citep{Bourrier_2018}. The main difference is that with a semi-major axis of 0.036~au,  GJ3470~b is about three times closer to its star. Therefore TOI-1759~b is farther to the evaporation limit \citep{Lecavelier_2007}.   

To calculate the evaporation rate of TOI-1759~b, we use the hydrodynamic escape model from \cite{Allan2019}. This model calculates the optical depth for the XUV (EUV plus X-ray) photons of the star, which penetrate in the upper atmosphere of the planet. For simplicity, we assume that the EUV photon energy is concentrated at 20~eV. These photons can then locally ionize neutral hydrogen atoms and the excess energy above the ionization threshold (i.e., $>13.6$ eV) is then used to heat the atmosphere, which expands and more easily evaporates. This bulk atmospheric outflow can potentially be detected in Lyman-$\alpha$ transit observations. One of the key inputs for the photoevaporation model is the high-energy XUV flux from the star incident on the planet. Given that the XUV flux is unknown for TOI-1759, we use empirical relations between magnetism and X-ray flux from \cite{Vidotto2014} to first estimate the X-ray flux of this star. Using the average magnetic fields reported in Table 2 and Figure 6 from \cite{Vidotto2014}, we estimate the X-ray flux of TOI-1759 to be  $\sim 6 \times 10^{5}$  and $\sim 1.5 \times 10^{6}$ erg/cm$^2$/s, if we consider an average field of 10 or 18~G, respectively. With these X-ray fluxes, we then use the empirical relations from \citep[][see their equations 19 and 21]{Johnstone2021} to estimate the EUV flux of this star to be  $\sim 1.1 \times 10^{6}$ and $\sim 2.1 \times 10^{6}$ erg~cm$^{-2}$~s$^{-1}$, which results in a total XUV flux of $\sim 1.7 \times 10^{6}$ and $\sim 3.6 \times 10^{6}$ erg/cm$^2$/s at the stellar surface. We note here that our values should be regarded as estimates, as the spread in empirical relations can be significant \citep[see][]{Vidotto2014}.

Given the model dependence on the XUV stellar flux, it is worth comparing our estimated values with that from GJ3470. Using values from \cite{Bourrier_2018}, we find an XUV surface flux for GJ3470 of $7.7\times 10^5$ erg/cm$^2$/s, which is comparable to that of TOI-1759, albeit a factor of few lower. Naively, we would expect that the faster rotation of GJ3470 (about 20-day period) and its likely younger age ($\sim 2$ Gyr, \citealt{Bourrier_2018}) would have implied in a larger XUV surface flux of GJ3470, compared to that of TOI-1759, in contrast to what we found. This discrepancy could be due to the scatter in the relations we used here (see previous paragraph). 

Using the lower-bound XUV flux value we found ($\sim 1.7 \times 10^{6}$ erg/s/cm$^2$), at the orbit of TOI-1759~b, the incident stellar flux is 940 erg/s/cm$^2$, which we adopt in our atmospheric escape model for TOI-1759~b.  Assuming the planet has a hydrogen atmosphere, we find an escape rate of $1.4 \times 10^{10}$~g~s$^{-1}$, which is remarkably similar to the rate derived for GJ3470~b \citep{Bourrier_2018}. Given these similarities with GJ3470~b, TOI-1759~b might be an interesting target to observe neutral hydrogen escape.

In conclusion, the discovery of TOI-1759~b provides an interesting target with prospects for the observation of the deep atmosphere. However given the possible high XUV flux, it is a potentially extremely interesting target in a search for escaping upper atmosphere in Lyman-$\alpha$ similar to GJ3470~b. Even the detection of the deep atmosphere will be in the capabilities of forthcoming facilities, which will aim at observing dozens of exoplanets atmospheres like the ESA space mission Ariel. In the sub-Neptune mass domain, it will be an interesting planet to be included in the first priority targets list.

\section{Conclusions}
\label{sec:conclusions}

We presented a detection of the transiting exoplanet TOI-1759~b and the characterization of the TOI-1759 system using TESS photometry and SPIRou/CFHT spectropolarimetry observations.  The planet has a radius of $3.06\pm0.22$~\RE\ and a mass of $6.8\pm2.0$~\ME, and therefore belongs in the sub-Neptune class, with a mean density of $1.3\pm0.5$~g\,cm$^{-3}$. It orbits at a distance of $0.1176\pm0.0013$~au from a moderately active cool dwarf star and its equilibrium temperature is $433\pm14$~K.

We measured the Doppler velocity shift of the star to a few \ms\ from our high resolution near infrared SPIRou spectra through both the CCF and LBL methods. These observations constrain the velocity semi-amplitude of the planet's orbit to within 3~$\sigma$. In a joint Bayesian MCMC analysis of the TESS photometry and SPIRou RVs we fitted a Keplerian model of the planet's RV orbit and a transit model to constrain the system's parameters. 

In addition, the SPIRou circularly polarized spectra detect the Zeeman signature of the photospheric magnetic field of TOI-1759, which allowed us to characterize the magnetic properties of this star. We found that the longitudinal magnetic field is modulated by the star rotation, providing a star rotation period of $35.65^{+0.17}_{-0.15}$~d. We reconstructed the magnetic map of the star with the Zeeman Doppler imaging technique, where we found a predominantly poloidal field  with an intermediate axisymmetry level. The mean magnetic field remained constant between 2020 and 2021, with a strength of $B_{\rm mean}=18\pm4$~G. However, we detected a change in the axisymmetric component from $75\pm5$\% in 2020 to $54\pm5$\% in 2021, and a change in the tilt and strength of the dipole field from $22\pm5$~deg and $-23\pm5$~G in 2020 to $46\pm5$~deg and $-27\pm5$~G in 2021. 
 
Finally, we use our measurements of the stellar magnetic field and system's properties to estimate the photo-evaporation rate in TOI-1759~b, which gives $>1.4 \times 10^{10}$~g~s$^{-1}$. This makes it a promising exoplanet to search for escaping upper atmosphere in Lyman-$\alpha$ transit observations and a potential target for the observation of the deep atmosphere.  Therefore, TOI-1759~b is an important exoplanet for the understanding of the mechanisms that underlie the observed sub-Neptune radius desert. 

\begin{acknowledgements}
     We acknowledge funding from the French National Research Agency (ANR) under contract number ANR\-18\-CE31\-0019 (SPlaSH). This work is supported by the ANR in the framework of the Investissements d'Avenir program (ANR-15-IDEX-02), through the funding of the ``Origin of Life'' project of the Grenoble-Alpes University.  TV and CC acknowledge funding from the Technologies for Exo-Planetary Science (TEPS) CREATE program and from the Fonds de Recherche du Qu\'ebec - Nature et technologies. JFD acknowledges funding from the European Research Council (ERC) under the H2020 research \& innovation programme (grant agreements 740651 New Worlds). AAV acknowledges funding from the European Research Council (ERC) under the European Union's Horizon 2020 research and innovation programme (grant agreement n. 817540, ASTROFLOW). This paper includes data collected with the TESS mission, obtained from the MAST data archive at the Space Telescope Science Institute (STScI). We acknowledge the use of public TESS data from pipelines at the TESS Science Office and at the TESS Science Processing Operations Center. Resources supporting this work were provided by the NASA High-End Computing (HEC) Program through the NASA Advanced Supercomputing (NAS) Division at Ames Research Center for the production of the SPOC data products. Some of the observations in the paper made use of the High-Resolution Imaging instrument ‘Alopeke. ‘Alopeke was funded by the NASA Exoplanet Exploration Program and built at the NASA Ames Research Center by Steve B. Howell, Nic Scott, Elliott P. Horch, and Emmett Quigley. Data were reduced using a software pipeline originally written by Elliott Horch and Mark Everett. ‘Alopeke was mounted on the Gemini North telescope of the international Gemini Observatory, a program of NSF’s OIR Lab, which is managed by the Association of Universities for Research in Astronomy (AURA) under a cooperative agreement with the National Science Foundation. on behalf of the Gemini partnership:  the National Science Foundation (United States), National Research Council (Canada), Agencia Nacional de Investigación y Desarrollo (Chile), Ministerio de Ciencia, Tecnolog'{i}a e Innovaci\'{o}n (Argentina), Ministério da Ciência, Tecnologia, Inovações e Comunicações (Brazil), and Korea Astronomy and Space Science Institute (Republic of Korea). 
     
\end{acknowledgements}

%
%

\bibliographystyle{aa}

\bibliography{bibliography}

\begin{appendix}

\section{$B_\ell$ GP posteriors}
\label{app:blonggpposteriors}

\begin{figure*}
   \centering
       \includegraphics[width=1.0\hsize]{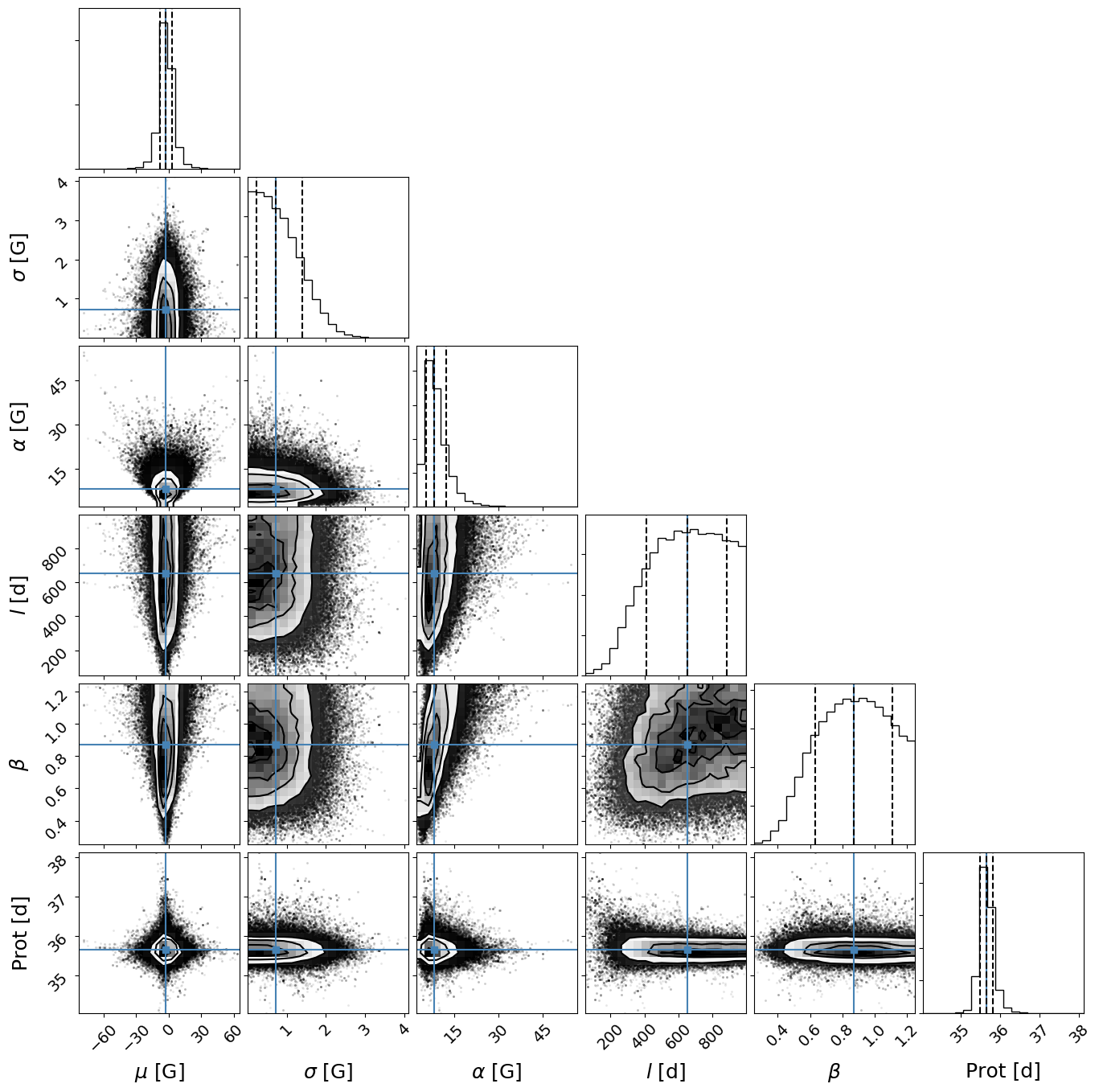}
      \caption{MCMC samples and the posterior distributions of parameters in the quasi-periodic GP analysis of the stellar activity in the SPIRou $B_\ell$ data.}
        \label{fig:Blong_QPgp_pairsplot}
  \end{figure*}

In this appendix we present the MCMC samples and posterior probability distributions of the quasi-periodic GP activity model parameters, as defined in Sect. \ref{sec:stellaractivity}, obtained for the longitudinal magnetic field ($B_\ell$) time series. Fig. \ref{fig:Blong_QPgp_pairsplot} illustrates these results.

\section{Comparison between CCF and LBL RVs}
\label{app:ccfvslbl}

We use two independent methods to calculate radial velocities, the CCF and the LBL. In this appendix we present a comparison between these two methods to verify the consistency between their results and if there is any residual systematics.

\begin{figure*}
   \centering
   \includegraphics[width=1.0\hsize]{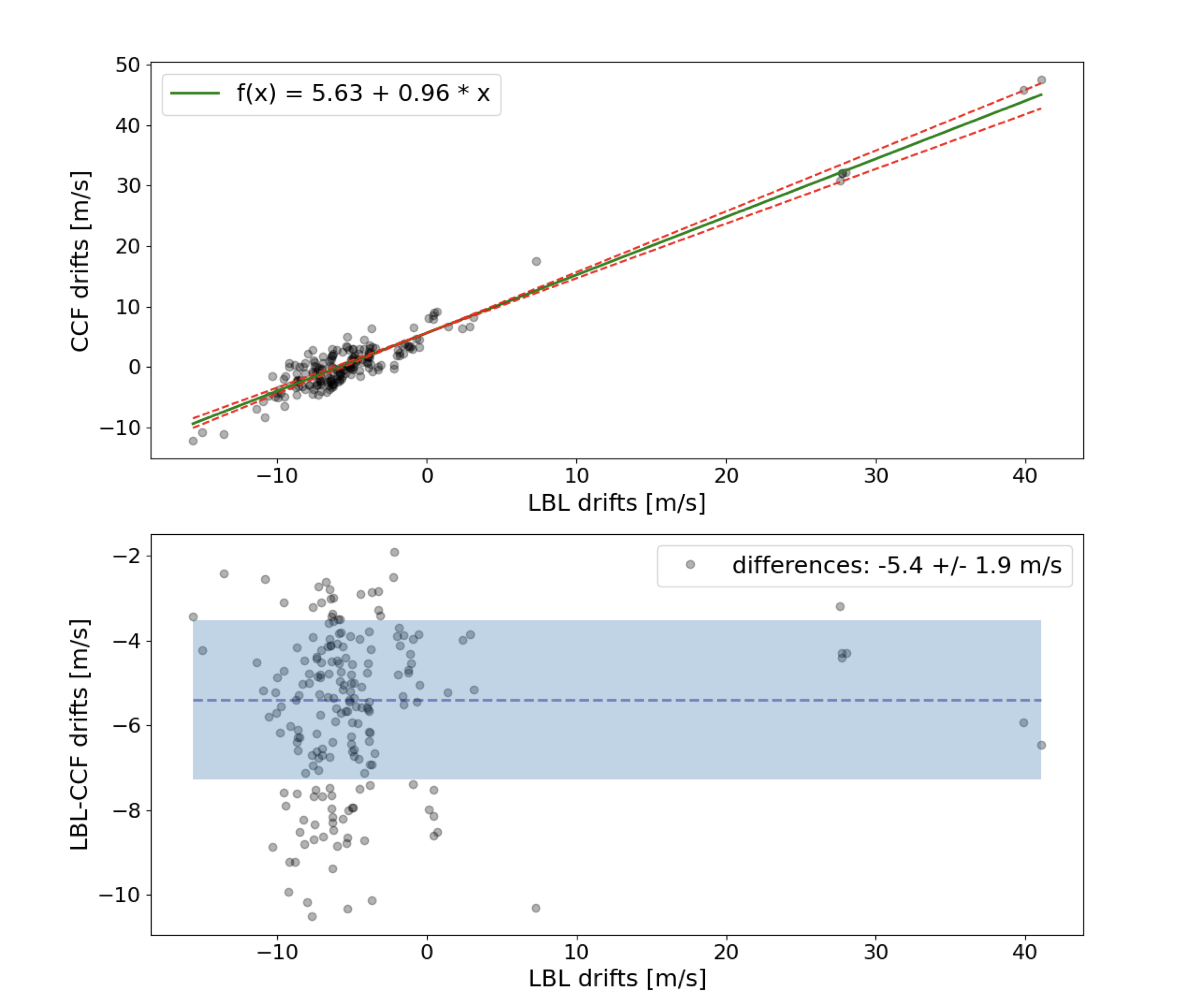}
      \caption{Comparison between CCF and LBL RV drifts measured from the Fabry-Perot spectra obtained by the simultaneous calibration fiber of SPIRou observations of TOI-1759. The top panel shows the CCF drifts versus LBL drifts (black circles), a linear fit to these quantities (green solid line) and its $1\sigma$ uncertainties (red dashed lines). The linear fit (parameters presented in the legend) shows a correlation of 0.96 between the CCF and LBL drifts. The bottom panel shows the differences between LBL and CCF drifts (black circles), the median (blue dashed line), and the $1\sigma=1.9$~\ms\ dispersion (blue shaded region). 
      }
        \label{fig:TOI-1759_ccf_vs_lbl_drifts}
  \end{figure*}

First, we compare the SPIRou instrumental drift measured from the simultaneous Fabry-Perot spectrum in the calibration fiber `C'.  Fig. \ref{fig:TOI-1759_ccf_vs_lbl_drifts} shows the drifts measured by the CCF method versus the drifts measured by the LBL method. Notice that there is a significant correlation of 0.96 between the drifts calculated by the two methods, with a median offset of $-5.4\pm1.9$~\ms\ of the LBL with respect to the CCF drifts. These measurements show that SPIRou has an absolute drift that varies within a typical range of $\pm4$~\ms\ (80th percentile), having much larger values ($\sim30-40$~\ms) on some occasions due to sporadic jumps of the instrument. The RMS dispersion of 1.9~\ms\ for the differences between the two measurements is in good agreement with the expected internal uncertainties, which is on the order of 1~\ms. The CCF drifts have a median error of 1.85~\ms and the LBL drift errors are not computed in the current version.

\begin{figure*}
   \centering
   \includegraphics[width=1.0\hsize]{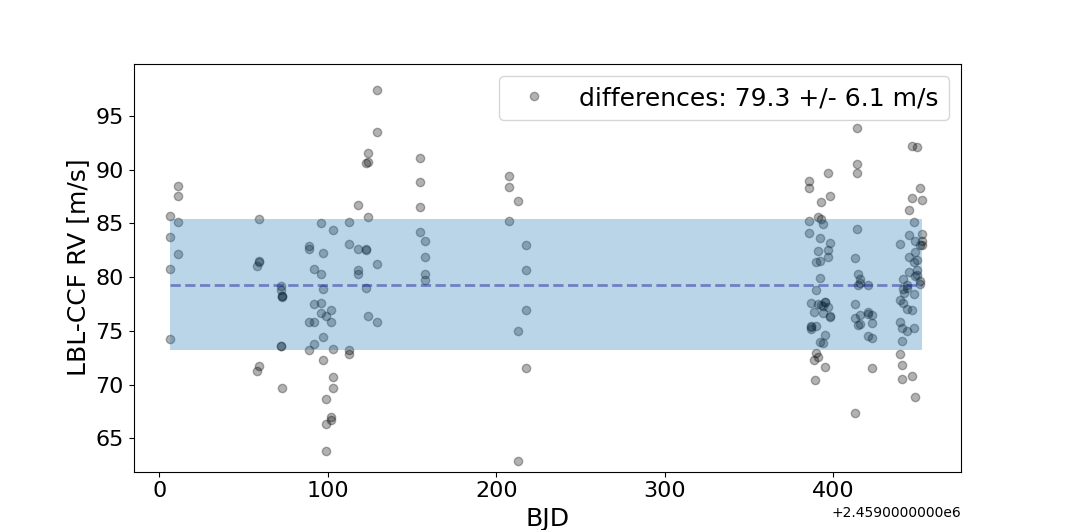}
      \caption{Comparison between CCF and LBL radial velocities of TOI-1759. The black circles show the differences between LBL and CCF RVs, the blue dashed line shows the median difference, and the blue shaded region represents the $1\sigma$ dispersion.}
        \label{fig:TOI-1759_ccf_vs_lbl_rvs}
  \end{figure*}

Now we compare the RV measurements of TOI-1759 performed by both methods.  Fig. \ref{fig:TOI-1759_ccf_vs_lbl_rvs} shows the time series of the difference between LBL and CCF RVs, where we did not detect any obvious systematics. The two methods provide RVs with a median offset of 79~\ms\ and a RMS dispersion of 6~\ms. The latter is also in agreement with the expected dispersion of 7~\ms, derived from the final RMS of 5.4~\ms\ and 4.6~\ms, assuming uncorrelated errors. 

\section{Fit models for CCF and LBL RVs}
\label{app:orbit+activitygpmodels}

\begin{figure*}
   \centering
   \includegraphics[width=0.75\hsize]{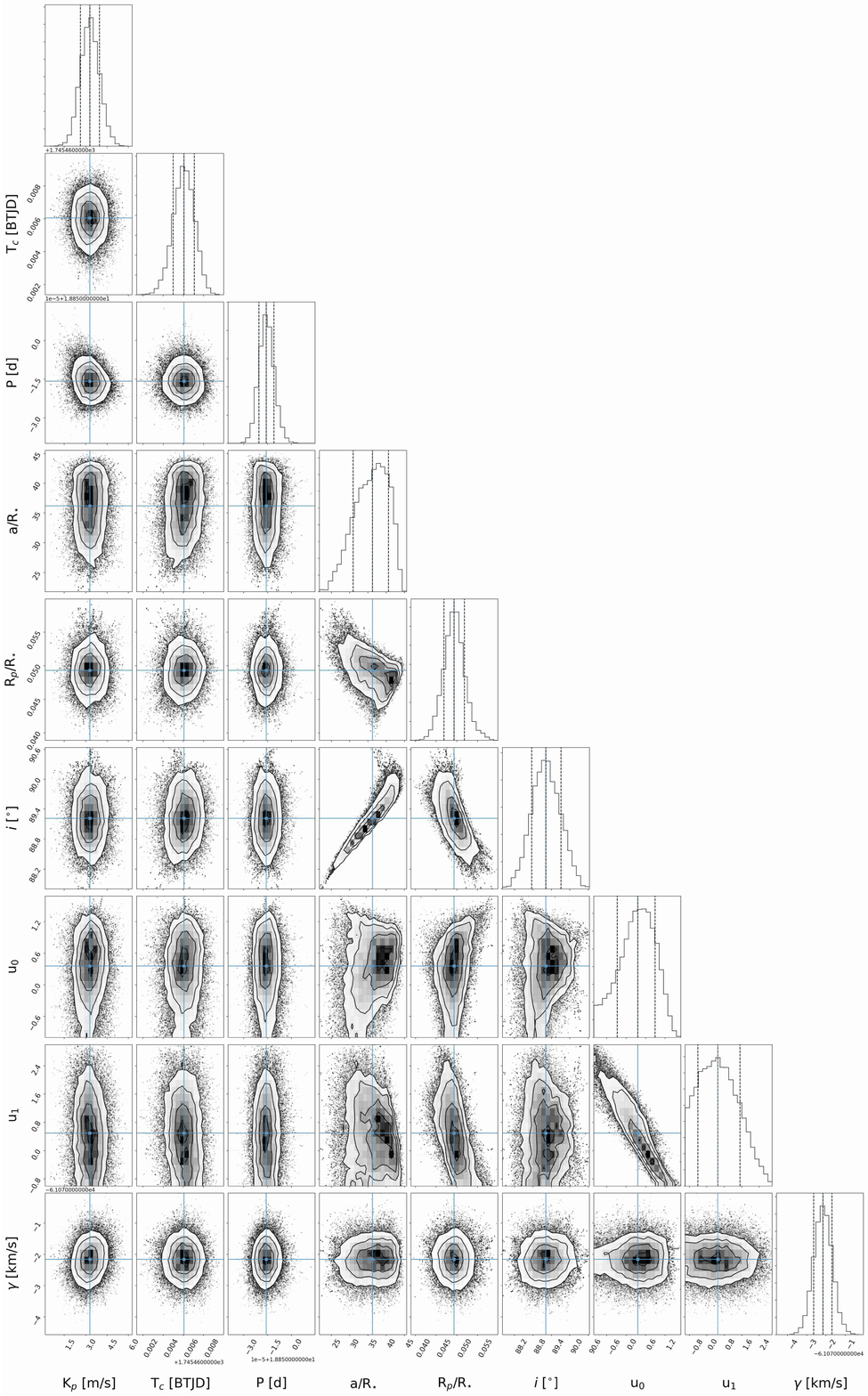}
      \caption{Pairs plot showing the MCMC samples and posterior distributions of the free parameters in our joint analysis of the TESS photometry and the SPIRou CCF RV data.  The contours mark the 1$\sigma$, 2$\sigma$, and 3$\sigma$ regions of the distribution. The blue crosses indicate the best fit values for each parameter and the dashed vertical lines in the projected distributions show the median values and the 1$\sigma$ uncertainty (34\% on each side of the median).
      }
        \label{fig:toi-1759_pairsplot_ccf}
  \end{figure*}

\begin{figure*}
   \centering
   \includegraphics[width=0.75\hsize]{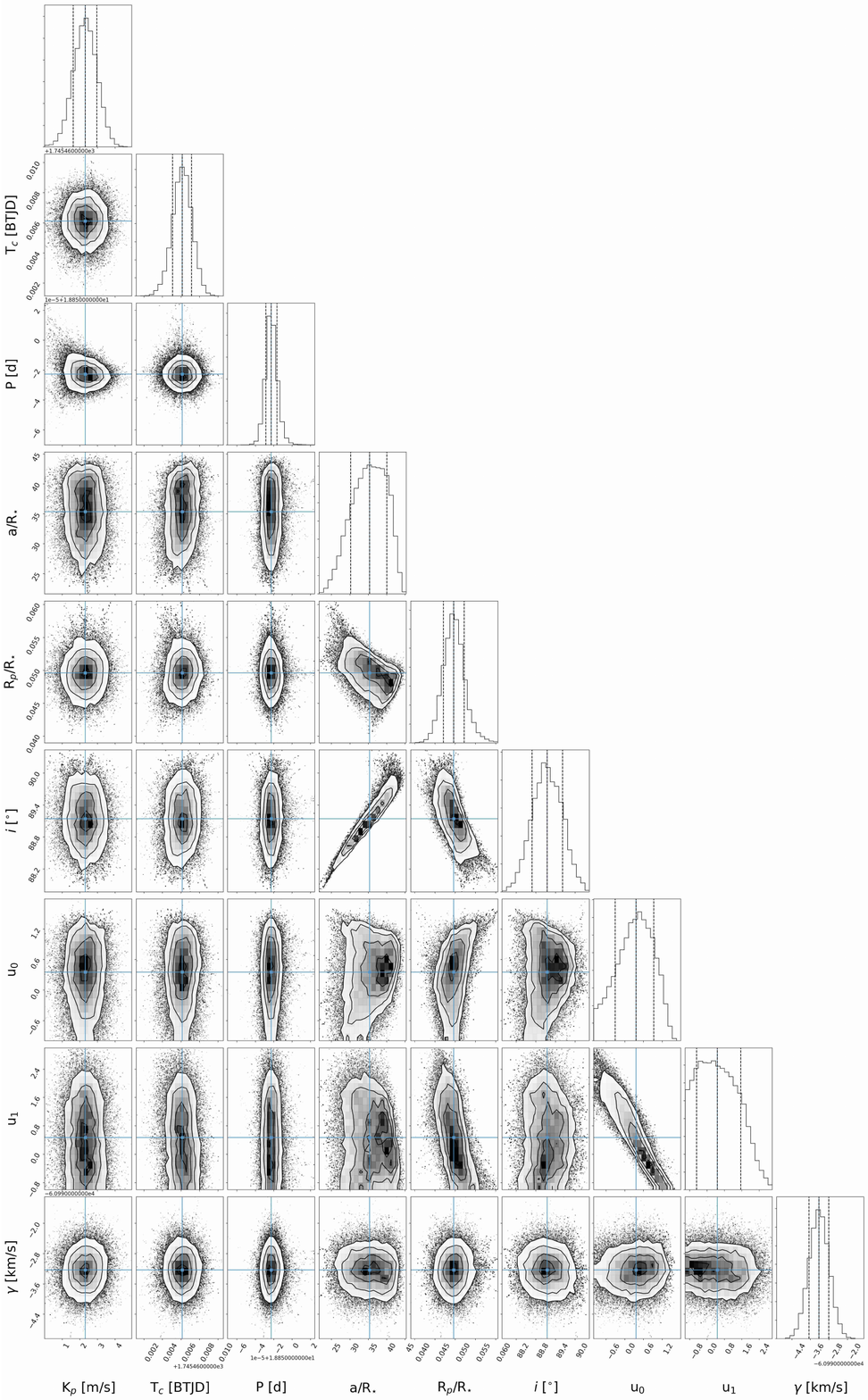}
      \caption{Same as Fig. \ref{fig:toi-1759_pairsplot_ccf} for the LBL RV data.
      }
        \label{fig:toi-1759_pairsplot_lbl}
  \end{figure*}

\begin{figure*}
   \centering
       \includegraphics[width=1.0\hsize]{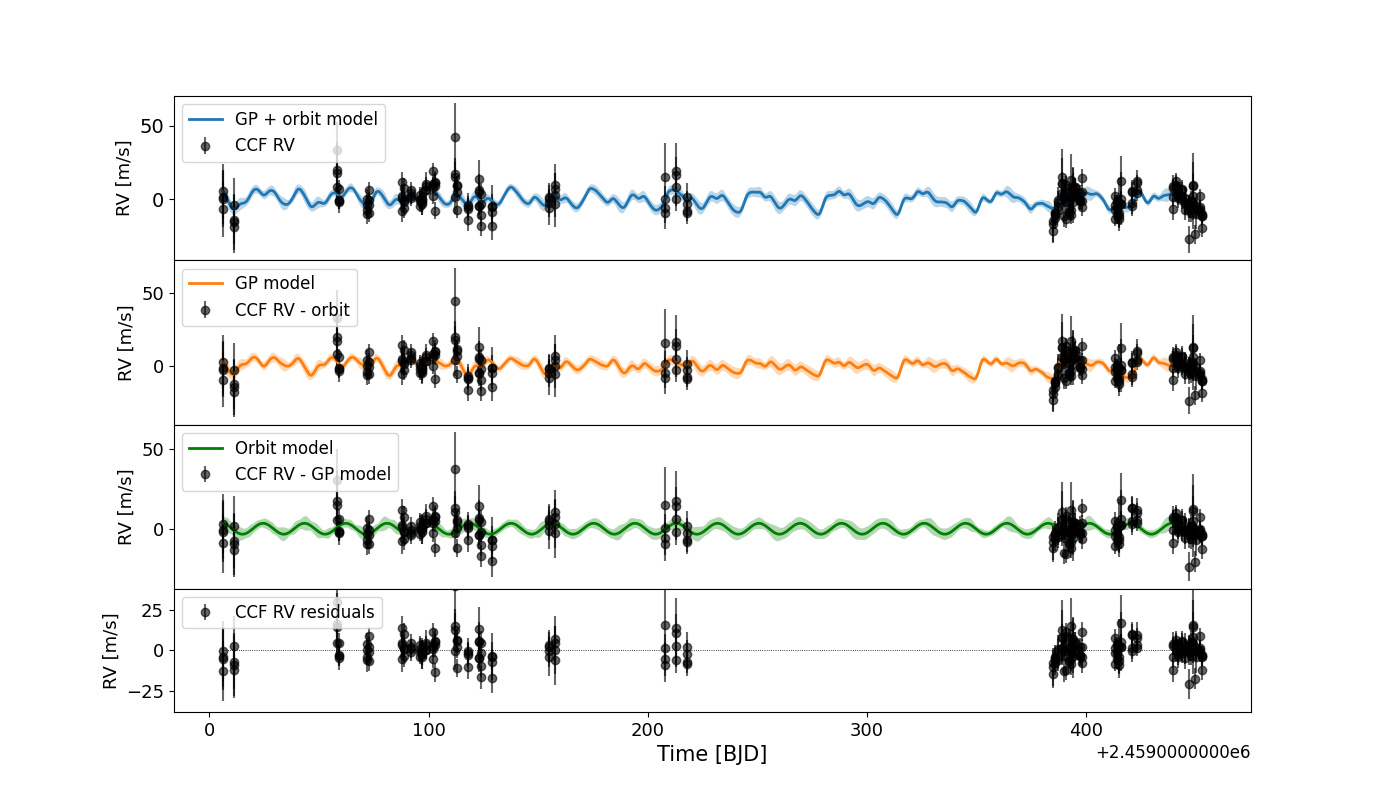}
      \caption{The best-fit models for the orbit of TOI-1759~b and the quasi-periodic GP model for the stellar activity obtained from our MCMC joint analysis of the TESS photometry and the SPIRou CCF RV data. From top to bottom, panels show: (1) the orbit+GP model and the CCF RV data; (2) the GP model and the CCF RV data minus the orbit model; (3) the orbit model and the CCF RV data minus the GP model; and (d) the residuals, that is, the CCF RV data minus the orbit + GP model.}
        \label{fig:toi1759_ccf_rv_fit}
  \end{figure*}

\begin{figure*}
   \centering
       \includegraphics[width=1.0\hsize]{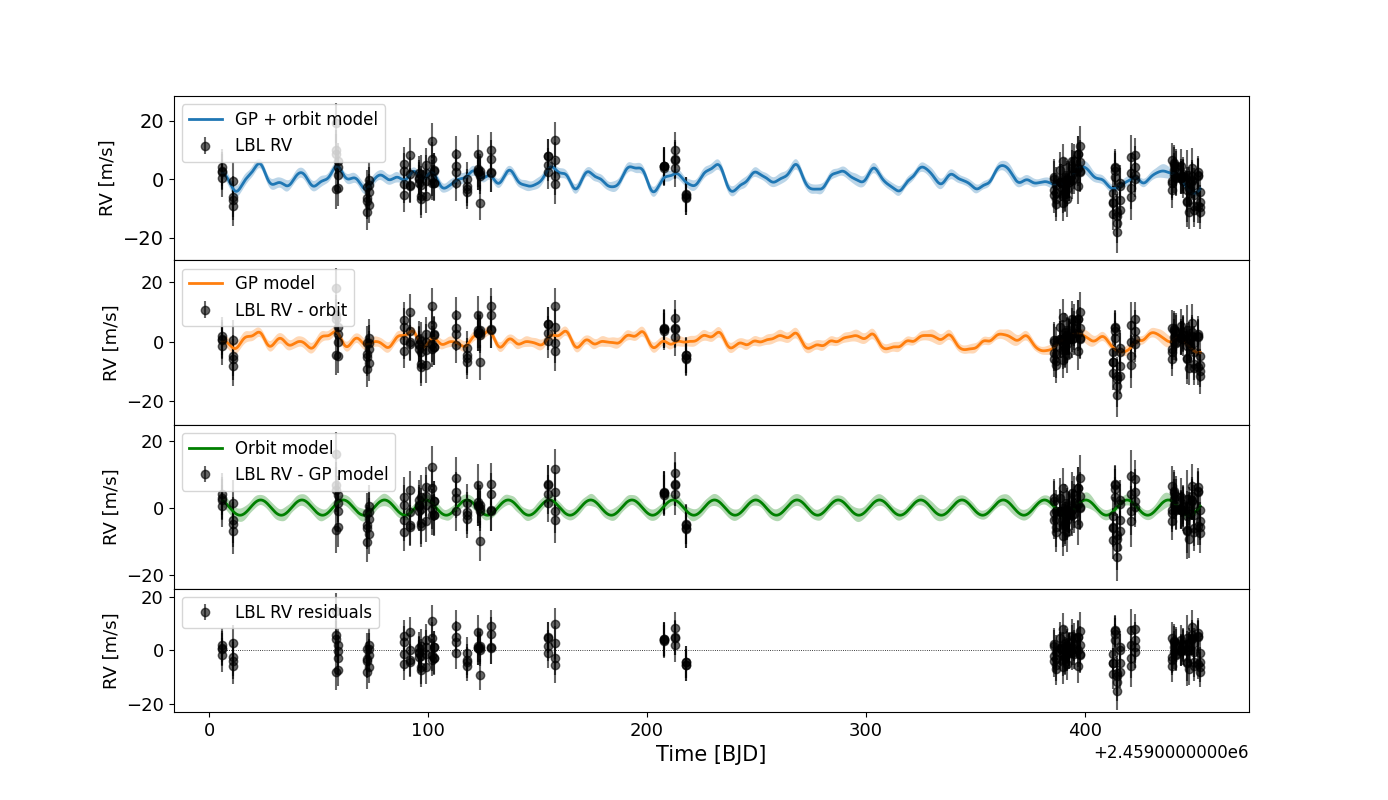}
      \caption{Same as Fig. \ref{fig:toi1759_ccf_rv_fit} for the LBL RV data.}
        \label{fig:toi1759_lbl_rv_fit}
  \end{figure*}

In this appendix we present the final results of our joint analysis of TESS photometry and SPIRou RV data, where we included a quasi-periodic Gaussian Process component to account for stellar activity in each dataset as described in Sect. \ref{sec:analysis}. Figs. \ref{fig:toi-1759_pairsplot_ccf} and \ref{fig:toi-1759_pairsplot_lbl} show the MCMC samples and posterior distributions of the transit and RV model parameters, and Figs. \ref{fig:toi1759_ccf_rv_fit} and \ref{fig:toi1759_lbl_rv_fit} show the fit models for the CCF and LBL RV data, respectively.

\section{Periodogram analysis of SPIRou RVs}
\label{app:periodogram-analysis-rvs}

\begin{figure*}
   \centering
   \includegraphics[width=1.0\hsize]{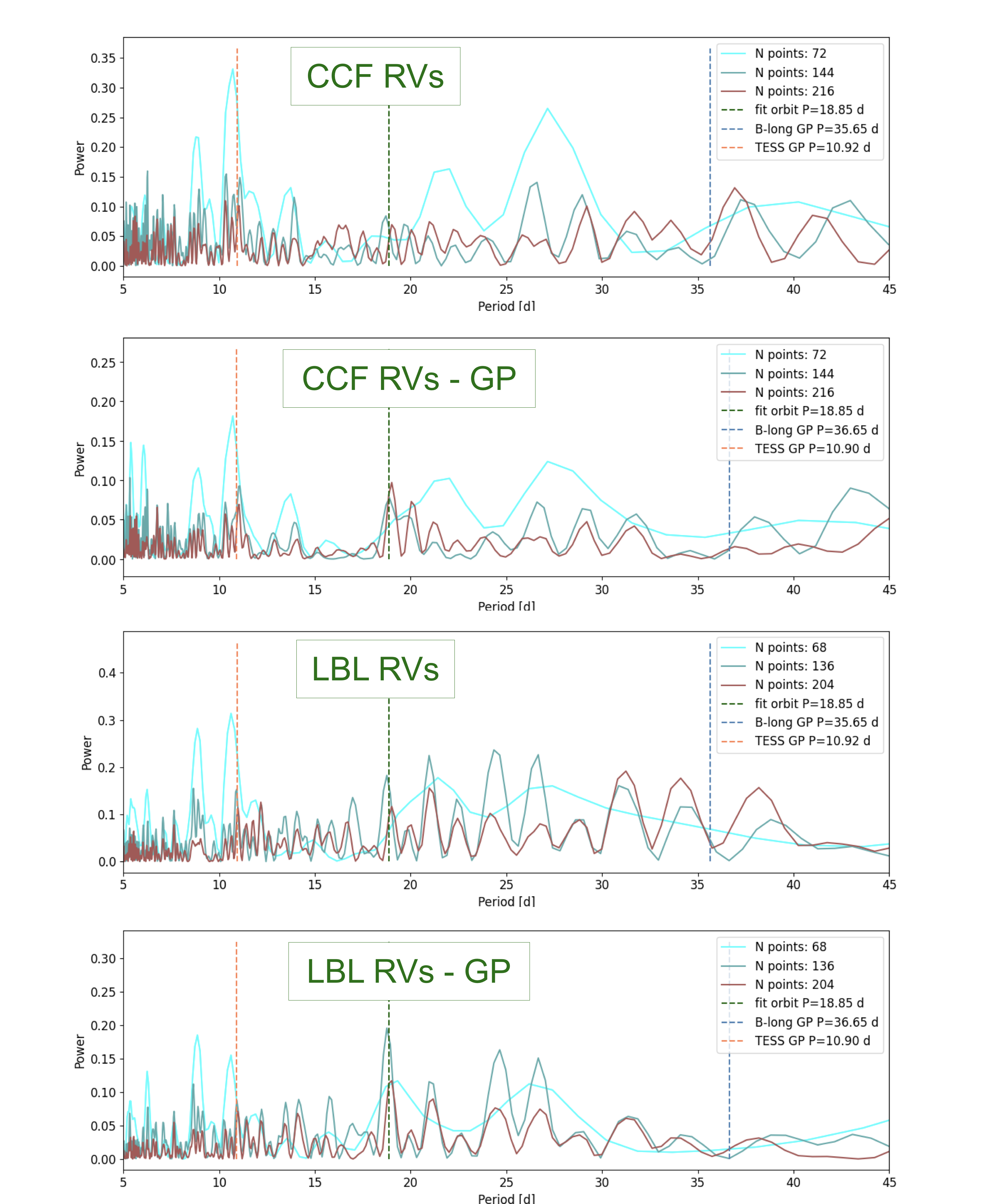}
      \caption{GLS periodogram analysis of SPIRou RVs of TOI-1759. The cyan, blue, and red lines show the GLS periodogram calculated for subsets containing 1-third, 2-thirds, and the entirety of the SPIRou RV data of TOI-1759, respectively. From top to bottom, panels show this analysis for CCF RVs, CCF RVs minus GP model, LBL RVs, and LBL RVs minus GP model. The removal of the GP model shows an improvement in the signature of the planet, expected at period of 18.850~d (green dashed line). The orange and blue dashed lines show the periods found by our GP analysis of the TESS data and $B_\ell$ data, respectively.
      }
        \label{fig:toi-1759_rv_periodogram}
  \end{figure*}

In this appendix we present a periodogram analysis of the SPIRou RV data to inspect the significance of detection of the orbital RV signal of TOI-1759~b.  Fig. \ref{fig:toi-1759_rv_periodogram} shows the conventional Generalized Lomb-Scargle periodogram (GLSP), where we calculated the GLSP for an increasing number of data points in a subset of our time series, starting at one-third, then two-thirds and then all data points. Both the CCF and LBL RVs show a peak at 18.85~d after removing the GP model, but a less significant peak when considering the data without subtracting the GP model. This low power reflects the marginal detection of the velocity semi-amplitude that we obtained in our analysis.

\begin{figure*}
   \centering
   \includegraphics[width=1.0\hsize]{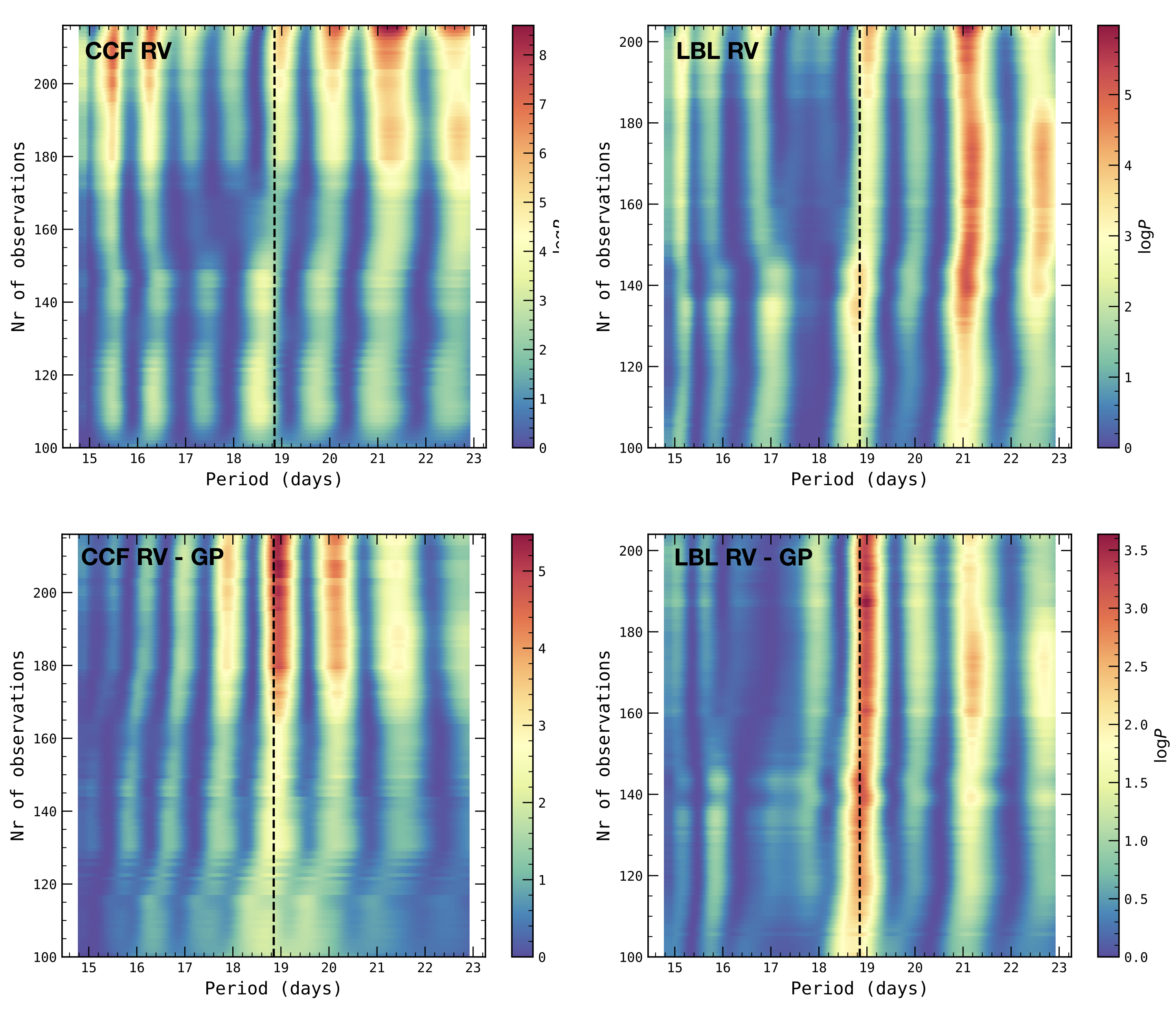}
      \caption{SBGLS periodogram analysis of the SPIRou RVs of TOI-1759. Each panel shows a color map of the power in the SBGLSP for each RV dataset. The vertical dashed lines show the orbital period of TOI-1759~b.
      }
        \label{fig:toi-1759_sbgls_periodogram}
  \end{figure*}

\begin{figure*}
   \centering
   \includegraphics[width=1.0\hsize]{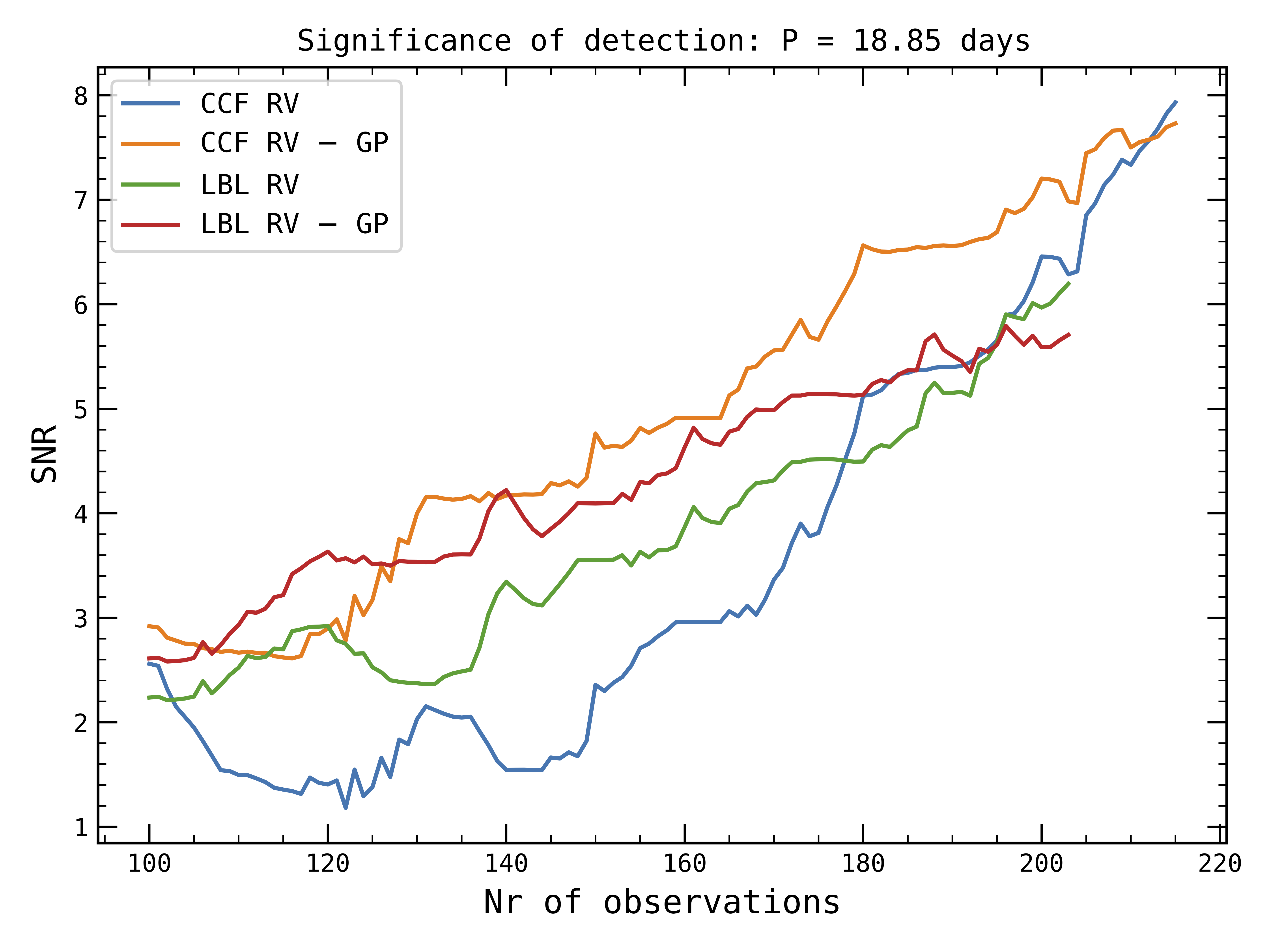}
      \caption{Significance of detection of the TOI-1759~b orbit RV signal at 18.85~d in the SBGLS periodogram analysis. Each curve represents the signal-to-noise ratio (SNR) as a function of the number of observations. The SNR is calculated by the ratio between the SBGLSP power at 18.85~d and the noise in the periodogram. There is a monotonic increase in the SNR with Nobs for all data sets, showing a coherent nature of the signal detected.
      }
        \label{fig:TOI-1759_sbgls_snr_vs_nobs}
  \end{figure*}

To further inspect the statistical significance of this faint signal, we employed the Stacked Bayesian General Lomb-Scargle periodogram (SBGLSP) analysis of \cite{MortierAndCameron2017}. The result is illustrated in Fig. \ref{fig:toi-1759_sbgls_periodogram}, where it shows that our RV data indeed presents a faint but coherent signal at 18.85~d, showing an increasing power with the number of observations. The signal becomes stronger and more evident after removing the GP model from both the CCF and LBL RVs. Finally, we calculate the signal-to-noise ratio of the 18.85~d peak in the SBGLSP as a function of the number of observations. Fig. \ref{fig:TOI-1759_sbgls_snr_vs_nobs} shows that both data sets (CCF and LBL), with and without removing the GP model, present a monotonic increase of the SNR with the number of observations, which confirms the coherence of the detected signal.

 Those analyses show that we detect the planet in the SPIRou data. The significance remains low, with detection of the RV semi-amplitude $K_p$ between 3 and 5\,$\sigma$ according the values reported in Table \ref{tab:separateanalysisfitparameters}. Still, the coherence of that signal and its agreement with both the period and the phase predicted from the TESS photometry allow us to conclude we significantly detect the planet.

\section{Bisector analysis}
\label{app:bisectoranalysis}

\begin{figure*}
   \centering
   \includegraphics[width=0.9\hsize]{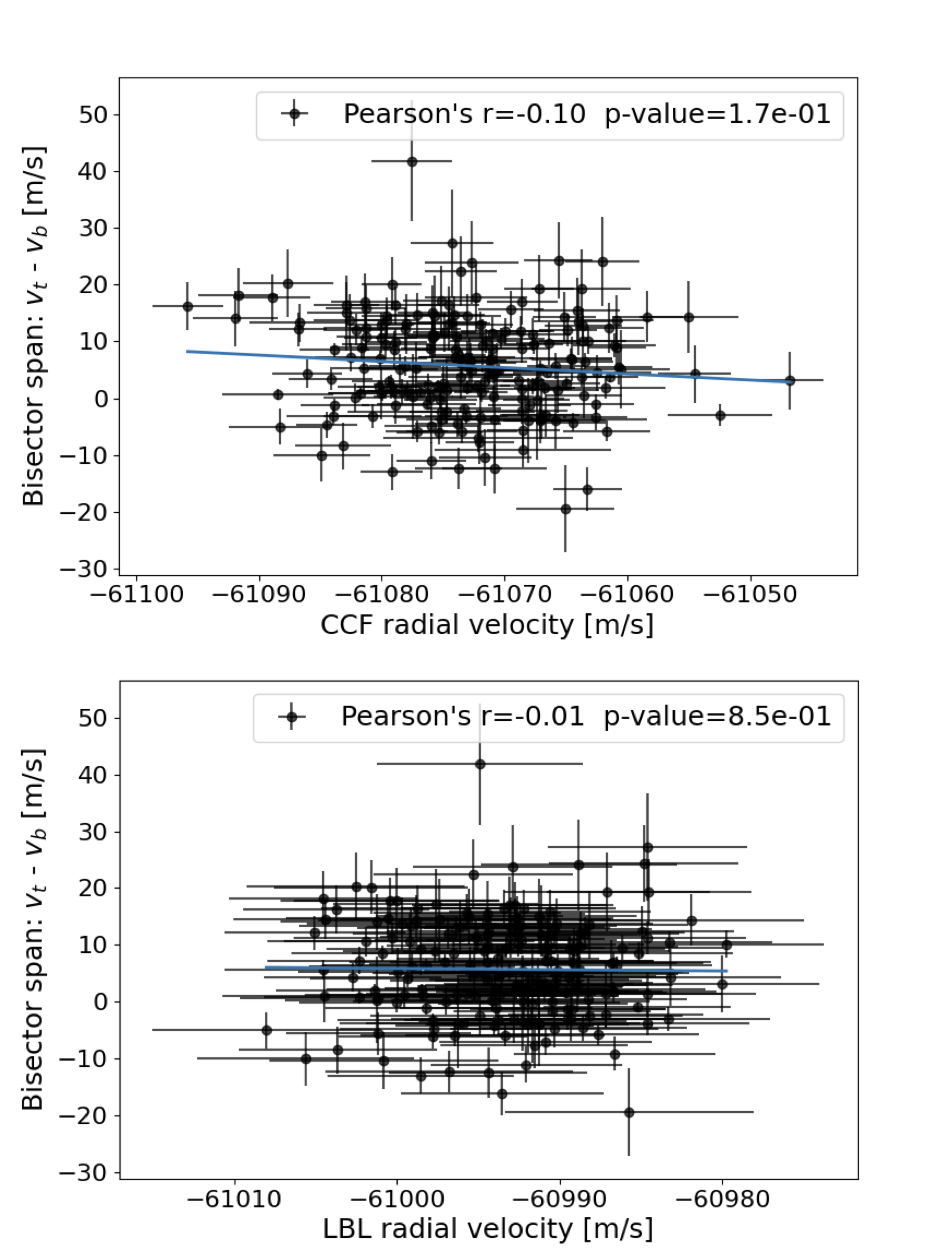}
      \caption{Bisector span versus SPIRou RVs. Top panel shows the CCF RVs and the bottom panel shows the LBL RVs, where we verify that there is no correlation between these quantities and the bisector span. The blue lines show a linear fit and the legends show the Pearson-r coefficients and their corresponding p-values.
      }
        \label{fig:bisector_rv_correlation}
  \end{figure*}

To check whether the RVs are correlated with some changes in the line profiles we use the bisector analysis as in \cite{Queloz2001} and \cite{Boisse2009}, where we calculate the bisector span, given by $B_{\rm span} = v_{t} - v_{b}$, where $v_{t}$ is the velocity shift of the bisector at the top of the CCF ($55\%  < {\rm depth} < 80\% $) and $v_{b}$ is the velocity shift of the bisector at the bottom of the CCF ($20\% < {\rm depth} < 40\% $).  Fig. \ref{fig:bisector_rv_correlation} shows that there is no significant correlation between $B_{\rm span}$ and the RVs, with a global Pearson-r coefficient of -0.10 and -0.01 and a p-value of 0.17 and 0.85, for CCF and LBL respectively. Therefore, the RVs do not vary significantly with the shape of the line profiles. Although our bisector analysis indicates the absence of an activity-related signal in the RVs of TOI-1759, further validation of this method in active stars is yet to be done with SPIRou.

\section{SPIRou log of observations, RVs, and $B_\ell$ data}
\label{app:spiroudata}

This appendix presents the log of SPIRou observations of TOI-1759 in Table \ref{tab:spirouobservations}, the CCF and LBL radial velocities in Table \ref{tab:spirouRVSCCF+LBL}, and the longitudinal magnetic field ($B_\ell$) data in Table \ref{tab:toi1759spiroublongdata}. 

\onecolumn

\begin{longtable}{ccccccccccc}
\caption{\label{tab:spirouobservations} Log of SPIRou observations of TOI-1759.} \\
\hline\hline
Epoch & UT Date & BJD & BERV & Exp. time & airmass & SNR & H$_{2}$O & Pol. seq. \\
 &  &  & (\kms) & (s) &  & at 1670~nm & rel. abs. & number \\
\hline
\endfirsthead
\caption{continued.}\\
\hline\hline
Epoch & UT Date & BJD & BERV & Exp. time & airmass & SNR & H$_{2}$O & Pol. seq. \\
 &  &  & (\kms) & (s) &  & at 1670~nm & rel. abs. & number \\
\hline
\endhead
\hline
\endfoot
1 & 2020-06-05T13:16:41 & 2459006.047578 & 9.9774 & 903 & 1.51 & 172 & 1.11 &  1 \\
1 & 2020-06-05T13:32:11 & 2459006.058353 & 9.9667 & 903 & 1.48 & 174 & 1.11 &  2 \\
1 & 2020-06-05T13:47:42 & 2459006.069133 & 9.9556 & 903 & 1.45 & 174 & 1.08 &  3 \\
1 & 2020-06-05T14:03:13 & 2459006.079901 & 9.9441 & 903 & 1.42 & 171 & 1.05 &  4 \\
2 & 2020-06-10T13:28:06 & 2459011.055684 & 10.4612 & 903 & 1.45 & 161 & 3.43 &  1 \\
2 & 2020-06-10T13:43:42 & 2459011.066528 & 10.4495 & 903 & 1.42 & 162 & 3.28 &  2 \\
2 & 2020-06-10T13:59:13 & 2459011.077292 & 10.4376 & 903 & 1.40 & 157 & 3.16 &  3 \\
2 & 2020-06-10T14:14:44 & 2459011.088067 & 10.4254 & 903 & 1.39 & 164 & 3.23 &  4 \\
3 & 2020-07-27T14:10:17 & 2459058.086711 & 11.1745 & 903 & 1.50 & 155 & 4.92 &  1 \\
3 & 2020-07-27T14:25:47 & 2459058.097483 & 11.1628 & 903 & 1.54 & 150 & 3.30 &  2 \\
3 & 2020-07-27T14:41:18 & 2459058.108256 & 11.1518 & 903 & 1.58 & 153 & 5.05 &  3 \\
3 & 2020-07-27T14:56:49 & 2459058.119030 & 11.1413 & 903 & 1.63 & 153 & 4.76 &  4 \\
4 & 2020-07-28T11:27:47 & 2459058.973896 & 11.2529 & 903 & 1.38 & 169 & 0.94 &  1 \\
4 & 2020-07-28T11:43:12 & 2459058.984605 & 11.2388 & 903 & 1.37 & 168 & 0.93 &  2 \\
4 & 2020-07-28T11:58:37 & 2459058.995313 & 11.2246 & 903 & 1.37 & 168 & 0.92 &  3 \\
4 & 2020-07-28T12:14:02 & 2459059.006021 & 11.2105 & 903 & 1.37 & 169 & 0.88 &  4 \\
5 & 2020-08-10T10:42:09 & 2459071.942673 & 10.2010 & 903 & 1.37 & 159 & 2.86 &  1 \\
5 & 2020-08-10T10:57:34 & 2459071.953381 & 10.1865 & 903 & 1.37 & 172 & 2.85 &  2 \\
5 & 2020-08-10T11:12:59 & 2459071.964091 & 10.1719 & 903 & 1.37 & 169 & 2.81 &  3 \\
5 & 2020-08-10T11:28:24 & 2459071.974799 & 10.1574 & 903 & 1.37 & 171 & 2.85 &  4 \\
6 & 2020-08-11T11:27:54 & 2459072.974483 & 10.0531 & 903 & 1.37 & 173 & 1.97 &  1 \\
6 & 2020-08-11T11:43:20 & 2459072.985191 & 10.0387 & 903 & 1.38 & 171 & 1.95 &  2 \\
6 & 2020-08-11T11:58:50 & 2459072.995964 & 10.0245 & 903 & 1.39 & 175 & 1.93 &  3 \\
6 & 2020-08-11T12:14:15 & 2459073.006672 & 10.0105 & 903 & 1.41 & 167 & 1.90 &  4 \\
7 & 2020-08-26T09:17:43 & 2459087.884529 & 8.2563 & 903 & 1.39 & 131 & 1.71 &  1 \\
7 & 2020-08-26T09:33:08 & 2459087.895238 & 8.2414 & 903 & 1.38 & 129 & 1.70 &  2 \\
7 & 2020-08-26T09:37:18 & 2459087.902075 & 8.2319 & 223 & 1.37 & 43 & 1.65 &  3 \\
7 & 2020-08-26T09:53:23 & 2459087.909300 & 8.2217 & 903 & 1.37 & 110 & 1.70 &  3 \\
7 & 2020-08-26T10:09:16 & 2459087.920331 & 8.2063 & 903 & 1.37 & 126 & 1.69 &  4 \\
8 & 2020-08-27T09:07:22 & 2459088.877370 & 8.1172 & 903 & 1.39 & 156 & 1.26 &  1 \\
8 & 2020-08-27T09:22:53 & 2459088.888143 & 8.1022 & 903 & 1.38 & 171 & 1.25 &  2 \\
8 & 2020-08-27T09:38:18 & 2459088.898851 & 8.0873 & 903 & 1.37 & 166 & 1.21 &  3 \\
8 & 2020-08-27T09:53:43 & 2459088.909560 & 8.0722 & 903 & 1.37 & 183 & 1.16 &  4 \\
9 & 2020-08-30T09:01:30 & 2459091.873371 & 7.6626 & 903 & 1.39 & 184 & 0.53 &  1 \\
9 & 2020-08-30T09:16:55 & 2459091.884080 & 7.6476 & 903 & 1.38 & 183 & 0.52 &  2 \\
9 & 2020-08-30T09:32:20 & 2459091.894788 & 7.6325 & 903 & 1.37 & 179 & 0.51 &  3 \\
9 & 2020-08-30T09:47:45 & 2459091.905497 & 7.6174 & 903 & 1.37 & 179 & 0.51 &  4 \\
10 & 2020-09-03T11:22:04 & 2459095.971094 & 6.8851 & 903 & 1.46 & 186 & 0.63 &  1 \\
10 & 2020-09-03T11:37:35 & 2459095.981867 & 6.8717 & 903 & 1.49 & 184 & 0.65 &  2 \\
10 & 2020-09-03T11:53:06 & 2459095.992640 & 6.8589 & 903 & 1.53 & 176 & 0.68 &  3 \\
10 & 2020-09-03T12:08:37 & 2459096.003412 & 6.8466 & 903 & 1.57 & 179 & 0.68 &  4 \\
11 & 2020-09-04T10:44:36 & 2459096.945095 & 6.7531 & 903 & 1.41 & 187 & 6.94 &  1 \\
11 & 2020-09-04T11:00:07 & 2459096.955868 & 6.7388 & 903 & 1.43 & 186 & 7.03 &  2 \\
11 & 2020-09-04T11:15:32 & 2459096.966576 & 6.7250 & 903 & 1.46 & 188 & 6.96 &  3 \\
11 & 2020-09-04T11:30:57 & 2459096.977284 & 6.7117 & 903 & 1.49 & 180 & 7.09 &  4 \\
12 & 2020-09-06T09:13:57 & 2459098.882192 & 6.5029 & 903 & 1.37 & 185 & 0.80 &  1 \\
12 & 2020-09-06T09:29:23 & 2459098.892901 & 6.4877 & 903 & 1.37 & 181 & 0.81 &  2 \\
12 & 2020-09-06T09:44:53 & 2459098.903674 & 6.4724 & 903 & 1.37 & 187 & 0.82 &  3 \\
12 & 2020-09-06T10:00:18 & 2459098.914382 & 6.4573 & 903 & 1.38 & 186 & 0.86 &  4 \\
13 & 2020-09-09T08:45:31 & 2459101.862503 & 6.0088 & 903 & 1.37 & 192 & 1.20 &  1 \\
13 & 2020-09-09T09:01:02 & 2459101.873276 & 5.9934 & 903 & 1.37 & 187 & 1.22 &  2 \\
13 & 2020-09-09T09:16:33 & 2459101.884049 & 5.9780 & 903 & 1.37 & 191 & 1.26 &  3 \\
13 & 2020-09-09T09:31:57 & 2459101.894757 & 5.9627 & 903 & 1.37 & 193 & 1.28 &  4 \\
14 & 2020-09-10T08:58:34 & 2459102.871592 & 5.8178 & 903 & 1.37 & 190 & 1.11 &  1 \\
14 & 2020-09-10T09:14:00 & 2459102.882301 & 5.8024 & 903 & 1.37 & 190 & 1.11 &  2 \\
14 & 2020-09-10T09:29:31 & 2459102.893074 & 5.7870 & 903 & 1.37 & 189 & 1.13 &  3 \\
14 & 2020-09-10T09:45:01 & 2459102.903846 & 5.7718 & 903 & 1.38 & 191 & 1.15 &  4 \\
15 & 2020-09-19T07:37:59 & 2459111.815780 & 4.2113 & 903 & 1.39 & 111 & 3.87 &  1 \\
15 & 2020-09-19T07:53:24 & 2459111.826488 & 4.1959 & 903 & 1.38 & 119 & 3.78 &  2 \\
15 & 2020-09-19T08:08:55 & 2459111.837262 & 4.1803 & 903 & 1.37 & 86 & 3.67 &  3 \\
15 & 2020-09-19T08:24:20 & 2459111.847968 & 4.1648 & 903 & 1.37 & 63 & 3.70 &  4 \\
16 & 2020-09-20T07:19:45 & 2459112.803121 & 4.0337 & 903 & 1.40 & 173 & 1.05 &  1 \\
16 & 2020-09-20T07:35:15 & 2459112.813894 & 4.0184 & 903 & 1.39 & 176 & 1.05 &  2 \\
16 & 2020-09-20T07:50:46 & 2459112.824666 & 4.0029 & 903 & 1.38 & 176 & 1.03 &  3 \\
16 & 2020-09-20T08:06:11 & 2459112.835377 & 3.9874 & 903 & 1.37 & 181 & 1.02 &  4 \\
17 & 2020-09-25T08:25:19 & 2459117.848721 & 2.9731 & 903 & 1.37 & 191 & 1.00 &  1 \\
17 & 2020-09-25T08:40:50 & 2459117.859493 & 2.9576 & 903 & 1.38 & 190 & 0.98 &  2 \\
17 & 2020-09-25T08:56:21 & 2459117.870265 & 2.9422 & 903 & 1.39 & 187 & 0.97 &  3 \\
17 & 2020-09-25T09:11:46 & 2459117.880974 & 2.9271 & 903 & 1.40 & 187 & 0.99 &  4 \\
18 & 2020-09-30T06:46:01 & 2459122.779804 & 2.0570 & 903 & 1.40 & 196 & 2.92 &  1 \\
18 & 2020-09-30T07:01:32 & 2459122.790576 & 2.0415 & 903 & 1.38 & 194 & 3.02 &  2 \\
18 & 2020-09-30T07:17:03 & 2459122.801349 & 2.0259 & 903 & 1.38 & 195 & 3.03 &  3 \\
18 & 2020-09-30T07:21:35 & 2459122.808283 & 2.0158 & 251 & 1.37 & 94 & 3.10 &  4 \\
18 & 2020-09-30T07:38:41 & 2459122.816379 & 2.0040 & 903 & 1.37 & 193 & 3.03 &  4 \\
19 & 2020-10-01T08:48:52 & 2459123.865126 & 1.7285 & 903 & 1.40 & 185 & 1.65 &  1 \\
19 & 2020-10-01T09:04:23 & 2459123.875899 & 1.7136 & 903 & 1.42 & 185 & 1.53 &  2 \\
19 & 2020-10-01T09:19:54 & 2459123.886672 & 1.6990 & 903 & 1.44 & 184 & 1.51 &  3 \\
19 & 2020-10-01T09:35:19 & 2459123.897379 & 1.6850 & 903 & 1.47 & 186 & 1.60 &  4 \\
20 & 2020-10-06T09:43:57 & 2459128.903400 & 0.6440 & 903 & 1.53 & 155 & 3.72 &  1 \\
20 & 2020-10-06T09:59:23 & 2459128.914108 & 0.6314 & 903 & 1.57 & 172 & 3.93 &  2 \\
20 & 2020-10-06T10:14:48 & 2459128.924816 & 0.6193 & 903 & 1.62 & 175 & 4.03 &  3 \\
20 & 2020-10-06T10:30:13 & 2459128.935525 & 0.6079 & 903 & 1.68 & 173 & 4.07 &  4 \\
21 & 2020-11-01T06:28:57 & 2459154.767808 & -4.5657 & 903 & 1.39 & 202 & 2.11 &  1 \\
21 & 2020-11-01T06:44:27 & 2459154.778581 & -4.5809 & 903 & 1.40 & 203 & 2.01 &  2 \\
21 & 2020-11-01T06:59:53 & 2459154.789288 & -4.5957 & 903 & 1.42 & 203 & 1.98 &  3 \\
21 & 2020-11-01T07:15:18 & 2459154.799996 & -4.6101 & 903 & 1.44 & 205 & 2.08 &  4 \\
22 & 2020-11-04T07:38:13 & 2459157.815865 & -5.2182 & 903 & 1.50 & 198 & 6.57 &  1 \\
22 & 2020-11-04T07:53:44 & 2459157.826638 & -5.2312 & 903 & 1.54 & 201 & 6.65 &  2 \\
22 & 2020-11-04T08:09:14 & 2459157.837410 & -5.2436 & 903 & 1.59 & 170 & 6.63 &  3 \\
22 & 2020-11-04T08:24:40 & 2459157.848118 & -5.2554 & 903 & 1.64 & 192 & 6.74 &  4 \\
23 & 2020-12-24T04:48:50 & 2459207.696740 & -11.9027 & 903 & 1.57 & 190 & 3.90 &  1 \\
23 & 2020-12-24T05:04:21 & 2459207.707512 & -11.9133 & 903 & 1.62 & 189 & 4.05 &  2 \\
23 & 2020-12-24T05:19:52 & 2459207.718288 & -11.9231 & 903 & 1.68 & 187 & 4.24 &  3 \\
23 & 2020-12-24T05:35:17 & 2459207.728992 & -11.9322 & 903 & 1.74 & 186 & 4.50 &  4 \\
24 & 2020-12-29T04:41:13 & 2459212.691250 & -12.1435 & 903 & 1.61 & 174 & 1.87 &  1 \\
24 & 2020-12-29T04:56:44 & 2459212.702022 & -12.1533 & 903 & 1.67 & 172 & 1.95 &  2 \\
24 & 2020-12-29T05:12:14 & 2459212.712794 & -12.1625 & 903 & 1.73 & 170 & 2.04 &  3 \\
24 & 2020-12-29T05:27:40 & 2459212.723502 & -12.1709 & 903 & 1.80 & 164 & 2.18 &  4 \\
25 & 2021-01-03T04:53:28 & 2459217.699560 & -12.3062 & 903 & 1.73 & 179 & 0.57 &  1 \\
25 & 2021-01-03T05:09:05 & 2459217.710396 & -12.3145 & 903 & 1.80 & 178 & 0.59 &  2 \\
25 & 2021-01-03T05:24:35 & 2459217.721168 & -12.3219 & 903 & 1.89 & 187 & 0.64 &  3 \\
25 & 2021-01-03T05:40:01 & 2459217.731876 & -12.3285 & 903 & 1.98 & 187 & 0.68 &  4 \\
26 & 2021-06-19T11:47:11 & 2459384.985854 & 11.2191 & 903 & 1.60 & 129 & 3.22 &  1 \\
26 & 2021-06-19T12:02:36 & 2459384.996561 & 11.2091 & 903 & 1.56 & 126 & 2.87 &  2 \\
26 & 2021-06-19T12:18:02 & 2459385.007269 & 11.1986 & 903 & 1.52 & 114 & 2.66 &  3 \\
26 & 2021-06-19T12:28:09 & 2459385.016155 & 11.1896 & 585 & 1.48 & 10 & 3.54 &  4 \\
27 & 2021-06-20T11:32:56 & 2459385.975993 & 11.2903 & 903 & 1.64 & 195 & 1.54 &  1 \\
27 & 2021-06-20T11:48:21 & 2459385.986702 & 11.2807 & 903 & 1.59 & 197 & 1.54 &  2 \\
27 & 2021-06-20T12:03:47 & 2459385.997411 & 11.2706 & 903 & 1.54 & 196 & 1.48 &  3 \\
27 & 2021-06-20T12:19:12 & 2459386.008120 & 11.2599 & 903 & 1.51 & 199 & 1.42 &  4 \\
28 & 2021-06-21T11:20:22 & 2459386.967298 & 11.3572 & 903 & 1.67 & 190 & 2.38 &  1 \\
28 & 2021-06-21T11:35:52 & 2459386.978069 & 11.3479 & 903 & 1.62 & 183 & 2.28 &  2 \\
28 & 2021-06-21T11:51:23 & 2459386.988843 & 11.3379 & 903 & 1.57 & 177 & 2.23 &  3 \\
28 & 2021-06-21T12:06:54 & 2459386.999615 & 11.3274 & 903 & 1.53 & 181 & 2.22 &  4 \\
29 & 2021-06-23T11:17:24 & 2459388.965321 & 11.4687 & 903 & 1.65 & 204 & 4.03 &  1 \\
29 & 2021-06-23T11:32:50 & 2459388.976029 & 11.4591 & 903 & 1.60 & 199 & 3.96 &  2 \\
29 & 2021-06-23T11:48:15 & 2459388.986738 & 11.4490 & 903 & 1.55 & 197 & 3.83 &  3 \\
29 & 2021-06-23T12:03:40 & 2459388.997446 & 11.4383 & 903 & 1.52 & 190 & 3.73 &  4 \\
30 & 2021-06-24T11:00:56 & 2459389.953915 & 11.5287 & 903 & 1.70 & 186 & 2.90 &  1 \\
30 & 2021-06-24T11:16:21 & 2459389.964623 & 11.5196 & 903 & 1.64 & 196 & 2.86 &  2 \\
30 & 2021-06-24T11:31:46 & 2459389.975331 & 11.5099 & 903 & 1.59 & 188 & 2.64 &  3 \\
30 & 2021-06-24T11:47:11 & 2459389.986039 & 11.4996 & 903 & 1.55 & 188 & 2.62 &  4 \\
31 & 2021-06-25T11:22:27 & 2459390.968894 & 11.5631 & 903 & 1.61 & 173 & 1.71 &  1 \\
31 & 2021-06-25T11:37:52 & 2459390.979601 & 11.5530 & 903 & 1.56 & 180 & 1.65 &  2 \\
31 & 2021-06-25T11:53:22 & 2459390.990377 & 11.5422 & 903 & 1.52 & 179 & 1.62 &  3 \\
31 & 2021-06-25T12:08:48 & 2459391.001083 & 11.5311 & 903 & 1.49 & 186 & 1.58 &  4 \\
32 & 2021-06-26T11:16:33 & 2459391.964841 & 11.6111 & 903 & 1.61 & 206 & 1.61 &  1 \\
32 & 2021-06-26T11:31:58 & 2459391.975550 & 11.6010 & 903 & 1.57 & 204 & 1.52 &  2 \\
32 & 2021-06-26T11:47:23 & 2459391.986258 & 11.5904 & 903 & 1.53 & 204 & 1.51 &  3 \\
32 & 2021-06-26T12:02:49 & 2459391.996970 & 11.5793 & 903 & 1.49 & 205 & 1.49 &  4 \\
33 & 2021-06-27T10:55:09 & 2459392.950018 & 11.6656 & 903 & 1.68 & 209 & 3.07 &  1 \\
33 & 2021-06-27T11:10:34 & 2459392.960726 & 11.6562 & 903 & 1.62 & 209 & 2.95 &  2 \\
33 & 2021-06-27T11:25:59 & 2459392.971435 & 11.6462 & 903 & 1.57 & 210 & 2.91 &  3 \\
33 & 2021-06-27T11:41:25 & 2459392.982143 & 11.6356 & 903 & 1.53 & 208 & 2.97 &  4 \\
34 & 2021-06-28T13:00:03 & 2459394.036796 & 11.6124 & 903 & 1.39 & 194 & 3.02 &  1 \\
34 & 2021-06-28T13:15:28 & 2459394.047502 & 11.5996 & 903 & 1.38 & 193 & 2.96 &  2 \\
34 & 2021-06-28T13:30:59 & 2459394.058275 & 11.5865 & 903 & 1.37 & 189 & 2.98 &  3 \\
34 & 2021-06-28T13:46:24 & 2459394.068984 & 11.5735 & 903 & 1.37 & 195 & 3.01 &  4 \\
35 & 2021-06-29T13:25:50 & 2459395.054740 & 11.6250 & 903 & 1.37 & 186 & 1.05 &  1 \\
35 & 2021-06-29T13:41:21 & 2459395.065513 & 11.6119 & 903 & 1.37 & 184 & 1.05 &  2 \\
35 & 2021-06-29T13:56:46 & 2459395.076222 & 11.5988 & 903 & 1.37 & 182 & 1.05 &  3 \\
35 & 2021-06-29T14:12:11 & 2459395.086931 & 11.5857 & 903 & 1.37 & 184 & 1.08 &  4 \\
36 & 2021-07-01T13:25:12 & 2459397.054380 & 11.6844 & 903 & 1.37 & 167 & 1.77 &  1 \\
36 & 2021-07-01T13:40:37 & 2459397.065089 & 11.6712 & 903 & 1.37 & 164 & 1.91 &  2 \\
36 & 2021-07-01T13:56:03 & 2459397.075796 & 11.6580 & 903 & 1.37 & 162 & 1.97 &  3 \\
36 & 2021-07-01T14:11:28 & 2459397.086505 & 11.6450 & 903 & 1.37 & 169 & 2.05 &  4 \\
37 & 2021-07-02T13:01:51 & 2459398.038204 & 11.7288 & 903 & 1.38 & 145 & 1.40 &  1 \\
37 & 2021-07-02T13:17:16 & 2459398.048910 & 11.7157 & 903 & 1.37 & 140 & 1.39 &  2 \\
37 & 2021-07-02T13:32:47 & 2459398.059683 & 11.7024 & 903 & 1.37 & 146 & 1.40 &  3 \\
37 & 2021-07-02T13:48:12 & 2459398.070391 & 11.6892 & 903 & 1.37 & 152 & 1.43 &  4 \\
38 & 2021-07-17T12:13:34 & 2459413.005251 & 11.7312 & 903 & 1.37 & 188 & 0.68 &  1 \\
38 & 2021-07-17T12:28:59 & 2459413.015960 & 11.7175 & 903 & 1.37 & 188 & 0.68 &  2 \\
38 & 2021-07-17T12:44:29 & 2459413.026733 & 11.7036 & 903 & 1.37 & 176 & 0.68 &  3 \\
38 & 2021-07-17T12:59:55 & 2459413.037441 & 11.6899 & 903 & 1.37 & 174 & 0.69 &  4 \\
39 & 2021-07-18T12:08:58 & 2459414.002104 & 11.7050 & 903 & 1.38 & 182 & 1.00 &  1 \\
39 & 2021-07-18T12:24:24 & 2459414.012813 & 11.6913 & 903 & 1.37 & 176 & 0.98 &  2 \\
39 & 2021-07-18T12:39:54 & 2459414.023586 & 11.6774 & 903 & 1.37 & 171 & 1.00 &  3 \\
39 & 2021-07-18T12:55:19 & 2459414.034294 & 11.6637 & 903 & 1.37 & 176 & 1.02 &  4 \\
40 & 2021-07-19T09:49:53 & 2459414.905548 & 11.7856 & 903 & 1.60 & 164 & 4.44 &  1 \\
40 & 2021-07-19T10:05:18 & 2459414.916257 & 11.7746 & 903 & 1.56 & 178 & 4.53 &  2 \\
40 & 2021-07-19T10:20:43 & 2459414.926965 & 11.7630 & 903 & 1.52 & 155 & 4.75 &  3 \\
40 & 2021-07-19T10:36:08 & 2459414.937673 & 11.7510 & 903 & 1.48 & 142 & 4.67 &  4 \\
41 & 2021-07-20T11:41:46 & 2459415.983295 & 11.6592 & 903 & 1.39 & 198 & 4.94 &  1 \\
41 & 2021-07-20T11:57:12 & 2459415.994003 & 11.6455 & 903 & 1.38 & 191 & 4.84 &  2 \\
41 & 2021-07-20T12:12:37 & 2459416.004711 & 11.6317 & 903 & 1.37 & 196 & 4.71 &  3 \\
41 & 2021-07-20T12:28:02 & 2459416.015420 & 11.6178 & 903 & 1.37 & 196 & 4.73 &  4 \\
42 & 2021-07-25T11:33:08 & 2459420.977489 & 11.4384 & 903 & 1.38 & 141 & 2.51 &  1 \\
42 & 2021-07-25T11:48:33 & 2459420.988198 & 11.4245 & 903 & 1.37 & 126 & 2.49 &  2 \\
42 & 2021-07-25T12:03:59 & 2459420.998906 & 11.4105 & 903 & 1.37 & 127 & 2.51 &  3 \\
42 & 2021-07-25T12:19:24 & 2459421.009615 & 11.3965 & 903 & 1.37 & 134 & 2.54 &  4 \\
43 & 2021-07-27T12:44:52 & 2459423.027380 & 11.2608 & 903 & 1.38 & 145 & 2.39 &  1 \\
43 & 2021-07-27T13:00:23 & 2459423.038152 & 11.2470 & 903 & 1.39 & 178 & 2.52 &  2 \\
43 & 2021-07-27T13:15:54 & 2459423.048926 & 11.2335 & 903 & 1.41 & 193 & 2.59 &  3 \\
43 & 2021-07-27T13:31:19 & 2459423.059633 & 11.2204 & 903 & 1.43 & 186 & 2.49 &  4 \\
44 & 2021-08-13T07:19:30 & 2459439.802032 & 10.0732 & 903 & 1.81 & 190 & 1.07 &  1 \\
44 & 2021-08-13T07:34:56 & 2459439.812741 & 10.0635 & 903 & 1.74 & 189 & 1.04 &  2 \\
44 & 2021-08-13T07:50:21 & 2459439.823448 & 10.0531 & 903 & 1.68 & 183 & 1.00 &  3 \\
44 & 2021-08-13T08:05:46 & 2459439.834157 & 10.0420 & 903 & 1.62 & 190 & 0.98 &  4 \\
45 & 2021-08-14T07:57:00 & 2459440.828101 & 9.9351 & 903 & 1.64 & 197 & 1.25 &  1 \\
45 & 2021-08-14T08:12:25 & 2459440.838810 & 9.9236 & 903 & 1.59 & 195 & 1.21 &  2 \\
45 & 2021-08-14T08:27:50 & 2459440.849517 & 9.9115 & 903 & 1.54 & 192 & 1.18 &  3 \\
45 & 2021-08-14T08:43:15 & 2459440.860226 & 9.8989 & 903 & 1.51 & 196 & 1.16 &  4 \\
46 & 2021-08-15T07:26:41 & 2459441.807078 & 9.8401 & 903 & 1.74 & 200 & 0.99 &  1 \\
46 & 2021-08-15T07:42:06 & 2459441.817787 & 9.8296 & 903 & 1.68 & 198 & 0.95 &  2 \\
46 & 2021-08-15T07:57:31 & 2459441.828494 & 9.8185 & 903 & 1.62 & 196 & 0.94 &  3 \\
46 & 2021-08-15T08:12:56 & 2459441.839203 & 9.8068 & 903 & 1.57 & 197 & 0.92 &  4 \\
47 & 2021-08-17T08:17:58 & 2459443.842764 & 9.5619 & 903 & 1.54 & 196 & 1.12 &  1 \\
47 & 2021-08-17T08:33:29 & 2459443.853537 & 9.5491 & 903 & 1.50 & 199 & 1.10 &  2 \\
47 & 2021-08-17T08:48:54 & 2459443.864246 & 9.5359 & 903 & 1.47 & 197 & 1.11 &  3 \\
47 & 2021-08-17T09:04:20 & 2459443.874954 & 9.5223 & 903 & 1.44 & 194 & 1.08 &  4 \\
48 & 2021-08-18T11:15:05 & 2459444.965794 & 9.2739 & 903 & 1.38 & 190 & 1.88 &  1 \\
48 & 2021-08-18T11:30:35 & 2459444.976560 & 9.2594 & 903 & 1.39 & 187 & 1.90 &  2 \\
48 & 2021-08-18T11:46:01 & 2459444.987268 & 9.2452 & 903 & 1.41 & 189 & 1.79 &  3 \\
48 & 2021-08-18T12:01:26 & 2459444.997977 & 9.2314 & 903 & 1.42 & 177 & 1.76 &  4 \\
49 & 2021-08-20T08:40:37 & 2459446.858585 & 9.1616 & 903 & 1.46 & 141 & 1.74 &  1 \\
49 & 2021-08-20T08:56:08 & 2459446.869358 & 9.1477 & 903 & 1.44 & 127 & 1.71 &  2 \\
49 & 2021-08-20T09:11:39 & 2459446.880131 & 9.1336 & 903 & 1.42 & 110 & 1.72 &  3 \\
49 & 2021-08-20T09:27:04 & 2459446.890842 & 9.1192 & 903 & 1.40 & 110 & 1.70 &  4 \\
50 & 2021-08-21T09:39:12 & 2459447.899294 & 8.9753 & 903 & 1.38 & 181 & 1.71 &  1 \\
50 & 2021-08-21T09:54:37 & 2459447.910002 & 8.9606 & 903 & 1.38 & 183 & 1.75 &  2 \\
50 & 2021-08-21T10:10:08 & 2459447.920780 & 8.9456 & 903 & 1.37 & 182 & 1.82 &  3 \\
50 & 2021-08-21T10:25:33 & 2459447.931483 & 8.9308 & 903 & 1.37 & 180 & 1.84 &  4 \\
51 & 2021-08-22T09:18:38 & 2459448.885043 & 8.8600 & 903 & 1.40 & 145 & 1.37 &  1 \\
51 & 2021-08-22T09:34:03 & 2459448.895751 & 8.8454 & 903 & 1.39 & 157 & 1.36 &  2 \\
51 & 2021-08-22T09:49:34 & 2459448.906523 & 8.8305 & 903 & 1.38 & 171 & 1.35 &  3 \\
51 & 2021-08-22T10:04:59 & 2459448.917232 & 8.8157 & 903 & 1.37 & 181 & 1.33 &  4 \\
52 & 2021-08-23T08:59:05 & 2459449.871499 & 8.7411 & 903 & 1.42 & 191 & 1.63 &  1 \\
52 & 2021-08-23T09:14:31 & 2459449.882208 & 8.7267 & 903 & 1.40 & 188 & 1.63 &  2 \\
52 & 2021-08-23T09:30:01 & 2459449.892980 & 8.7120 & 903 & 1.39 & 195 & 1.62 &  3 \\
52 & 2021-08-23T09:45:26 & 2459449.903689 & 8.6972 & 903 & 1.38 & 192 & 1.59 &  4 \\
53 & 2021-08-25T08:16:06 & 2459451.841702 & 8.4989 & 903 & 1.47 & 187 & 2.12 &  1 \\
53 & 2021-08-25T08:31:31 & 2459451.852410 & 8.4851 & 903 & 1.45 & 191 & 2.03 &  2 \\
53 & 2021-08-25T08:47:02 & 2459451.863183 & 8.4709 & 903 & 1.42 & 189 & 1.90 &  3 \\
53 & 2021-08-25T09:02:27 & 2459451.873892 & 8.4565 & 903 & 1.40 & 192 & 1.77 &  4 \\
54 & 2021-08-26T08:05:17 & 2459452.834222 & 8.3643 & 903 & 1.49 & 200 & 0.73 &  1 \\
54 & 2021-08-26T08:20:43 & 2459452.844930 & 8.3507 & 903 & 1.46 & 196 & 0.71 &  2 \\
54 & 2021-08-26T08:36:13 & 2459452.855703 & 8.3366 & 903 & 1.43 & 194 & 0.70 &  3 \\
54 & 2021-08-26T08:51:39 & 2459452.866414 & 8.3223 & 903 & 1.41 & 194 & 0.69 &  4 \\
\end{longtable}

\begin{longtable}{ccccc}
\caption{CCF and LBL SPIRou radial velocities minus the median velocities of -61072.6 and -60992.9~\ms, respectively.} \\
\hline\hline
\label{tab:spirouRVSCCF+LBL}
BJD & CCF RV & $\sigma_{\rm CCF}$ & LBL RV  & $\sigma_{\rm LBL}$\\
 & \ms & \ms & \ms & \ms \\
\hline
\endfirsthead
\caption{continued.}\\
\hline\hline
BJD & CCF RV & $\sigma_{\rm CCF}$ & LBL RV  & $\sigma_{\rm LBL}$\\
 & \ms & \ms & \ms & \ms \\
\hline
\endhead
\hline
\endfoot
2459006.047578 & -6.3 & 18.7 & 2.6 & 6.4 \\ 
2459006.058353 & 0.9 & 19.0 & 3.8 & 6.4 \\ 
2459006.069133 & 6.0 & 18.6 & 0.1 & 6.3 \\ 
2459006.079901 & 2.2 & 18.5 & 2.1 & 6.2 \\ 
2459011.055684 & -15.1 & 19.3 & -6.2 & 6.9 \\ 
2459011.066528 & -3.6 & 18.9 & -1.1 & 6.8 \\ 
2459011.077292 & -18.5 & 17.5 & -7.5 & 6.8 \\ 
2459011.088067 & -13.3 & 16.9 & -9.6 & 6.7 \\ 
2459058.086711 & 34.0 & 19.3 & 18.9 & 6.7 \\ 
2459058.097483 & 18.4 & 8.9 & 9.7 & 6.8 \\ 
2459058.108256 & 8.8 & 9.1 & 8.3 & 6.7 \\ 
2459058.119030 & 20.7 & 19.1 & -3.8 & 6.7 \\ 
2459058.973896 & 7.8 & 6.0 & 3.8 & 5.9 \\ 
2459058.984605 & 0.2 & 6.1 & 6.1 & 5.8 \\ 
2459058.995313 & -0.5 & 5.9 & 1.3 & 5.9 \\ 
2459059.006021 & -1.4 & 7.5 & -3.4 & 5.9 \\ 
2459071.942673 & -3.1 & 6.3 & -7.6 & 6.3 \\ 
2459071.953381 & -9.5 & 6.9 & -11.6 & 6.1 \\ 
2459071.964091 & -7.1 & 5.4 & -6.6 & 6.1 \\ 
2459071.974799 & 0.8 & 6.6 & -3.0 & 6.0 \\ 
2459072.974483 & -8.8 & 6.3 & -9.0 & 6.1 \\ 
2459072.985191 & -0.1 & 5.9 & -2.4 & 6.1 \\ 
2459072.995964 & -3.5 & 5.2 & -4.5 & 6.1 \\ 
2459073.006672 & 6.8 & 5.6 & -0.6 & 6.2 \\ 
2459087.884529 & 1.3 & 7.6 & - & - \\ 
2459087.895238 & 12.2 & 6.9 & - & -  \\ 
2459087.909300 & 2.5 & 7.8 & - & - \\ 
2459087.920331 & -7.6 & 7.7 & - & - \\ 
2459088.877370 & -5.4 & 7.0 & -2.0 & 6.3 \\ 
2459088.888143 & 8.4 & 6.9 & 4.8 & 6.1 \\ 
2459088.898851 & -1.1 & 6.8 & 2.5 & 6.1 \\ 
2459088.909560 & -0.4 & 8.9 & -5.6 & 5.8 \\ 
2459091.873371 & 1.5 & 5.8 & -2.2 & 5.9 \\ 
2459091.884080 & 6.9 & 5.6 & 8.1 & 5.8 \\ 
2459091.894788 & 4.5 & 4.1 & 1.7 & 5.8 \\ 
2459091.905497 & 2.9 & 5.4 & -2.8 & 5.8 \\ 
2459095.971094 & 3.1 & 5.7 & 1.8 & 5.9 \\ 
2459095.981867 & -4.1 & 4.3 & -0.0 & 5.7 \\ 
2459095.992640 & 0.9 & 6.7 & -1.5 & 6.0 \\ 
2459096.003412 & -1.0 & 5.4 & 0.2 & 5.9 \\ 
2459096.945095 & -2.6 & 7.9 & -7.1 & 6.3 \\ 
2459096.955868 & 0.5 & 8.4 & 1.6 & 6.3 \\ 
2459096.966576 & -1.2 & 8.8 & -6.1 & 6.3 \\ 
2459096.977284 & 0.1 & 7.8 & -2.1 & 6.3 \\ 
2459098.882192 & 6.9 & 5.6 & -5.9 & 6.1 \\ 
2459098.892901 & 11.1 & 7.8 & 1.8 & 6.1 \\ 
2459098.903674 & 9.4 & 6.1 & -0.9 & 5.9 \\ 
2459098.914382 & 6.6 & 4.5 & 4.5 & 5.8 \\ 
2459101.862503 & 19.9 & 5.6 & 12.9 & 6.0 \\ 
2459101.873276 & 11.2 & 4.6 & -1.5 & 6.0 \\ 
2459101.884049 & 9.0 & 4.9 & 6.5 & 5.9 \\ 
2459101.894757 & 3.0 & 5.6 & 0.2 & 6.1 \\ 
2459102.871592 & 9.6 & 7.2 & 2.6 & 6.0 \\ 
2459102.882301 & -7.3 & 5.7 & -1.5 & 6.2 \\ 
2459102.893074 & 12.5 & 7.2 & -1.5 & 6.0 \\ 
2459102.903846 & 11.7 & 6.3 & 2.7 & 6.0 \\ 
2459111.815780 & 17.9 & 10.8 & - & - \\ 
2459111.826488 & 2.2 & 11.6 & - & - \\ 
2459111.837262 & 15.6 & 10.1 & - & - \\ 
2459111.847968 & 42.7 & 23.0 & - & - \\ 
2459112.803121 & 5.8 & 6.6 & 8.3 & 6.1 \\ 
2459112.813894 & 10.4 & 6.3 & 4.1 & 6.0 \\ 
2459112.824666 & -7.0 & 5.6 & -1.6 & 6.0 \\ 
2459112.835377 & 9.3 & 7.1 & 2.2 & 5.9 \\ 
2459117.848721 & -3.5 & 7.9 & -2.6 & 5.8 \\ 
2459117.859493 & -4.8 & 7.1 & -3.5 & 5.8 \\ 
2459117.870265 & -5.9 & 7.6 & -0.1 & 5.8 \\ 
2459117.880974 & -13.4 & 7.6 & -4.8 & 5.8 \\ 
2459122.779804 & 5.6 & 6.1 & 8.4 & 6.3 \\ 
2459122.790576 & -4.3 & 8.6 & 2.4 & 6.2 \\ 
2459122.801349 & 14.0 & 13.5 & 3.4 & 6.2 \\ 
2459122.808283 & 6.7 & 11.2 & - & - \\ 
2459122.816379 & -3.9 & 6.1 & 2.3 & 6.2 \\ 
2459123.865126 & 3.8 & 7.4 & 0.7 & 5.9 \\ 
2459123.875899 & -10.4 & 7.1 & 1.6 & 6.0 \\ 
2459123.886672 & -2.5 & 6.4 & 2.5 & 6.0 \\ 
2459123.897379 & -17.7 & 7.0 & -8.3 & 5.9 \\ 
2459128.903400 & -17.8 & 9.5 & 2.1 & 6.6 \\ 
2459128.914108 & -4.3 & 8.6 & 9.6 & 6.3 \\ 
2459128.924816 & -8.1 & 8.2 & 6.8 & 6.2 \\ 
2459128.935525 & -4.6 & 14.5 & 1.7 & 6.2 \\ 
2459154.767808 & 1.7 & 9.8 & 7.8 & 5.7 \\ 
2459154.778581 & 0.7 & 9.6 & 4.7 & 5.7 \\ 
2459154.789288 & -1.8 & 9.6 & 2.2 & 5.7 \\ 
2459154.799996 & -5.4 & 11.7 & 7.7 & 5.7 \\ 
2459157.815865 & 7.6 & 10.9 & 13.2 & 6.0 \\ 
2459157.826638 & 10.4 & 14.3 & 6.3 & 6.2 \\ 
2459157.837410 & -2.5 & 15.8 & -2.0 & 6.7 \\ 
2459157.848118 & 3.2 & 15.0 & 0.7 & 6.4 \\ 
2459207.696740 & 1.0 & 8.9 & 4.4 & 6.4 \\ 
2459207.707512 & -9.1 & 10.7 & 4.1 & 6.4 \\ 
2459207.718288 & -5.5 & 10.5 & 4.2 & 6.4 \\ 
2459207.728992 & 15.5 & 23.5 & 3.5 & 6.2 \\ 
2459212.691250 & 20.0 & 9.0 & 3.4 & 6.3 \\ 
2459212.702022 & 8.6 & 10.4 & 6.2 & 6.3 \\ 
2459212.712794 & 0.5 & 8.1 & 9.7 & 6.3 \\ 
2459212.723502 & 17.2 & 21.6 & 6.5 & 6.5 \\ 
2459217.699560 & -8.6 & 7.6 & -5.5 & 6.0 \\ 
2459217.710396 & 1.8 & 9.4 & -6.4 & 6.0 \\ 
2459217.721168 & -2.5 & 8.9 & -5.6 & 5.9 \\ 
2459217.731876 & -7.3 & 7.0 & -6.7 & 5.9 \\ 
2459384.985854 & -20.9 & 8.7 & - & - \\ 
2459384.996561 & -14.0 & 8.3 & - & - \\ 
2459385.007269 & -16.6 & 11.8 & - & - \\ 
2459385.975993 & -8.9 & 7.3 & -0.2 & 6.0 \\ 
2459385.986702 & -12.6 & 7.5 & -3.9 & 5.9 \\ 
2459385.997411 & -9.2 & 6.2 & 0.6 & 5.9 \\ 
2459386.008120 & -10.5 & 6.6 & -5.8 & 5.9 \\ 
2459386.967298 & -1.2 & 7.3 & -4.8 & 5.9 \\ 
2459386.978069 & -6.5 & 6.9 & -7.1 & 6.0 \\ 
2459386.988843 & -0.0 & 8.9 & -2.3 & 6.3 \\ 
2459386.999615 & -5.0 & 6.6 & -8.7 & 6.0 \\ 
2459388.965321 & 4.5 & 6.7 & -1.2 & 6.2 \\ 
2459388.976029 & 10.8 & 8.6 & 2.7 & 6.1 \\ 
2459388.986738 & 15.7 & 18.9 & -1.1 & 6.1 \\ 
2459388.997446 & 11.3 & 17.3 & -1.4 & 6.2 \\ 
2459389.953915 & 2.6 & 7.3 & -1.5 & 6.2 \\ 
2459389.964623 & -11.1 & 7.0 & -8.4 & 6.0 \\ 
2459389.975331 & 4.4 & 7.3 & 5.8 & 6.0 \\ 
2459389.986039 & 1.3 & 6.2 & -5.3 & 6.4 \\ 
2459390.968894 & -1.3 & 6.3 & -4.7 & 6.3 \\ 
2459390.979601 & 10.4 & 5.8 & -3.6 & 6.3 \\ 
2459390.990377 & -13.2 & 5.7 & -5.3 & 6.2 \\ 
2459391.001083 & 8.2 & 5.2 & 0.5 & 6.3 \\ 
2459391.964841 & -1.3 & 4.4 & -0.9 & 6.2 \\ 
2459391.975550 & -7.2 & 5.2 & -3.5 & 6.1 \\ 
2459391.986258 & 5.6 & 5.3 & -0.1 & 6.1 \\ 
2459391.996970 & -8.6 & 5.0 & -7.0 & 6.1 \\ 
2459392.950018 & 13.8 & 17.3 & -1.2 & 6.6 \\ 
2459392.960726 & -10.0 & 6.6 & -1.4 & 6.3 \\ 
2459392.971435 & -6.5 & 7.5 & -0.5 & 6.3 \\ 
2459392.982143 & 2.2 & 6.9 & -0.6 & 6.1 \\ 
2459394.036796 & 8.5 & 14.1 & 5.8 & 6.3 \\ 
2459394.047502 & 7.1 & 14.5 & 1.0 & 6.2 \\ 
2459394.058275 & 3.0 & 15.2 & 3.6 & 6.1 \\ 
2459394.068984 & -1.4 & 14.1 & -0.6 & 6.3 \\ 
2459395.054740 & 9.0 & 6.3 & 6.1 & 5.7 \\ 
2459395.065513 & 3.9 & 4.1 & 2.1 & 5.8 \\ 
2459395.076222 & 9.1 & 6.0 & 3.6 & 5.8 \\ 
2459395.086931 & 4.4 & 5.5 & 0.3 & 5.6 \\ 
2459397.054380 & -0.6 & 6.2 & 8.3 & 6.1 \\ 
2459397.065089 & 0.0 & 5.0 & 3.5 & 6.2 \\ 
2459397.075796 & 4.6 & 5.6 & 8.3 & 6.3 \\ 
2459397.086505 & -0.7 & 5.4 & -2.6 & 6.2 \\ 
2459398.038204 & 15.0 & 5.8 & 11.0 & 6.9 \\ 
2459398.048910 & -0.8 & 6.3 & 5.8 & 7.2 \\ 
2459398.059683 & 5.4 & 4.9 & 2.5 & 6.8 \\ 
2459398.070391 & -4.6 & 6.3 & 2.1 & 6.7 \\ 
2459413.005251 & -6.6 & 6.4 & -8.3 & 5.6 \\ 
2459413.015960 & -13.2 & 4.6 & -12.1 & 5.6 \\ 
2459413.026733 & 2.8 & 5.7 & -8.2 & 5.7 \\ 
2459413.037441 & -3.3 & 6.3 & -4.9 & 5.8 \\ 
2459414.002104 & -2.0 & 5.3 & 3.7 & 5.8 \\ 
2459414.012813 & -6.1 & 6.2 & 4.0 & 5.8 \\ 
2459414.023586 & -10.1 & 6.2 & 2.7 & 6.0 \\ 
2459414.034294 & -0.9 & 4.5 & 0.3 & 5.9 \\ 
2459414.905548 & -10.9 & 7.7 & -12.7 & 6.7 \\ 
2459414.916257 & -8.8 & 7.7 & -11.5 & 6.3 \\ 
2459414.926965 & -13.3 & 8.0 & -15.1 & 7.0 \\ 
2459414.937673 & -2.1 & 16.9 & -18.1 & 7.3 \\ 
2459415.983295 & -2.5 & 8.2 & -2.8 & 6.2 \\ 
2459415.994003 & 12.8 & 17.3 & -1.4 & 6.1 \\ 
2459416.004711 & 1.0 & 18.3 & -10.7 & 6.1 \\ 
2459416.015420 & -2.6 & 8.6 & -7.3 & 6.1 \\ 
2459420.977489 & 5.5 & 7.0 & 1.7 & 7.0 \\ 
2459420.988198 & -2.2 & 7.3 & -6.3 & 7.7 \\ 
2459420.998906 & 6.1 & 7.4 & 7.2 & 7.6 \\ 
2459421.009615 & -4.1 & 7.2 & -3.3 & 7.3 \\ 
2459423.027380 & 12.8 & 7.8 & 4.0 & 6.9 \\ 
2459423.038152 & 6.6 & 4.8 & -0.3 & 6.0 \\ 
2459423.048926 & 5.0 & 5.4 & 1.3 & 5.8 \\ 
2459423.059633 & 11.3 & 5.7 & 8.0 & 5.7 \\ 
2459439.802032 & 9.8 & 7.1 & 6.3 & 5.9 \\ 
2459439.812741 & -6.3 & 7.5 & -4.1 & 5.9 \\ 
2459439.823448 & 1.7 & 6.2 & -1.7 & 5.8 \\ 
2459439.834157 & 8.4 & 5.7 & -0.8 & 5.8 \\ 
2459440.828101 & 10.4 & 6.7 & 5.3 & 6.2 \\ 
2459440.838810 & 8.1 & 5.2 & 1.1 & 6.1 \\ 
2459440.849517 & 9.8 & 5.9 & 4.7 & 6.1 \\ 
2459440.860226 & 12.8 & 5.6 & 2.8 & 6.0 \\ 
2459441.807078 & 1.2 & 6.9 & 0.0 & 6.1 \\ 
2459441.817787 & 2.5 & 6.0 & 0.4 & 6.0 \\ 
2459441.828494 & 3.5 & 5.0 & 4.2 & 6.0 \\ 
2459441.839203 & 4.6 & 6.8 & 3.4 & 6.0 \\ 
2459443.842764 & 2.5 & 5.8 & 0.7 & 5.8 \\ 
2459443.853537 & 6.7 & 5.7 & 4.2 & 5.8 \\ 
2459443.864246 & -0.1 & 6.0 & 0.3 & 5.7 \\ 
2459443.874954 & 7.1 & 6.5 & 0.2 & 5.6 \\ 
2459444.965794 & -1.6 & 7.3 & 0.5 & 5.9 \\ 
2459444.976560 & 1.1 & 7.7 & 1.4 & 5.8 \\ 
2459444.987268 & -2.6 & 7.9 & 0.9 & 5.9 \\ 
2459444.997977 & -6.8 & 6.8 & -0.7 & 5.9 \\ 
2459446.858585 & -4.6 & 7.8 & -7.9 & 7.1 \\ 
2459446.869358 & -3.2 & 6.8 & -3.9 & 7.6 \\ 
2459446.880131 & -7.1 & 8.0 & 1.2 & 8.6 \\ 
2459446.890842 & -26.7 & 9.2 & -8.2 & 8.5 \\ 
2459447.899294 & -0.4 & 6.8 & -3.2 & 5.7 \\ 
2459447.910002 & -3.6 & 5.7 & -1.7 & 5.6 \\ 
2459447.920780 & -6.2 & 5.9 & -11.4 & 5.7 \\ 
2459447.931483 & -8.8 & 5.9 & -4.0 & 5.8 \\ 
2459448.885043 & 2.8 & 22.8 & 3.7 & 6.5 \\ 
2459448.895751 & 9.3 & 22.7 & -0.4 & 6.1 \\ 
2459448.906523 & 10.1 & 8.0 & -0.8 & 5.8 \\ 
2459448.917232 & -1.1 & 7.4 & 0.7 & 5.6 \\ 
2459449.871499 & -23.4 & 6.5 & -10.8 & 5.7 \\ 
2459449.882208 & -3.6 & 5.9 & -2.3 & 5.7 \\ 
2459449.892980 & -6.4 & 6.1 & -5.6 & 5.8 \\ 
2459449.903689 & -9.7 & 10.0 & -9.4 & 5.6 \\ 
2459451.841702 & -5.0 & 4.7 & 1.0 & 5.8 \\ 
2459451.852410 & 2.6 & 5.1 & 0.7 & 5.9 \\ 
2459451.863183 & -8.0 & 6.0 & -9.4 & 5.8 \\ 
2459451.873892 & -6.5 & 6.6 & 2.0 & 5.8 \\ 
2459452.834222 & -9.7 & 5.1 & -4.7 & 5.7 \\ 
2459452.844930 & -19.1 & 6.2 & -11.6 & 5.8 \\ 
2459452.855703 & -11.1 & 5.6 & -9.8 & 5.4 \\ 
2459452.866414 & -10.8 & 5.9 & -7.9 & 5.6 \\ 
\hline
\end{longtable}

\begin{longtable}{ccrr}
\caption{Longitudinal magnetic field data of TOI-1759 measured by SPIRou calculated both from APERO and Libre-Esprit reduction.} \\
\hline\hline
\label{tab:toi1759spiroublongdata}
epoch & BJD &\multicolumn{1}{c}{$B_\ell$ (APERO)} &\multicolumn{1}{c}{$B_\ell$ (Libre-Esprit)}\\
    & \multicolumn{1}{c}{G}& \multicolumn{1}{c}{G}\\
\hline
\endfirsthead
\caption{continued.}\\
\hline\hline
epoch & BJD &\multicolumn{1}{c}{$B_\ell$ (APERO)} &\multicolumn{1}{c}{$B_\ell$ (Libre-Esprit)}\\
    & \multicolumn{1}{c}{G}& \multicolumn{1}{c}{G}\\
\hline
\endhead
\hline
\endfoot

1 & 2459006.0637411 & $-3.7\pm2.8$ & $-2.5\pm1.5$\\
2 & 2459011.0718929 & $-8.6\pm2.9$ & $-5.0\pm1.6$\\
3 & 2459058.1028702 & $0.8\pm3.1$ & $-0.5\pm1.6$\\
4 & 2459058.9899588 & $1.2\pm2.6$ & $2.0\pm1.4$ \\
5 & 2459071.9587360 & $0.1\pm2.7$ & $-4.0\pm1.4$\\
6 & 2459072.9905777 & $-5.2\pm2.7$ & $-2.3\pm1.4$\\
7 & 2459087.9004949 & - & $-1.0\pm2.1$ \\
8 & 2459088.8934810 & $4.3\pm2.5$ & $1.0\pm1.4$ \\
9 & 2459091.8894338 & $10.4\pm2.4$ & $5.7\pm1.3$ \\
10 & 2459095.9872534 & $-6.2\pm2.4$ & $-1.0\pm1.3$\\
11 & 2459096.9612056 & $2.5\pm3.1$ & $-3.4\pm1.4$\\
12 & 2459098.8982873 & $-6.6\pm2.7$ & $-5.0\pm1.3$ \\
13 & 2459101.8786459 & $-2.2\pm2.9$ & $-5.1\pm1.3$\\
14 & 2459102.8877032 & $-4.6\pm2.9$ & $-3.5\pm1.3$ \\
15 & 2459111.8268090 & - & $-0.9\pm3.5$ \\
16 & 2459112.8192644 & $-5.0\pm2.6$ & $-3.7\pm1.3$\\
17 & 2459117.8648631 & $-8.3\pm2.5$ & $-6.4\pm1.2$ \\
18 & 2459122.7950031 & $1.4\pm2.8$ & $-4.5\pm1.6$\\
19 & 2459123.8812691 & $-2.3\pm2.5$ & $-0.9\pm1.3$ \\
20 & 2459128.9194623 & $4.4\pm3.1$ & $4.3\pm1.5$ \\
21 & 2459154.7839181 & $-7.5\pm2.1$ & $-3.2\pm1.2$ \\
22 & 2459157.8320079 & $-5.4\pm2.8$ & $-4.1\pm1.4$  \\
23 & 2459207.7128831 & $-5.7\pm2.7$ & - \\
24 & 2459212.7073920 & $-3.1\pm2.9$ & - \\
25 & 2459217.7157500 & $-5.0\pm2.5$ & - \\
27 & 2459385.9920565 & $-4.8\pm2.6$  & $-2.0\pm1.3$ \\
28 & 2459386.9834565 & $-4.4\pm2.6$ & $-2.7\pm1.6$ \\
29 & 2459388.9813834 & $-8.1\pm2.7$ & $-5.5\pm1.3$ \\
30 & 2459389.9699772 & $-5.3\pm2.7$ & $-2.8\pm1.3$ \\
31 & 2459390.9849890 & $-7.5\pm2.7$ & $-7.0\pm1.3$ \\
32 & 2459391.9809046 & $-4.0\pm2.8$ & $-6.2\pm1.3$ \\
33 & 2459392.9660802 & $-12.1\pm2.9$ & $-8.7\pm1.3$ \\
34 & 2459394.0528894 & $-6.6\pm2.9$ & $-5.5\pm1.3$ \\
35 & 2459395.0708516 & $-7.7\pm2.7$ & $-8.1\pm1.3$ \\
36 & 2459397.0704424 & $-11.4\pm2.6$ & $-10.4\pm1.5$ \\
37 & 2459398.0542970 & $-10.1\pm3.1$ & - \\
38 & 2459413.0213462 & $5.4\pm2.4$ & $2.3\pm1.3$ \\
39 & 2459414.0181992 & $7.4\pm2.6$ & $5.1\pm1.4$ \\
40 & 2459414.9216105 & $4.6\pm2.8$ & $4.5\pm1.6$ \\
41 & 2459415.9993571 & $4.1\pm2.6$ & $3.9\pm1.3$ \\
42 & 2459420.9935519 & $0.5\pm3.3$ & $-0.8\pm2.0$ \\
43 & 2459423.0435229 & $-3.7\pm2.5$ & $-3.5\pm1.4$\\
44 & 2459439.8180944 & $-7.3\pm2.4$ & $-4.4\pm1.3$\\
45 & 2459440.8441636 & $-2.7\pm2.2$ & $-3.6\pm1.3$ \\
46 & 2459441.8231406 & $-4.2\pm2.5$ & $-1.0\pm1.3$ \\
47 & 2459443.8588750 & $1.4\pm2.2$ & $-0.4\pm1.2$  \\
48 & 2459444.9818996 & $2.0\pm2.8$ & $1.9\pm1.3$ \\
49 & 2459446.8747289 & $6.3\pm4.0$ & $-0.7\pm2.1$ \\
50 & 2459447.9153898 & $7.9\pm2.4$ & - \\
51 & 2459448.9011373 & $1.2\pm2.8$ & $1.0\pm1.6$  \\
52 & 2459449.8875939 & $3.8\pm2.6$ & $4.0\pm1.2$\\
53 & 2459451.8577968 & $3.1\pm2.6$ & $1.6\pm1.3$ \\
54 & 2459452.8503171 & $0.8\pm2.4$ & $-1.3\pm1.2$  \\
\hline
\end{longtable}

\end{appendix}

\end{document}